\documentclass[twocolumn,twoside,fleqn,showpacs,showkeys,pra,aps,10pt,superscriptaddress,final]{article}

\usepackage{hep}
\graphicspath{{figs/}}      

\def\refname{References}

\def\journalname{??}

\def\@pacs@name{PACS numbers: }%
\def\@keys@name{Keywords: }%

\def\Dated@name{Dated: }%
\def\Received@name{Received }%
\def\Revised@name{Revised }%
\def\Accepted@name{Accepted }%
\def\Published@name{Published }%
\def\address{\replace@command\address\affiliation}%
\def\altaddress{\replace@command\altaddress\altaffiliation}%

\definecolor{hs}{gray}{1}
\definecolor{orangec}{cmyk}{.15,.7,.96,.0}
\definecolor{oorangec}{cmyk}{.8,.2,.5,.4}
\definecolor{ooorangec}{cmyk}{1,.9,0.08,.04}      

\definecolor{orangecc}{cmyk}{.88,.49,.29,.0}
\definecolor{orangeccc}{cmyk}{.68,.60,.57,.07}

\newcommand{\HZsanhao}{\fontsize{15.5bp}{\baselineskip}\selectfont}
\newcommand{\liuhao}{\fontsize{8.5pt}{\baselineskip}\selectfont}
\newcommand{\Yihao}{\fontsize{18pt}{13.5pt}\selectfont}
\newcommand{\xiaosi}{\fontsize{11.5pt}{13.5pt}\selectfont} 

\newfont{\xbt}{cmb10 at 12pt}

\newcommand{\catchline}{%
\vspace*{-22.5mm}
\noindent
\hspace*{-3mm}
\begin{tabular*}{176mm}{lcr}
\parbox[t]{44mm}{\includegraphics{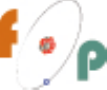}}
&
\raisebox{5.2mm}[0cm][0cm]{\parbox[t]{83mm}{\centering{\xiaosi\bf Frontiers of Physics}\\[1mm]
\liuhao\color{orangecc}{\doi{\doiurl}}}}
&
\raisebox{5mm}[0cm][0cm]{\parbox[t]{44mm}{\hfill \color{orangeccc}{{\journalname}}\par
\hfill {\color{orangeccc}{\volumenumber,~\issuenumber~(\volumeyear)}}}}\\[-1.4mm]
\multicolumn{3}{l}{\color{orangec}{\rule{\textwidth}{.5pt}}}\\[1mm]
\multicolumn{3}{l}{\Yihao\color{orangec}{\Papertype}}\\[5mm]
\end{tabular*}
}

\newcommand{\doi}[1]{https://doi.org/{#1}}
\renewcommand{\title}[1]
{\vspace*{-5mm}\begin{center}
{\HZsanhao\bf #1}
\end{center}
}

\renewcommand{\author}[1]
{\vspace*{0mm}
\begin{center}
{\bf #1}
\end{center}
}
\newcommand{\add}[1]{\begin{center}{\small\it #1}\end{center}}

\newcommand{\abs}[1]{
\begin{center}
\parbox[t]{156mm}{\noindent\color{oorangec}#1}
\end{center}}

\newcommand{\keywords}[1]{
\begin{center}
\parbox[t]{156mm}{\noindent{\bf\color{ooorangec}Keywords}\ \ #1}
\end{center}}

\newcommand{\acknowledgements}[1]{\vspace*{4mm}\noindent{\renewcommand{\baselinestretch}{1.05}\footnotesize{\color{ooorangec}\bf Acknowledgements}\quad{#1}}}

\def\journalname{??}
\def\volumenumber#1{\gdef\@volumenumber{#1}}%
\def\@volumenumber{}%
\def\issuenumber#1{\gdef\@issuenumber{#1}}%
\def\@issuenumber{}%
\def\volumeyear#1{\gdef\@volumeyear{#1}}%
\def\@volumeyear{}%

\renewcommand\thesection{\arabic{section}}
\renewcommand\thesubsection{\arabic{section}.\arabic{subsection}}
\renewcommand\thesubsubsection{\arabic{section}.\arabic{subsection}.\arabic{subsubsection}}

\titleformat{\section}[hang]{\color{ooorangec}\vspace*{-1.2mm}\titlerule\vspace{1mm}\large\usefont{T1}{fradmcn}{m}{n}\xbt}{\thesection}{1em}{}
\titlespacing{\section}{0mm}{8mm}{5mm}

\titleformat{\subsection}{\normalfont\normalsize\color{ooorangec}}{\thesubsection}{1em}{}
\titlespacing{\subsection}{0mm}{5mm}{3mm}

\titleformat{\subsubsection}{\normalfont\normalsize\it\color{ooorangec}}{\thesubsubsection}{1em}{}
\titlespacing{\subsubsection}{0mm}{3mm}{3mm}

\titlecontents{section}
[0em] {\normalfont} {\thecontentslabel\quad} {\thecontentslabel}
{\titlerule*[.7pc]{}\contentspage}

\titlecontents{subsection}
[1.5em] {\normalfont} {\thecontentslabel\quad} {\thecontentslabel}
{\titlerule*[.7pc]{}\contentspage}

\titlecontents{subsubsection}
[3.8em] {\normalfont} {\thecontentslabel\quad} {\thecontentslabel}
{\titlerule*[.7pc]{}\contentspage}

\usepackage[labelfont=bf,textfont={bf,it}]{caption}
\usepackage[textfont={small},justification=justified,singlelinecheck=off]{caption}
\newlength{\halfpagewidth}
\divide\halfpagewidth by 1

\makeatletter
\newskip\@footindent
\renewcommand\footnoterule{\kern6\p@ \hrule width 0.4\columnwidth \kern4\p@}
\@addtoreset{footnote}{page}

\long\def\@makefntext#1{\@setpar{\@@par\@tempdima \hsize
\advance\@tempdima-\@footindent
\parshape \@ne \@footindent \@tempdima}\par
\noindent\hspace*{-1.2mm} \hbox to
\z@{\hss\@thefnmark\hspace{0em}}#1}
\renewcommand\thefootnote{\myfootnotestyle{\arabic{footnote}}}

\renewcommand\thefootnote{\pinumber{\arabic{footnote}}}
\def\@makefnmark{\hbox{\textsuperscript{\@thefnmark}}}

\newcommand\pinumber[1]{$^{\arabic{footnote}}$}
\makeatother

\usepackage{upgreek}
\usepackage{amsfonts}
\usepackage{mathrsfs}
\usepackage{wasysym}
\usepackage{amsmath}
\usepackage{amssymb}
\usepackage{color}
\usepackage{graphicx}
\usepackage{xspace}
\usepackage{enumerate}
\usepackage{tikz}
\usepackage{pstricks}
\usepackage{hyperref}
\usepackage{dsfont}
\usepackage{lipsum,widetext}
\usepackage{cuted}

\usepackage{colortbl}
\definecolor{mygray}{cmyk}{0,0,0,0.3}

\usepackage{ntheorem}
\theoremstyle{plain}
{
\theoremseparator{.}

}

\begin{document}

\newcommand{\Papertype}{\sc Research article} 
\def\volumeyear{2021} 
\def\volumenumber{} 
\def\issuenumber{} 
\def\journalname{Front. Phys.} 
\newcommand{\doiurl}{10.1007/s11467-021-1048-y} 
\newcommand{\allauthors}{Gaole Dai} 
\catchline{}

\thispagestyle{firstpage}
\begin{strip}

\title{Designing nonlinear thermal devices and metamaterials under the Fourier's law: A route to nonlinear thermotics}

\author{Gaole Dai$^{\dag}$}

\add{School of Sciences, Nantong University, Nantong 226019, China\\
	Corresponding author.\ E-mail: $^\dag$gldai@ntu.edu.cn}

\abs{Nonlinear heat transfer can be exploited to reveal novel transport phenomena and thus enhance people's ability to manipulate heat flux at will. However, there hasn't been a mature discipline called nonlinear thermotics like its counterpart in optics or acoustics to make a systematic summary of relevant researches. In the current review, we focus on recent progress in an important part of nonlinear heat transfer, i.e., tailoring nonlinear thermal devices and metamaterials under the Fourier's law, especially with temperature-dependent thermal conductivities. We will present the basic designing techniques including solving the equation directly and the transformation theory. Tuning nonlinearity coming from multi-physical effects, and how to calculate effective properties of nonlinear conductive composites using the effective medium theory are also included.  Based on these theories, researchers have successfully designed various functional materials and devices such as the thermal diodes, thermal transistors, thermal memory elements, energy-free thermostats, and intelligent thermal materials, and some of them have also been realized in experiments. Further, these phenomenological works can provide a feasible route for the development of nonlinear thermotics.}



\end{strip}

\renewcommand{\thefootnote}{\fnsymbol{footnote}}
\renewcommand{\thefootnote}{$^{\arabic{footnote})}$}

\tableofcontents



\section{Introduction}

\noindent Nonlinear phenomena are ubiquitous and also very important in our world. A system is nonlinear means the feedback is not proportional to the input variables, which usually can be  governed by a or a set of nonlinear differential (or difference) equations, while some simple nonlinear systems just need an expression standing for the nonlinear relationship between input and output variables.  A well-known case of nonlinear system is a pendulum, whose dynamics can be linearized under small amplitude approximation~\cite{1}. Different from their linear counterparts, the superposition property of solutions ceases to hold for nonlinear equations~\cite{2}. Besides, nonlinear equations can be sensitive to initial conditions and thus difficult to be solved analytically or numerically. But, at the same time, new phenomena can occur due to the nonlinearity, including but not limited to chaos, bifurcation, shock waves, and solitons~\cite{1,2}. Exploiting nonlinearity has been quite successful in many basic fields of physics, such as optics~\cite{3}, mechanics and acoustics~\cite{4,5}, and in particular, semiconductor physics with the invention of fundamental electronic devices including diodes and transistors~\cite{6}.

Thermodynamics is another important field of physics, as heat implies the energy transport and conversion, which provides the material foundation for human life and social production. However, nonlinear phenomena in heat transfer haven't been systematically summarized as a mature discipline (called nonlinear thermotics) in theory like its optical/acoustical counterpart, and lack already well-established practical applications like the fundamental electronic devices which have changed people's life profoundly, although quite a lot efforts have been put into this field. Considering the increasing demand to manipulate heat transfer efficiently from macroscopic building temperature control to chip cooling at the nanoscale, nonlinearity in thermal phenomena needs more attention for the development of nonlinear thermotics in both theory and applications.

So, the first question might be what should be the central topic of nonlinear thermotics? In other words, what the nonlinearity means here? Since there are three basic mechanisms for heat transfer, i.e., thermal conduction, thermal convection, and thermal radiation, it's quite difficult to give a comprehensive answer at once. Instead, we could start from heat conduction and the conductive thermal diode (rectifier) as thermal rectification might be the most widely studied nonlinear thermal phenomenon especially in the conduction regime.

\subsection{Thermal conduction: The Fourier's law and beyond}

\noindent It's well known that heat conduction is driven by a spatial thermal bias, and usually can be described by the famous Fourier's law in bulk materials, i.e.,
\begin{equation}\label{fuliye}
\mathbf j=-\kappa \nabla T.
\end{equation}
The Fourier's law shows a relationship between the heat flux density vector $\mathbf j$ on a cross-section and temperature (denoted by $T$) gradient in space. The coefficient $\kappa$ in front of the temperature gradient is the thermal conductivity (conductance). With the law of continuity for heat flux and in the absence of an internal source, the evolution equation of the temperature is
\begin{equation}\label{fuliye2}
\rho C\frac {\partial  T}{\partial t}-\nabla\cdot\left(\kappa \nabla T\right)=0.
\end{equation}
Here $\rho$ is the density and $C$ is the specific heat capacity.
In steady cases, we can see thermal conductivity is the only material property to influence heat transfer if we don't consider inhomogeneous shapes which could change the surface area through which the heat is flowing. The nonlinearity of the heat conduction 
equation can come from a temperature-dependent density $\rho(T)$, specific heat capacity $C(T)$  and thermal conductivity $\kappa(T)$. We should notice that the nonlinearity of the equation is not equal to construct a nonlinear form (between $\mathbf j$ and $\nabla T$) of Eq.~(\ref{fuliye}). The change of $\rho(T)$ actually corresponds to thermal expansion and needs to take continuum mechanics into consideration. $C$ is usually taken as a constant, although the Debye model predicts $C \propto T^3$ at low temperatures~\cite{7}. Finally, $\kappa(T)$ might be the most used nonlinear property in researches. The specific expression of $\kappa(T)$ can have various forms. If we use the nonlinear polarization theory in nonlinear optics as an analogy, then $\kappa$ should depend on $\nabla T$ since the temperature/negative temperature gradient can corresponds to the electric potential/electric field. However, the conductivity of natural nonlinear materials usually shows a direct dependence on temperature, like $\kappa \propto T^{n}$ ($n$ is a real number). For example, physical kinetics gives $\kappa \propto T^{-1}$ (the Eucken's law) or $\kappa \propto T^3$ for dielectric materials at high/low temperatures, and $\kappa \propto \sqrt{T}$ for metals at low-temperatures~\cite{8,9}.

Fourier's law describes the heat conduction as a diffusive process in which the heat carriers (phonons in semiconductors and insulators, or electrons in metals) are able to collide multiple times, and implies an instantaneous response of thermal signals whose speed of propagation is infinite. There are two common types of deviation from Fourier's law: the anomalous heat conduction mainly about size effects, and the non-Fourier conduction with intrinsic wave nature.
When the mean free-path of the heat carriers is much smaller than the size of the material, Fourier's law is valid and $\kappa$ is not extensive as the size increases. However, in low-dimensional micro and nanoscales systems, the size's influence on $\kappa$ can be significant due to classical (e.g., effusion of highly rarefied gas) or quantum (e.g., tunneling) effects~\cite{10,11,12,13,111d,13b}. Harmonic/linear lattice is a simple model whose thermal conductivity is divergent under infinite periodicity since there's no phonon scattering and the energy transport is resistance-free~\cite{8}. Taking lattices as classical dynamic systems, researchers have been seeking for the condition to obtain a finite thermal conductivity in nonlinear/anharmonic lattices~\cite{11}. Here the nonlinearity can origin from the phonon scattering (e.g., the famous Fermi-Pasta-Ulam-Tsingou model~\cite{14}) or an external potential added to the Hamiltonian (e.g., the Frenkel-Kontorova lattice~\cite{15}). On the other hand, non-Fourier conduction  
happens when the time scale of heating (usually a pulse) is smaller than the relaxation time of heat carriers~\cite{10}. In these cases, modifications of classical Fourier's law include the Maxwell-Cattaneo equation, the Guyer-Krumhansl equation, the phase lag models and so on~\cite{16,17}, 
which change the heat equation from parabolic to hyperbolic and allow a temperature wave traveling at a finite speed, e.g., the second sound in helium II~\cite{18}. 

As will be seen below, nonlinear thermal transport like rectification are mainly studied under the Fourier's law (normal conduction) and the anomalous heat conduction. In anomalous heat conduction, an effective thermal conductivity as a function of the size and other shape parameters can be used in the heat equation, and sometimes the temperature-dependent conductivity $\kappa(T)$ is also seen as a kind of anomalous conduction. Anyway, a nonlinear heat equation should mainly refer to $\kappa(T)$ because other variables in $\kappa$ can't violate the superposition property although they can bring other effects such as spatial asymmetry.

\subsection{From thermal diodes to nonlinear thermal metamaterials}

\noindent A thermal diode or a thermal rectifier is a two-terminal device in which the magnitude of the heat flow should change when reversing the direction of the thermal bias (from the forward mode to the reverse mode). Such a rectification effect can be applied to solar energy collecting, passive cooling for buildings~\cite{19}, or constructing thermal logic circuits for a phonon computer with other components~\cite{12}.  Though the definition of the rectification ratio (or diodicity) $\gamma$ has some different versions, here we use 
\begin{equation}\label{ratio}
\gamma=\frac{\left|Q_{x}^{+}+Q_{x}^{-}\right|}{\text{Max}\left\{\left|Q_{x}^{+}\right|,\left|Q_{x}^{-}\right|\right\}},
\end{equation}
where $Q_{x}^{+}$ and $Q_{x}^{-}$ denote the directional fluxes in the forward and reverse modes respectively. Here the subscript $x$ implies the temperature bias is applied on the $x$ direction, and both $Q_{x}^{+}$ and $Q_{x}^{-}$ can be positive or negative according to their directions. How to calculate $Q_{x}^{\pm}$ also can cause some controversy. For a one-dimensional system in which the cross-section area vertical to the $x$ direction is the same everywhere, we can simply use the spatial average heat flux densities $\bar j_{x}^{+}$ and $\bar j_{x}^{-}$. In particular, if no internal heat source exists, it's obvious that $\bar j_{x}^{\pm}=j_{x}^{\pm}$ due to the continuity of heat flux.  In some other cases with a good symmetry about the $x$ axis, the structures can have a varying or graded cross-sectional area $S(x)$ and it's convenient to use the average flux volume $Q^{\pm}_x=\int_{0}^{L}  j^{\pm}_x(x)S(x) {\rm d}x/L $ in the calculation of rectification ratio. Here $L$ is the length of the whole system along the $x$ direction. For more general two/three-dimensional cases, it's even not easy to say ``applying the thermal bias along the $x$ direction" because the two terminals/ports to which the heat sources are attached can have different cross-section areas and might even be non-parallel. In these cases, the flux flowing out of the  system from one terminal might be another choice, which equals to the flux flowing into the system form the other terminal in the absence of internal sources. For simplicity, in the following contents, we prefer the notation $\bar j_{x}^{\pm}$ used in   one-dimensional systems if no extra statement about the structure is needed. 

\begin{figure}[!ht]
	\centering
	\includegraphics[width=1.\linewidth]{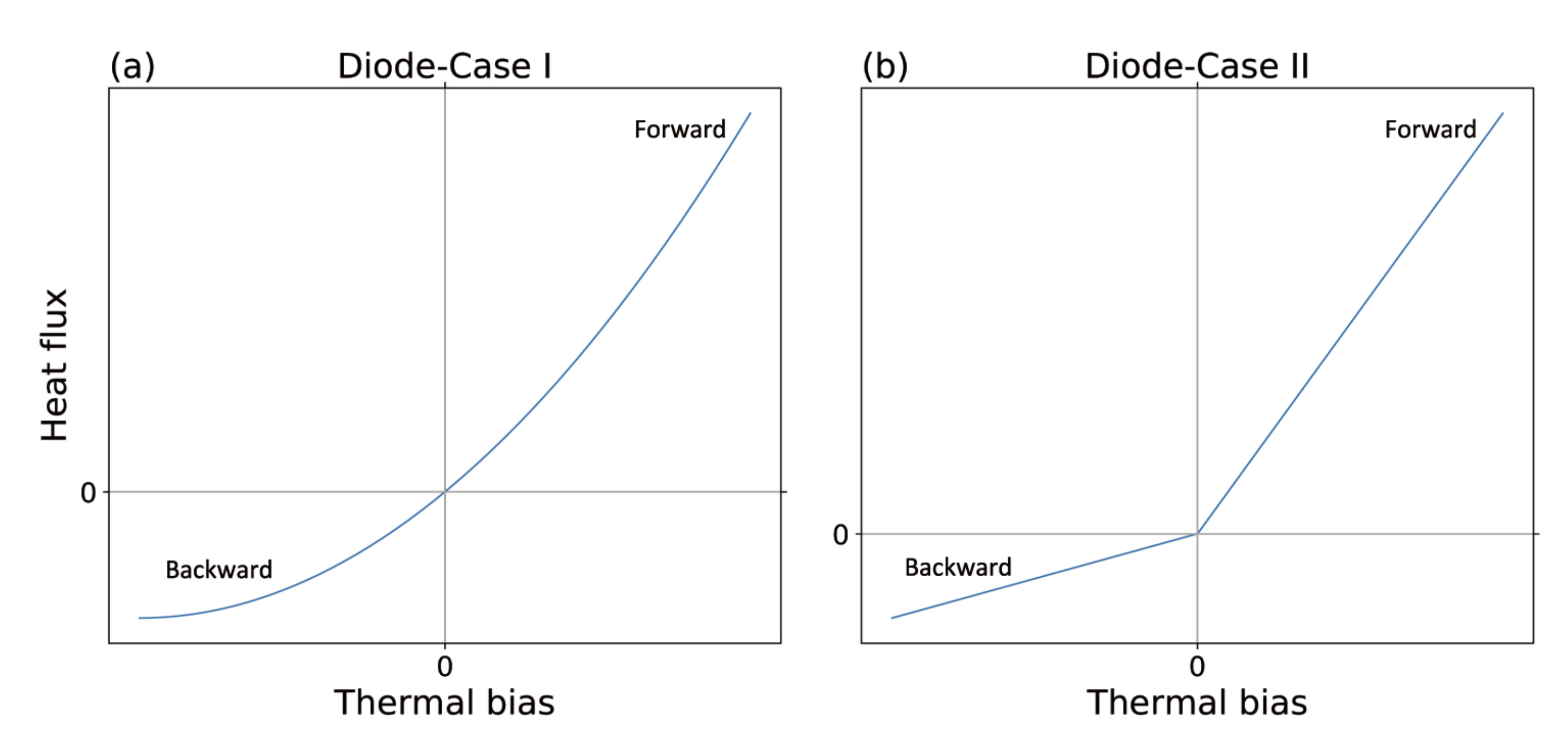}
	\caption{(a--b) Two typical heat flux--thermal-bias curves for thermal diodes. In (a), the curve is nonlinear and asymmetric. In (b), the curve is also asymmetric and only piecewise linear.}\label{1-1}
\end{figure}

Obviously, $\gamma=1$ means no rectification. For a thermal diode, $\gamma$ is usually smaller than 1,  and a greater-than-one $\gamma$ should be accounted for by a mode pumping heat from the cold source to the hot one driven by external energies.
According to the definition, a thermal diode must has an asymmetric relationship between the average directional heat flux density  $\bar j_x$ and the temperature bias $\Delta_x T$ on the two terminals, i.e., $\bar j_x(\Delta_x T)\neq -\bar j_x(-\Delta_x T)$ (at least for a given $\Delta_x T$). Usually there is no heat flux in the absence of an external thermal bias, then the asymmetry actually prohibits the global linearity. In other words, the $\bar j_x$--$\Delta_x T$ curve can't be a straight line as a whole in forward and reverse modes. Fig.~\ref{1-1} shows two typical flux--thermal-bias curves. In particular, the curve in Fig.~\ref{1-1}(b) consists of a two-segment polyline.  Since the slope of the flux--thermal-bias curve (usually positive) can be seen as the effective thermal conductivity of the system, thermal rectification is usually related to nonlinear thermal conductivities (making the governing equations also nonlinear) and asymmetric structures. However, the analysis seeking for the necessary or sufficient condition can be more complicated and we would leave the detailed discussion in Section 2.1. Actually, when heat shuttling exists, which means heat flux can be driven with a zero thermal bias, the asymmetric flux--thermal-bias curve might be a straight line and then the material can be linear.

Though heat rectification was firstly found in a copper oxide--copper heterojunction as early as 1936~\cite{25}, the enduring and widespread attention to this phenomenon and its derivatives actually began with the discovery of phononic heat rectification in low-dimensional nonlinear lattices two decades ago with the development of nanotechnology~\cite{12,26}. Nonlinear lattices with inhomogeneous parameters are used to construct a three/two-segment asymmetric structure coupled with a hot/cold bath at the two terminals~\cite{27,28}. At the interface, the temperature-dependent phonon spectra of different lattices which comes from the nonlinearity would match or mismatch with each other when reversing the thermal bias, so the interface resistance can change dramatically and generate the rectification effect~\cite{28,29}. Later, a thermal transistor and a thermal logic gate based on it~\cite{30,31}, a thermal memory~\cite{32}, and thermal shuttling~\cite{33} are also realized in nonlinear lattices, which can provide powerful tools to manipulate heat transfer and the possibility to do phonon computing like electronic computers~\cite{12}.
In addition, it's found that thermal rectification can exist in linear lattices with nonlinear system-bath
coupling~\cite{34}, and complex networks with anharmonic interactions~\cite{35,36} as well. 
In experiments, Chang {\it et al.}~\cite{37} fabricated a thermal diode based on the solitonic transport behavior of phonons in carbon nanotubes or boron nitride nanotubes with mass gradient, and the rectification ratio can reach 7\%. These nonlinear devices have also been designed or fabricated in various individual or hybrid quantum systems for thermoelectrics and heat transfer~\cite{38,39,40,41}, such as single quantum dot or coupled ones~\cite{46,47}, molecular junctions~\cite{42,43,44,45}, and tunnel-junctions~\cite{48,49}. Recently, there is a growing research interest in tuning nonlinear heat transport with Josephson junctions~\cite{50,51,52,53,54}. It is notable that the coherent phase dynamics of the Josephson junction satisfies a nonlinear equation having the same form as the motion of a pendulum~\cite{1}. 

However, realistic materials can hardly meet the theoretical requirements exactly and a more experimentally-friendly route could be exploiting nonlinear heat transfer under the phenomenological Fourier's law. Since the ratio $\bar j_x/\Delta_x T$ can be seen as the effective conductivity/conductance, we can directly focus on the temperature-dependent thermal conductivity to design thermal diodes and other devices. Besides rectification, how to realize other functions for controlling heat fluxes inspires the researchers to combine nonlinearity with thermal metamaterials. Thermal metamaterial~\cite{55,56,57,58} (and metadevice, a term from Ref.~\cite{59}) is a fast growing field in the past decade, which uses the transformation theory and other extended methods to realize novel heat transport phenomena not existing in naturally occurring materials, such as a thermal cloak that shields the thermal signals of an obstacle, a concentrator to amplify the heat flux, and a rotator to change the direction of heat fluxes in the given area~\cite{60}. The fundamental methods of thermal metamaterials
are firstly applied on linear conductive systems under the Fourier's law~\cite{61}, and until now only a relatively smaller part of researches involve nonlinearity like temperature-dependent thermal conductivity~\cite{62}. However, these nonlinear metamaterials can reveal intelligent applications that can rarely be achieved in metamaterials only made of temperature-independent materials, like devices capable of switching functions according to the ambient temperature and macroscopic thermal diodes~\cite{62}. Therefore, tuning nonlinearity in thermal metamaterials should be a subject worthy of attention for their promising potential in constructing tunable and adaptive devices for heat modulation. In addition, the nonlinearity here can come from the thermal properties themselves or other applied physical fields as heat transport are often driven by or can drive other physical phenomena.

This review focuses on the recent process on nonlinear thermal (meta)-devices and thermal metamaterials which are able to work when the Fourier's law is still valid and are mainly made of macroscopic temperature-responsive materials. For simplicity, we use the term ``nonlinear elements'' in the following part to refer to these nonlinear artificial functional materials and devices. These researches can provide a route for the development of nonlinear thermotics, and form an analogue to the nonlinear polarization theory of nonlinear optics in the classical regime, while the works based on microscopic classical/quantum heat carriers might be taken as the counterpart of nonlinear photonics handled totally by quantum theories~\cite{63}, and contribute another important part of nonlinear thermotics. We will first talk about solving and transforming the nonlinear Fourier's law for designing nonlinear elements in conductive systems in Section 2 and Section 3 respectively. In Section 4, we will review the works in which other physical fields are considered simultaneously with the Fourier's law. These coupled multiphysics can demonstrate some nonlinear behaviors. Effective medium theory for nonlinear conductive composites is included in Section 5, which is important for designing and fabricating realistic elements. Finally, a summary and some perspectives will be given in Section 6.

\section{Solving the nonlinear heat equation}

\noindent Solving the nonlinear heat equation analytically for the temperature or heat flux distribution is the most direct way to seek for the condition of a certain thermal phenomenon. This method can also be called the scattering-cancellation method/technique in metamaterials~\cite{64}, i.e., analyzing the general solution of the governing equation to seek for transparency-like and other functions, such as cloaking, camouflage, and expanding the flux~\cite{65,66,67,68}. In this section, we will review the recent efforts to tailor thermal rectification, negative differential thermal resistance, thermal hysteresis, and temperature trapping using nonlinear bulk materials. Exact or approximate solutions and qualitative analysis are all included, and related experimental verification is also presented. We can find these works focus on (quasi)-one-dimensional systems since heat flux should be the same everywhere without a internal source when $\kappa(T)$ is a continuous for the spatial variable (denoted as $x$), which can simplify the solving procedure a lot.

\subsection{Thermal rectification}

\noindent First we continue the discussion on thermal diodes or thermal rectification. Under the Fourier's law for steady states, some experiments more than 40 years ago have shown thermal rectification can be realized in contacting bulk materials with strongly different temperature-responsive thermal
conductivities, such as a quartz--graphite junction and a monocrystalline nonhomogeneous GaAs sample~\cite{69,70}. Early in this century, such a mechanism were re-found or re-examined by solving the nonlinear Fourier's law with $\kappa(T(x),x)$. Hu {\it et al.}~\cite{71} studied the different temperature dependences of $\kappa$ in the Frenkel-Kontorova lattice and $\phi^4$ model, and used  two segments of these lattices to construct a thermal diode. They also checked this idea with corresponding realistic materials like quartz and diamond. Later, Peyrard~\cite{72} pointed out that in the thermal diode, thermal conductivity should be a function of space (asymmetric) and temperature (nonlinear) at the same time, and the two variables can't be separable, meaning $\kappa(T,x)\neq \Pi(T)\Lambda(x)$. This conclusion coincides with the previous two-segment diodes which might be the most simplified design. Further, Go and Sen~\cite{22} found such an inseparability is not a sufficient condition for rectification while the separability $\kappa(T,x)= \Pi(T)\Lambda(x)$ must result in no rectification, where $\Lambda(x)$ can be a discontinuous function.
A simple inference is that nonlinear thermal conductivity plus spatial asymmetry can't guarantee the existence of rectification, as it's easy to construct a two-segment asymmetric structure with a discontinuity in $\Pi(T)$ and check the heat flux have the same magnitude in two modes.  We also note that the  works mentioned above all have symmetric (or more exactly, homogeneous) shapes along the $x$ direction, which means the systems are (quasi)-one-dimensional. It has been reported that the asymmetric shape with a nonlinear thermal conductivity can also induce thermal rectification in bulk materials, whether $\kappa(T,x)$ is separable or not~\cite{23}. However, as we have discussed in Section 1.2, the definition of rectification ratio in arbitrary shapes should be treated carefully.  

Anyway, until now, such a two-segment heterojunction is still the most used structure due to its simplicity in simulation and analysis.
Dames~\cite{73} calculated the rectification ratio of a two-segment bar with different power-law thermal conductivities for the segments, i.e., $\kappa_1\propto T^{n_1}$ while $\kappa_2\propto T^{n_2}$, in a low temperature environment (100--200~K). In the low-bias limit, the perturbation theory gives an analytical expression for the maximum thermal rectification ratio, writing 
\begin{equation}\label{dames}
\gamma_{\text{max}}\approx \frac{1}{2}\left|n_1-n_2\right|\frac{T_{h}-T_{c}}{T_{h}+T_{c}}.
\end{equation}
Here $T_{h}$ and $T_{c}$ are the temperatures of the hot and cold sources respectively, and the two segments have the same shape and size as rectangles (which is a default structure condition if there's no other statement in th following paragraphs). We can see a higher $\gamma$ can be achieved if $n_1 n_2<0$ which means the conductivities of the two materials behave in the opposite directions when the temperature increases. 
Cases in which the thermal conductivities of the two materials can be expressed as different polynomials of the temperature have also been studied~\cite{74,80}. It's found that  as the highest order of the polynomial expression increases, a larger $\gamma$ can be achieved, which also coincides with Dames's conclusion in Eq.~(\ref{dames}).    Actually, all the cases mentioned above have the same characteristic that $\gamma$ also changes with the value of thermal bias (and actually the temperatures of sources), which corresponds to the flux--bias curve drawn in Fig.~\ref{1-1}(a). In addition, when $\kappa$ is a natural exponential function of $T$ ($\kappa\propto\exp(T)$ for a material and $\kappa\propto\exp(-T)$ for another), $\gamma$ is found only depends on the thermal bias $\Delta_x T$, i.e., $\gamma=1-\exp(-\Delta_x T)$~\cite{74}.

\begin{figure}[!ht]
	\centering
	\includegraphics[width=.9\linewidth]{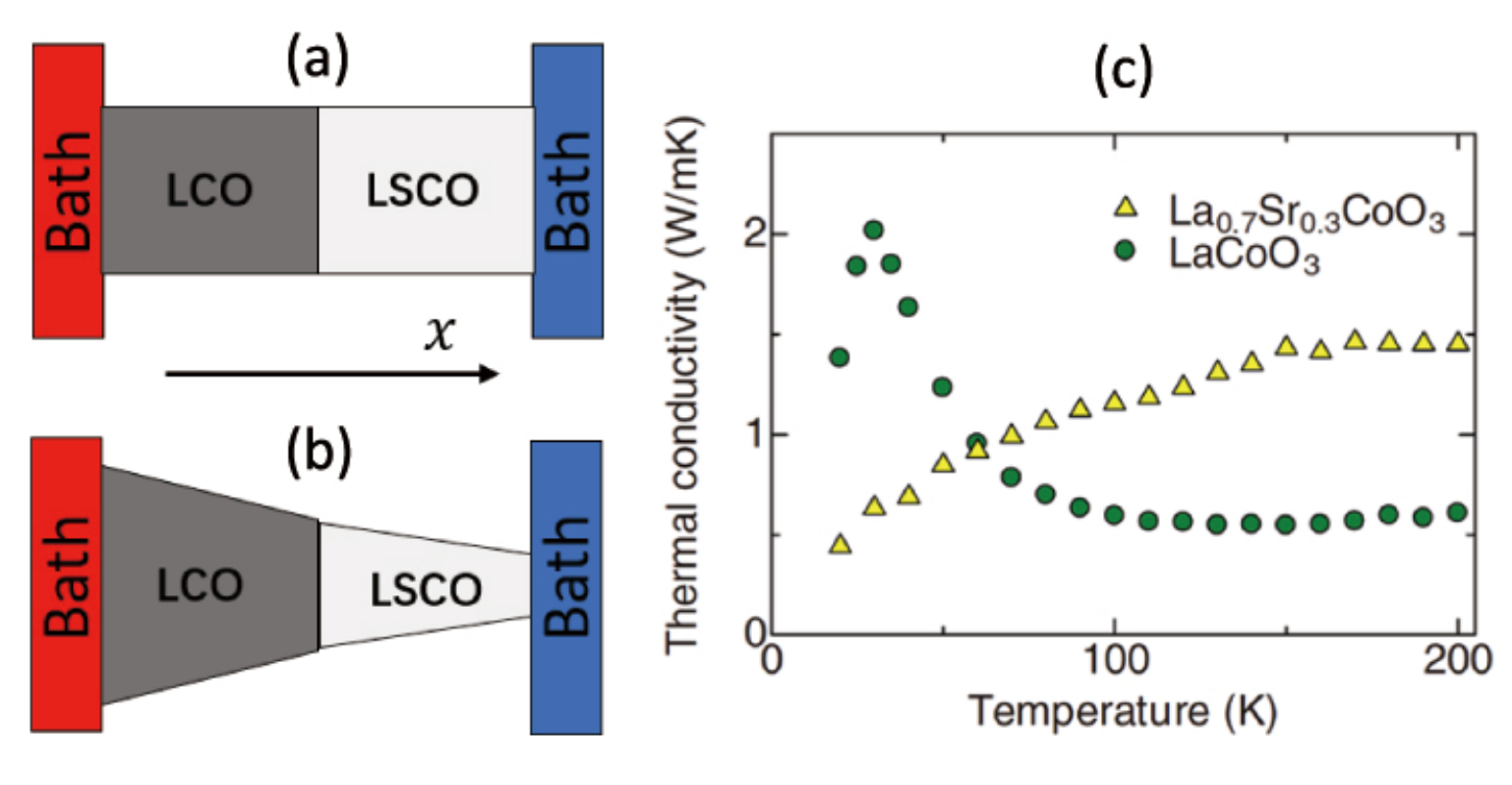}
	\caption{(a--b) Two-segment diodes made of LaCoO$_{3}$ (LCO) and La$_{0.7}$Sr$_{0.3}$CoO$_{3}$ (LSCO) with rectangular and pyramid shapes, respectively. (c) Thermal conductivity against the temperature for LCO and LSCO. (c) is reproduced with permission from Ref.~\cite{76}. Copyright 2011 AIP Publishing.}\label{2-0}
\end{figure}

In experiments, Kobayashi {\it et al.}~\cite{75} used two cobalt oxides LaCoO$_{3}$ (LCO) and La$_{0.7}$Sr$_{0.3}$CoO$_{3}$ (LSCO) to make a heterojunctions thermal diode. 
The two bars of LCO/LSCO are rectangular (see Fig.~\ref{2-0}(a)) with identical cross-sections perpendicular to the $x$ direction and have only a tiny difference in the length (6.3~mm and 6.1~mm). We can see from Fig.~\ref{2-0}(c) that the thermal conductivity of LCO increases while that of LSCO decreases when $T$ is higher than about 40~K, so they kept the cold source at this temperature. When the thermal bias takes 60~K, they obtained a rectification of 0.3 (recalculated according to the definition given by Eq.~(\ref{ratio})).
Further, a more asymmetric binary material system 
is fabricated using LCO/LSCO with pyramid shapes~\cite{76} instead of homogeneous bars with a total length of 19~mm; see Fig.~\ref{2-0}(b). Though they obtained a smaller $\gamma$ which is 0.26, the impact of the geometry factor on rectification should be judged more carefully since the samples have different lengths in the two experiments. It's found that the pyramid structure can be equivalent to a rectangle-rectangle heterojunction with the same length in which the two rectangles may have different lengths and cross-section areas. Their calculation results show the maximized $\gamma$ exists in asymmetric shapes thus the rectification performance can be enhanced using pyramid structures~\cite{76}.
Other nonlinear materials like Al-based alloys~\cite{77,78,79} (working at high temperatures above 300~K using Al-based alloys and another material; $\gamma_{\text{max}}$ can exceed 0.5~\cite{79}) are also used to make asymmetric two-segment structures. 

\begin{figure}[!ht]
	\centering
	\includegraphics[width=.7\linewidth]{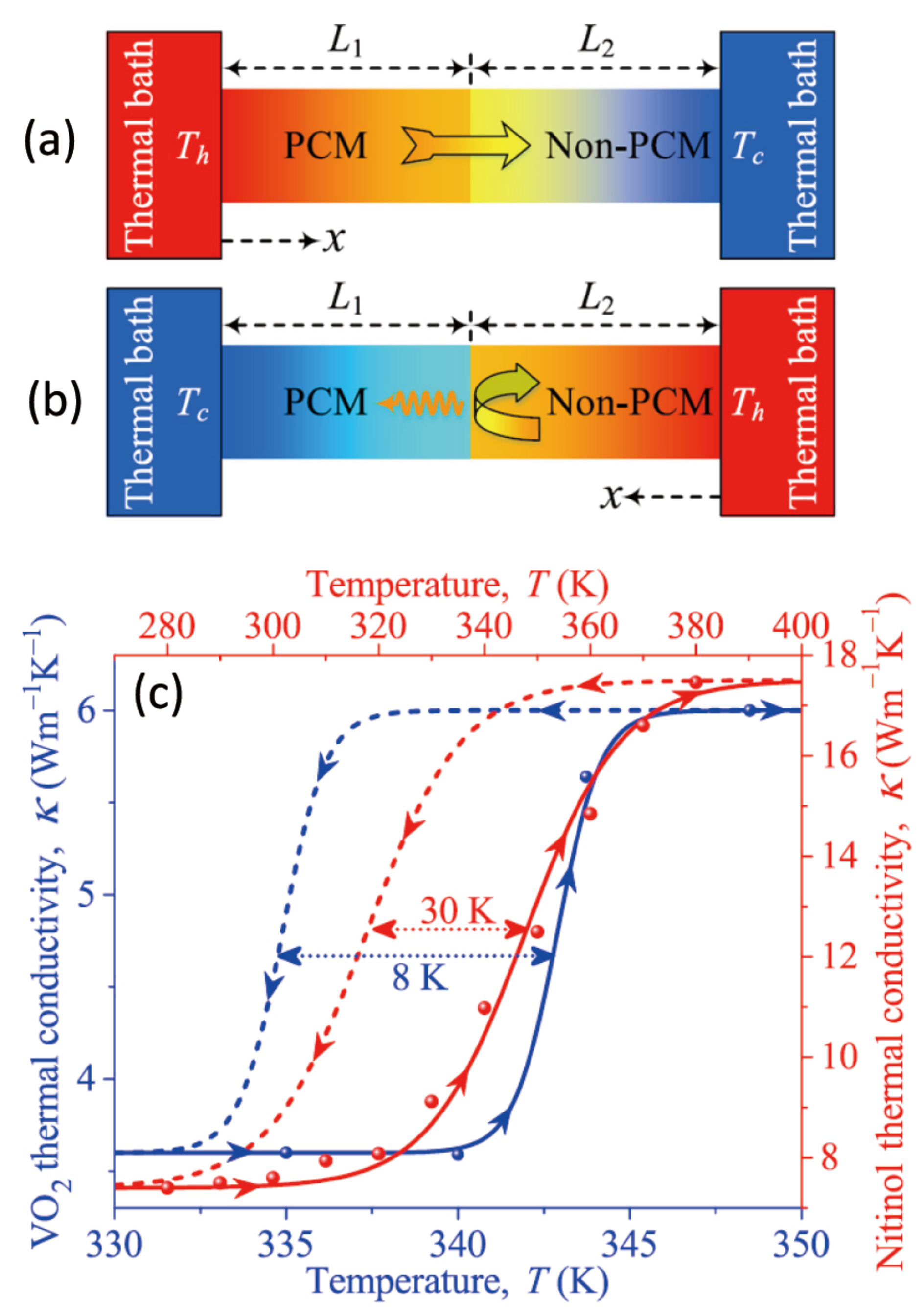}
	\caption{(a--b) Design sketch of a single--phase-change thermal diode in the forward/reverse mode. (c) Thermal hysteresis loops on thermal conductivity for VO$_2$ and Nitinol, and the dot symbols are corresponding experimental measurements. Reproduced with permission from Ref.~\cite{86}. Copyright 2018 AIP Publishing.}\label{2-1}
\end{figure}

To enhance the rectification performance, phase change materials (PCMs) are also introduced since phase change can be seen as an extreme nonlinearity. For example, the maximized $\gamma$ in Refs.~\cite{77,78,79} is achieved through the lattice transition of Ag$_{2}$Te.
PCMs have been widely used in thermal management, and energy conversion or storage~\cite{81,82,83} as their physical properties can change greatly when a environment parameter evolves to a certain point, usually accompanied by the absorption or release a large amount of latent heat (first-order transition). Here we focus on the sudden change of the thermal conductivity when the material is being heated/cooled, and the works involving the latent heat transfer will be discussed in Section 4.

$\kappa(T)$ of ideal PCMs can be described with a simple step-function or a smooth logistic function. The logistic function used here can be written as
\begin{equation}\label{pcm}
\kappa(T)=\kappa_h+\frac{\kappa_c-\kappa_h}{1+e^{\beta(T-T_p)}},
\end{equation}
which shows a S-curve, and thus it belongs to the sigmoid functions. It's easy to check that $\kappa_h$ and $\kappa_c$ are the thermal conductivities (far) above and below the transition tempearature $T_p$ respectively, and $\beta$ is a scaling factor to control the quickness of the jump from high conductance to low conductance. The larger $\beta$ is, the faster the transition happens. A single phase change material can naturally be a thermal switch which can have two different flux--thermal-bias curves (or straight lines) representing the on/off states when the temperatures of the two sources are higher or lower than the transition temperature simultaneously. For thermal diodes,  
based on the general principle of nonlinearity plus asymmetry, the familiar two-segment structure can be used again, composed of a phase change material (Material A) and another different material (Material B).  The effect of the presence of Material B is to guarantee the whole phase change material are in the same phase when $T_h>T_p>T_c$. Material B can be a non-phase change material (temperature-dependent or temperature-independent) or another phase change material. Anyway, it's better that the two segments have opposite temperature coefficients $\partial \kappa/\partial T$ within the working temperature ranges if we make a simple extension of Eq.~(\ref{dames}) for power-law conductivities.

When the thermal conductivity of Material B is temperature-independent or can approximately be seen as a constant(see Fig.~\ref{2-1}(a--b)), Kobayashi {\it et al.}~\cite{84} used the step-function and found
\begin{equation}
\gamma_{\text{max}}=1-\sqrt{\text{Mim}\{\kappa_h,\kappa_c\}/\text{Max}\{\kappa_h,\kappa_c\}}
\end{equation}
when the thermal  conductivity of Material B is $\sqrt{\kappa_h\kappa_c}$. In addition, the length ratio of the two materials can also be considered to optimize $\gamma$~\cite{85}. Ordonez-Miranda {\it et al.}~\cite{86} compared the performance of vanadium dioxide (VO$_2$; metal--insulator transition) and Nitinol (nickel-titanium alloy; martensite--austenite transition) as Material A respectively, using logistic functions in their analysis. They found that a larger $\beta$ and a higher thermal conductivity contrast $\left|\kappa_h-\kappa_c\right|$ in Eq.~(\ref{pcm}) and can enhance the rectification effect. In addition, the two materials show thermal hysteresis which means the transition points are different in the cooling and heating processes, and we will discuss this phenomenon specifically later in this section. In experiments, Kobayashi {\it et al.}~\cite{84} fabricated a sample with MnV$_{2}$O$_{4}$
as Material A and La$_{1.98}$Nd$_{0.02}$CuO$_{4}$ as Material B working below 60~K with a small thermal bias of 2~K ($\gamma=0.26$). 
Garcia-Garcia and Alvarez-Quintana~\cite{87} compared the performances of Nitinol--Fe/Cu/Graphite diodes and the biggest $\gamma$ is found in the Nitinol--Graphite heterojunction with 0.32 at $\Delta_x=160$~K above 65~$^{\circ}$C. Nitinol is actually an important shape memory alloy (SMA)~\cite{88} and 65~$^{\circ}$C is close to its martensite-to-austenite transition point (or about 340~K as shown in Fig.~\ref{2-1}(c)) when being heated.
Another experiment conducted by Pallecchi {\it et al.}~\cite{89} employed the PCM poly(N-isopropylacrylamide) (PNIPAM) and the non-PCM polydimethylsiloxane (PDMS) in the diode structure. The transition point of PNIPAM is nearly 300~K so the device can work above the room temperature, and $\gamma$ is in the order of 0.5 for a wide thermal bias range. Other single--phase-change diode includes
PCMs like a solid paraffin-polystyrene foam hybrid (particularly given a oscillating temperature input)~\cite{90}, and a composite consisting of first-order PCM neopentylglycol (NPG) and second-order PCM gadolinium (Gd) which both show more complicated $\kappa$--$T$ relationships than a simple jump in the S-curve~\cite{91}.

When Material B is also phase-variant, Kang {\it et al.}~\cite{92} gave a general theory to predict $\gamma$ of this kind of structures using step functions. Their results show that the optimized parameters should satisfy $(\kappa_{h,A}-\kappa_{c,A})(\kappa_{h,B}-\kappa_{c,B})<0$ (i.e., opposite temperature coefficients of $\kappa$) and high thermal conductivity contrast for both PCMs. Here the subscripts $A$ and $B$ added to the notations $\kappa_h$ or $\kappa_c$ means Materal A's and Material B's thermal conductivities at high/low temperatures. Further, based on the similar analysis, Cottrill {\it et al.}~\cite{93} built a dual--phase-change junction of highly
porous paraffin-polystyrene foam hybrid
and a PNIPAM aqueous solution, and obtained a relatively high rectification ratio of 0.62. 
It is worth noting that the PNIPAM aqueous solution used here is a liquid and the natural convection is also taken into consideration in their design. We will focus on the effect of convection as another source of nonlinearity in Section 4. Recently, Kasali {\it et al.}~\cite{94} studied the case with two phase change materials, i.e., VO$_2$ and polyethylene (PE), having close transition temperatures but opposite temperature dependencies of $\kappa$. Their analytical calculations based on $\kappa$ with the logistic-function form finally gives an optimal rectification ratio of 0.6.

Another mechanism to get a sudden change in thermal conductance is creating moving contact structures. Tso and Chao~\cite{95} built such a thermal diode with a SMA-based actuation system. When being heated, the lattice transition of SMA can cause shape changes in addition to the change in its intrinsic thermal conductivity as mentioned above. In particular, the name ``shape memory'' effect means SMA which suffered a plastic deformation at a low temperature can return to its original shape above the activating temperature~\cite{88}. In their experiment, two SMA springs made of Ni--Ti--Cu will elongate when the inlet temperature is higher than
the activating temperature, and drive copper blocks through a lever system. The moving copper blocks can touch the fixed copper blocks attached to the heat sources and provide an effective heat conduction path in the forward mode, while the interface resistance should increase dramatically when the copper blocks are not in contact in the reverse mode. A high $\gamma$ close to 1 can be obtained according to the insulation performance in the reverse mode. Due to its special thermal and mechanical properties, SMA have many well-established or potential applications in manipulating heat transfer such as a thermal regulator in Li-ion batteries~\cite{96}, and heat switches in spacecrafts~\cite{97,98}. Some works to be introduced in the following paragraphs also depend on the shape change of SMA. Similarly, Gaddam {\it et al.}~\cite{99} proposed a liquid thermal diode using a chamber not completely filled with mercury, which undergoes a sudden change in thermal conductance when the heated mercury expands and fills the gap in the forward mode, while the gap insulates the heat transfer in the reverse mode. Overall, the sudden change of $\kappa$ induced by PCM or moving contact structures corresponds to the flux--thermal-bias curve drawn in Fig.~\ref{1-1}(b), which is a two-piece polyline.

Although the experimental samples in the works mentioned above are usually in millimeter size, the basic ideas to realize thermal rectification based on the different temperature trends of $\kappa$~\cite{100,101,102,103} and asymmetric/inhomogeneous geometry (mass-gradient, porous structure, etc.)~\cite{103,104,105,106,107,108,109,111a,111b,111c} in the two segments can also work at the nanoscale (especially the one/two-dimensional materials like carbon nanomaterials~\cite{13,111d,13b}) as long as the heat flux is expressed by the temperature gradient and the (effective) thermal conductivity, in spite of their various phonon scattering mechanisms. In addition to the mismatch of phonon power spectra we have mentioned in the Introduction part, standing wave and local resonance effects of phonons~\cite{111a,111b}), and phonon localization from defect engineering~\cite{111c}) are also important mechanisms that have been revealed.
Recently, Zhang {\it et al.}~\cite{110} developed a perturbation theory for thermal rectification by considering a comprehensive expression $\kappa(T(x),L,W(x))$ in the heat equation. Here $W(x)$ is a local physical quantity varying
with position like porosity, mass, and the characteristic length in other directions, and their results could provide general explanations for several previous experimental and numerical observations from normal conduction to nanoscale anomalous conduction. 

Finally, we want to talk about the relationship between thermal diode/rectification and nonreciprocity which are both related to one-way transport. In wave systems, the counterpart of a thermal diode can be a Faraday isolator~\cite{24-1,huyi2,20,21} having different transmission coefficients in the forward/backward modes. The term diode and rectifier are also used frequently (e.g., for acoustic waves~\cite{24-00,24-0}) while they might have different meanings in other situations. What's more, the wave isolator should break the Lorentz reciprocity (or reciprocal theorem)~\cite{24-1,huyi2,20,21}, meaning the coupling relationship between the source and the observed field keeps the same when exchanging the positions of source and observation point. Though we can see the definition of a thermal diode requires the existence of two sources with different temperatures, which is quite different from wave isolators, the reciprocal theorem and its breaking mechanism can also be studied in heat conduction, a diffusive system~\cite{huyi1,huyi3,23}. It's revealed that the absence of global steady-state reciprocity can generate the diode effect in a two-port system (which can be a more general case than the one-dimensional two-terminal cases we have focused on in this review), while intrinsic directional heat generation can also make a diode in a reciprocal system when only the heat flux flowing out of the system is focused on~\cite{23}.
Also, wave and diffusive systems share some common points in breaking Lorentz reciprocity. We have mentioned that nonlinear material (temperature-dependent thermal conductivity) with spatial asymmetry is not a sufficient condition for thermal rectification. In fact, it's also not a necessary condition~\cite{23}. In Section 4.2, we will give an example that linear material can also makes a thermal diode if convection is considered. These two points on necessary and sufficient conditions are similar to the conclusions on breaking reciprocity in wave systems such as electromagnetics~\cite{20}, acoustics~\cite{21}, and quantum mechanics~\cite{24}.

\subsection{Negative differential thermal resistance and thermal hysteresis}

\noindent Now we turn to another possible characteristic pattern in the flux--bias curve, which is crucial for building thermal transistors and other devices like a logic gate to do Boolean operations~\cite{30,31}, i.e.,
negative differential thermal resistance (NDTR) or negative differential thermal conductance (NDTC). NDTR corresponds to a decline or a discontinuous drop in the flux--thermal-bias curve, saying the flux decrease as the external temperature bias increases, and the derivative measuring the effective differential conductance $\partial j/\partial \Delta T$ (or the discrete form $\delta j/\delta \Delta T$) is negative.

\begin{figure}[!ht]
	\centering
	\includegraphics[width=.7\linewidth]{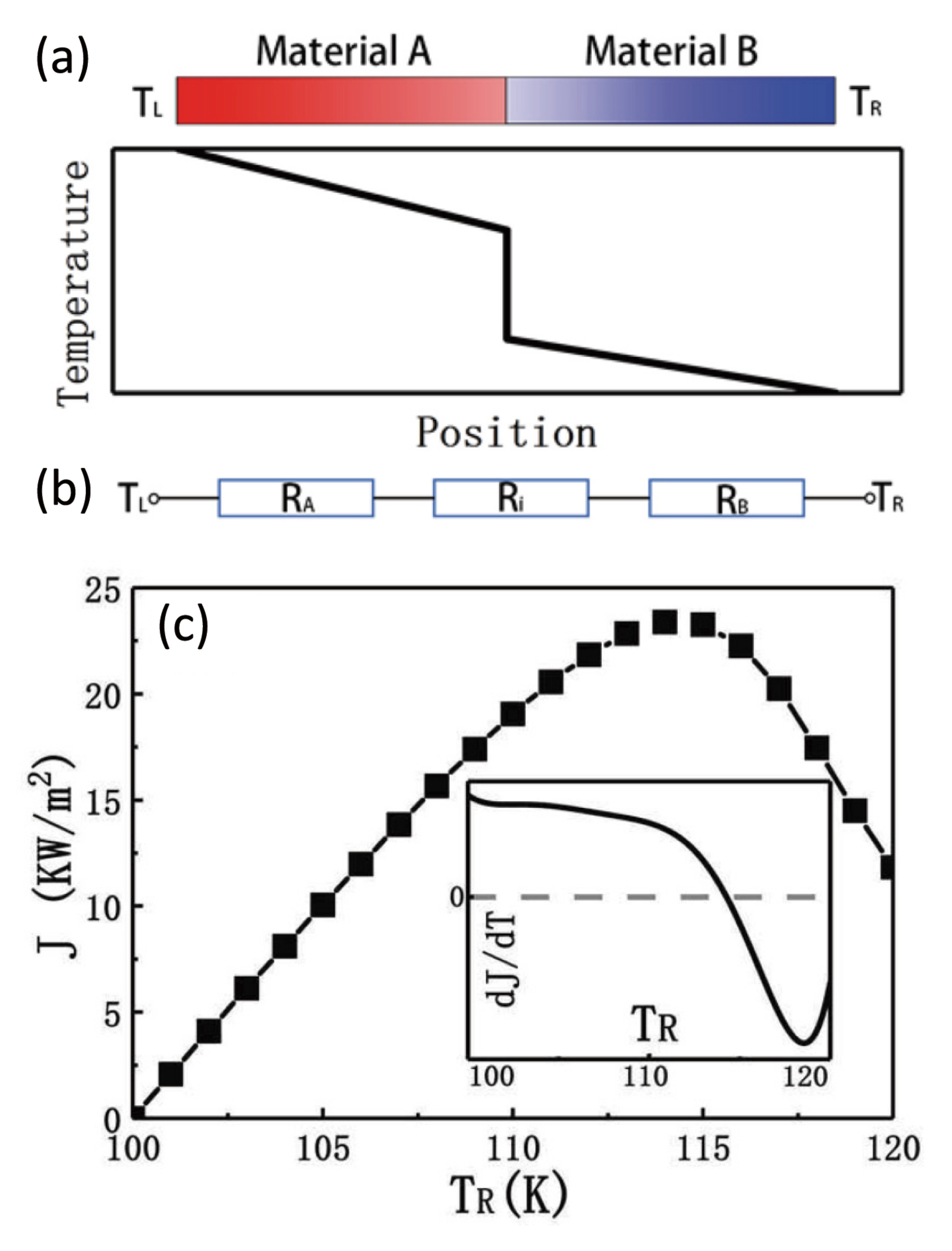}
	\caption{(a) Schematic of a homojunction with an temperature jump at the interface. (b) The effective series model for the thermal resistances of Material A, Material B and the interface. (c) Thermal flux versus the temperature of the right heat source when the temperature of the left one is invariant. The subplot in it show the differential thermal conductance and the decreasing line corresponds to NDTR. Reproduced with permission from Ref.~\cite{112}. Copyright 2020 AIP Publishing.}\label{2-15}
\end{figure}

Various microscopic mechanisms have been revealed to generate NDTR at nanoscale, from mismatch of phonon spectra in a one-dimensional heterogeneous chain~\cite{104,104bb} or heterostructures based on two-dimensional materials like graphene~\cite{111ee,111e}, to nonlinear quantum systems like Josephson junctions~\cite{111ab}, which can refer to the introduction parts of Refs.~\cite{111,112}. However, can NTRC be realized under the Fourier's law with bulk materials? Recently, Yang {\it et al.}~\cite{112} proposed a new mechanism to achieve NDTR in a macroscopic homojunction. Here the point of working principle is modulating the thermal resistance and temperature jump at the interface. Particularly, the interface thermal resistance (ITR)  can be divided into two parts according to its generation mechanism. One is the intrinsic ITR, or Kapitza resistance~\cite{113,114}, which comes from the mismatch of heat carriers' energy spectra at the interface and has been deeply exploited in asymmetric nonlinear lattices. 
On the other hand, the extrinsic ITR is caused by the incomplete contact of the two segments at the interface and usually should be reduced by the increasing interface pressure. 
They established a two-segment model shown in Fig.~\ref{2-15}(a)~\cite{112}. Material A and Material B can have temperature-dependent thermal conductivities, and their resistances ($R_A,R_B$) along with the ITR ($R_i$) form a three-component series resistance thermal circuit illustrated in Fig.~\ref{2-15}(b). Heat sources with a temperature of $T_L$ or $T_R$ are put on the left/right boundary. The temperature jumps from $T_{mL}$ on the left side of the interface to $T_{mR}$ on the right side. 
The total ITR can be expressed with a phenomenological formula~\cite{112}
\begin{equation}\label{interface}
R_{i}=\frac{L_{i}}{C_{1} \tanh (P / 100)+C_{2}},
\end{equation}
and the heat flux across the interface with a temperature jump form $T_{iA}$ on the left to $T_{iB}$ on the right is $(T_{iA}-T_{iB})/R_{i}$.
Here, $L_{i}$ is the length of the middle segment (taken as a constant in their numerical validation), and $P$ is the interface pressure. The coefficients $C_1$ and $C_2$ are two close positive numbers. $R_{i}$ ia actually a function of the temperature since $P$ is influenced by the thermal expansion of the materials. For the existence of NDTR, $\text{d} R_{i}/\text{d} T_R=(\partial R_{i}/\partial P)(\text{d} P/\text{d} T_R)$ should be positive if $T_L$ is fixed, so the materials must shrink when being heated and form a gap and a much smaller $P$. As a result, the materials should have negative thermal expansion coefficients, for example, the silicon (Si). Material A and Material B can be the same so the structure can be a homojunction. In fact, here the extrinsic ITR is dominant as the Kapitza resistance is mainly a monotonically decreasing function of temperature~\cite{112}. The conclusion is consistent with the case $L_i$ is variable since the gap makes $L_i$ larger too.
Fig.~\ref{2-15}(c) shows the simulation results of the flux--bias curve ($T_L$ is fixed at 100~K) through the finite element method (FEM), and NDTR happens when $T_R$ is larger than 114~K.

More generally, the negative thermal expansion and the moving contact statures mentioned above can all be classified as utilizing the thermo-mechanical effects, and more related researches including NDTR and thermal transistors will also be discussed in Section 4 when thermal radiation is added into the consideration.

Besides thermal transistors and thermal logic gates, NDTR can also be utilized to realize thermal bistability and then thermal memory. Bistability means the system can have two stable states under the same input parameters, and this concept can be naturally extended to multistability. In fact, bistability and multistability correspond to the bifurcation theory of nonlinear differential equations, which refers to the change in numbers of stable solutions~\cite{1}.
The most common and widely used bistability phenomenon might be magnetic hysteresis in a ferromagnet~\cite{115}, demonstrating the magnetization--magnetic-field curve as a loop under magnetization and demagnetization. The bistability here means the same magnetic-field can corresponds to two possible magnetization values. Magnetic hysteresis is crucial for constructing modern electronic device like magnetic disk storage (memory) as the two stable magnetization states can be taken as signal ``0'' or ``1''. Bistability, often accompanied by the hysteresis phenomena, has also been found in many other systems, such as optics~\cite{116}, the Schmitt trigger~\cite{117}, the anomalous Hall effect/quantum anomalous Hall effect~\cite{118,119}, and even biological systems~\cite{120,121}. 
Here, we take thermal bistability as a broad concept, mainly describing a physical quantity can have two different values for every $T$ within a certain range of temperature, e.g., the previously mentioned $\kappa$--$T$ hysteresis loops of SMA and VO$_2$. Then the different values usually reflected as the high/low temperature signal at a certain place can also be exploited to build a thermal memory.

\begin{figure}[!ht]
	\centering
	\includegraphics[width=.7\linewidth]{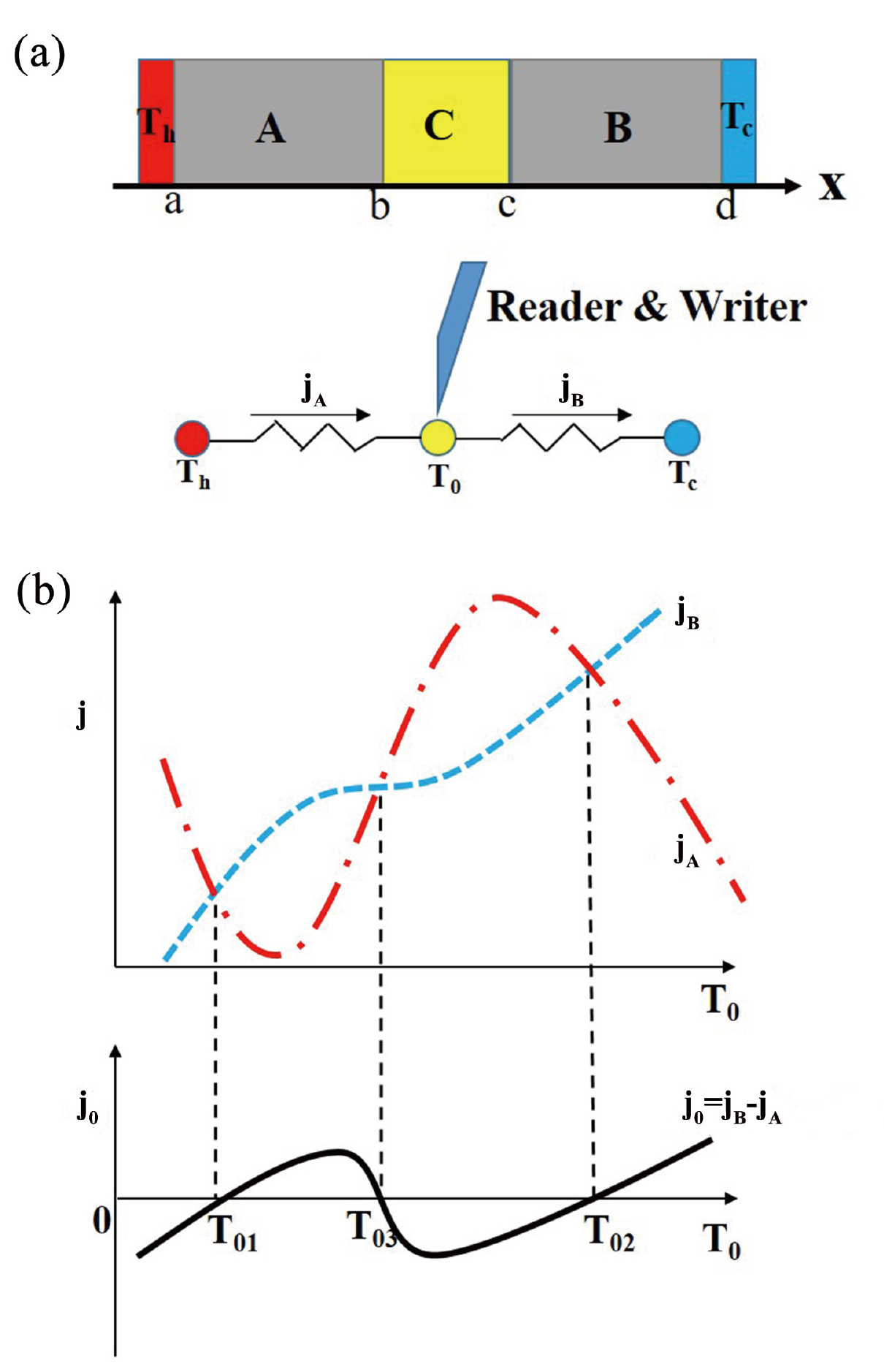}
	\caption{(a) Sketch of a thermal bistable system for thermal memory with functions of writing and reading temperature signals in Region C. (b) Reasoning for finding bistability. Heat fluxes in Region A ($j_A$) and Region B ($j_B$) versus the uniform temperature $T_0$ in Region C are plotted on the upper side. The intersections of the two curves correspond to the possible value of $T_0$ as no net flux $j_0=j_A-j_B$ should exist in Region C, illustrated on the downside as $j_0$ against $T_0$. Multivalued $T_0$ can't be obtained if the two curves are both straight lines. Reproduced with permission from Ref.~\cite{123}. Copyright 2020 American Physical Society.}\label{2-2}
\end{figure}

\begin{figure}[!ht]
	\centering
	\includegraphics[width=.8\linewidth]{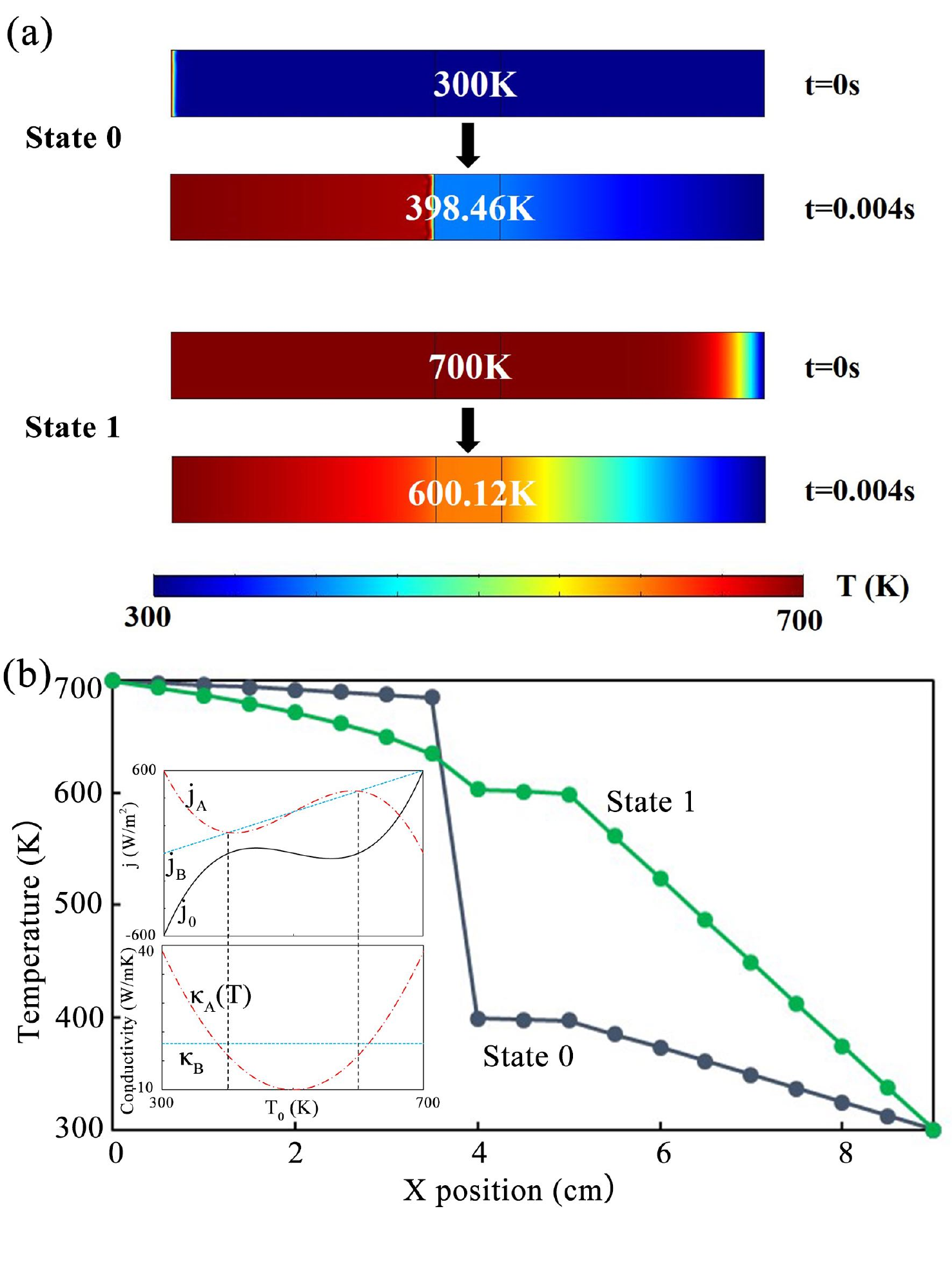}
	\caption{Simulation results for the two possible stable states of the thermal bistability device. (a) Two different initial conditions are applied. In state 0, the whole device is set as 300~K at the time $t=0$ and the stationary temperature in the middle is 398.46~K when $t=0.004$~s. In state 1, the initial temperature is 700~K and the stationary temperature in the middle is 600.12~K. (b) The spatial temperature distributions for the two states are plotted. The subplots show the corresponding relationships between thermal fluxes/conductivities and the temperature in Region C. Two vertical dashed lines frame the scope of local NDTR in Region A. Reproduced with permission from Ref.~\cite{123}. Copyright 2020 American Physical Society.}\label{2-3}
\end{figure}

Thermal memory based on thermal bistability have been first designed by nonlinear lattices~\cite{32} and later realized in labs using VO$_2$ again at the nanoscale, based on its thermal hysteresis for electrical conductance and the the Wiedemann–Franz to map this hysteresis to thermal conductance~\cite{122}. Inspired by these works, Wang {\it et al.}~\cite{123} proposed the macroscopic thermal bistability and thermal memory under the Fourier's law. As shown in Fig.~\ref{2-2}(a), they considered a three-segment model heated/cooled by sources at the two terminals. Region C, occupied by a good conductor of heat like copper, is narrower than Region A (with heat flux $j_A$) and Region B (with heat flux $j_B$), and the temperature signal shall be read or written in Region C because the net heat flux $j_0=j_B-j_A$ in it is close to zero. When the temperatures of the two sources are fixed, the existence of bistability requires the temperature $T_0$ in Region C can have different values according to the initial temperature conditions on the whole device. Further, they reasoned through the dependence of $j_0$ on $T_0$ to look for more than one solutions of $T_0$ for a given $j_0=0$; see Fig.~\ref{2-2}(b). They found that at least one of the materials in Region A and Region B should have a temperature-responsive thermal conductivity. One choice is that the material in Region B is linear while the thermal conductivity in Region A can have a parabolic relationship with $T$, meaning a quadratic function expression~\cite{123}
\begin{equation}
\kappa_A=\kappa_{A0}+\kappa_{A1}T+\kappa_{A2}T^2.
\end{equation}
The emergency of two stable states usually involves two competing mechanisms with the opposite evolution trends, so the coefficients $A_1$ and $A_2$ should have opposite signs. By doing the Kirchhoff transformation, they finally solved out the parameters for all the materials if the two stationary temperatures in Region C and the sources' temperatures ($T_h>T_c$) are given. Also, their conclusion was validated by numerical results shown in Fig.~\ref{2-3}. We can see the stationary temperatures achieved in Region C are indeed diffident under the heating (State 0) and cooling (State 1) processes. Also, local NDTR in Region A can be found in the subplot in Fig.~\ref{2-3}(b) when $\partial j_A/\partial T_0>0$ since the thermal bias here is $T_h-T_0$. In fact, the cubic parabola shape of $j_A$ against $T_0$ with two metastable points (the first and second derivatives are both zero) must have a NDTR piece to link the two stable states, and a hysteresis loop can be constructed in such a bistable system. They also used this structure to simulate a complete thermal storage process and proposed an experimental program using SMA~\cite{123}.

Another nonlinear element based on NDTR and hysteresis is the thermal memristor, the counterpart of the electrical memristor in the thermal domain.
In 1971, Chua~\cite{124} predicted the fourth fundamental passive electric circuit elements named ``memristor'' (memory resistor) besides the resistor, the capacitor and the inductor (an alternative for the fourth fundamental element termed
memtranstor was proposed in 2015~\cite{124sy}). The memristor is assumed to link the magnetic flux $\varphi$  and charge $q$ as $\text{d} \varphi=M\text{d}q$, by introducing the memristance $M$. Especially, when $M$ is charge-dependent, the nonlinear memristor can induces $V=M(q)I$ through Faraday's law of electromagnetic induction. Thus the resistance here has a memory effect related to the current previously flowed through the device. Strukov {\it et al.}~\cite{125} built the first electrical memistor using TiO$_2$ in labs showing a Lissajous-type Volt--Ampere characteristic curve, i.e., a kind of pinched hysteresis loop. 
In industry, resistive random-access memory (RRAM or ReRAM) is usually referred to as a memristor, which is a new type of non-volatile computer memory. Further, the concept of memristor has been generalized to memristive systems including memcapacitors and meminductors, whose capacitance or inductance can have memory effects too~\cite{126,127}.                                                                                                                                                                                                                                                                                     
Memristive systems are believed to have promising prospects in many fields~\cite{127}, such as constructing digital memories, digital logic circuits, and neuromorphic circuits~\cite{128,129}.

\begin{figure}[!ht]
	\centering
	\includegraphics[width=.95\linewidth]{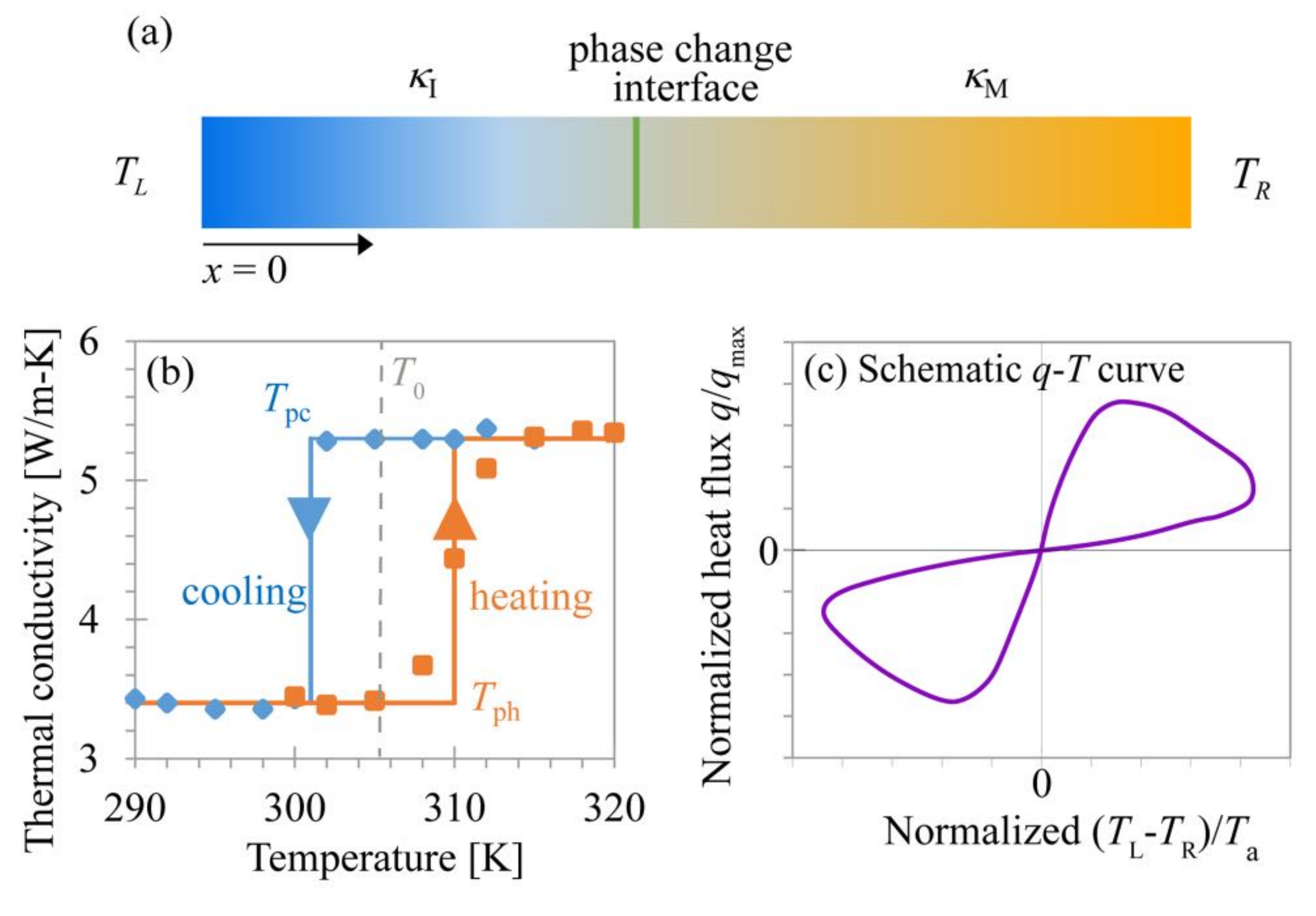}
	\caption{(a) Sketch of the phase-change thermal memristor made of a W$_{0.2}$V$_{1.98}$O$_{2}$ bar. Heat sources are applied on the left and right sides. The green line stands for the phase change interface which isolates the two phases of metal and insulator. (b) Thermal hysteresis on thermal conductivity of W$_{0.2}$V$_{1.98}$O$_{2}$, and the square and diamond symbols are corresponding experimental measurements. (c) The Lissajous-type characteristic curve for normalized flux against the normalized thermal bias. Adapted with permission from Ref.~\cite{131}. Copyright 2019 AIP Publishing.}\label{2-35}
\end{figure}

As an analogy from electric resistance to thermal resistance, Ben-Abdallah~\cite{130} proposed a theoretical framework of the thermal memristor in heat conduction, using the PCM VO$_2$, and then designed a neuronal AND gate with two memristors. Based on the definition of electrical memristor, its thermal counterpart should has a nonlinear thermal conductivity depending on the history of heat flux, which can be related to the thermal hysteresis phenomenon inherent to some PCMs mentioned above. However, his work didn't give the Lissajous-type flux--thermal-bias characteristic curve.
Later, Yang {\it et al.}~\cite{131} made the advance in revealing this important property of thermal memristor. In their design (see Fig.~\ref{2-35}(a)), 
a tungsten (W) doped vanadium dioxide, i.e., W$_{0.2}$V$_{1.98}$O$_{2}$, is used, whose $\kappa$--$T$ hysteresis loop is shown in Fig.~\ref{2-35}(b).  The 
thermal hysteresis is quite strong here as the loop is almost a rectangle or composed of two step-functions. The different transition points under heating an cooling are denoted by $T_{ph}$ and $T_{pc}$ respectively, and $\kappa$ can be taken as a constant above or below the corresponding transition points.
To obtain a Lissajous-type flux--thermal-bias characteristic curve in Fig.~\ref{2-35}(c), periodic temperature sources are applied, taking~\cite{131}
\begin{equation}
T_{L}  =  T_{0}+T_{a} \sin (2 \pi f t)
\end{equation}
on the left side and
\begin{equation}
T_{R}  =  T_{0}+T_{a} \cos (2 \pi f t)
\end{equation}
on the right. The reference temperature is $T_0=(T_{ph}+T_{pc})/2$ and the heating period $1/f$ should the larger than both the time scale of the phase transition along the whole W$_{0.2}$V$_{1.98}$O$_{2}$ bar and the relaxation time of heat conduction for the use of steady equation. Their calculations show that it does demonstrate a Lissajous-type flux--thermal-flux curve during a heating/cooling cycle on which NDTR must exist because a monotonic function can't be plotted as a loop (actually, a self-crossing loop); see the segment with negative slope in Fig.~\ref{2-35}(c). NDTR should correspond to the process when the high-temperature metallic phase turns into the low-temperature insulating phase or vice versa, just as the evolution path of two metastable states mentioned in thermal bistability. In other words, the thermal memristor is a special case of thermal bistability, and can be used in building high-performance thermal memories and artificial neural networks for thermal computation in the future. Although the W$_{0.2}$V$_{1.98}$O$_{2}$ bar is set to be 10~$\mu$m long (less than 0.5~mm as required) and 30~nm wide~\cite{131}, this framework can be extended to larger scales if the cycle frequency is carefully tuned since the Fourier's law doesn't break down in their research. Again, the work exploits the metal--insulator phase change of vanadium dioxide, especially the hysteresis on thermal conductivity. Besides thermal rectification and thermal memristor, this strong nonlinear material has a wide range of applications not restricted in the thermal domain~\cite{132}. In fact, the electrical memristor and other memory metamaterials have been fabricated with VO$_2$ in labs earlier~\cite{133,134}. The bipolar thermal transistor introduced in Section 4 would also use its thermal hysteresis but on emissivity.

\subsection{Temperature trapping}

\noindent Usually, maintaining the temperature near a desired target needs to consume a lot of energy, for example, using an air conditioner or a refrigerator. In these scenarios, thermostats which are a temperature sensor and actuator to switch the heating or cooling functions on or off, should be designed carefully to reduce energy consumption. Keeping the temperature 
invariant is also related to the thermal cloak as it's expected to prohibit the heat flux from flowing
into the cloak without external energy input, whether it's designed based on the transformation theory~\cite{60} or solving the equation directly~\cite{65,66}. It seems that the thermal cloak can be used as an energy-free thermostat. However, conventional designs require an absolute zero thermal conductivity in some components of the structures, which can't be obtained in both simulations and experiments. So, when the ambient temperature changes, the region in the cloak should still suffer a significant temperature change.

\begin{figure}[!ht]
	\centering
	\includegraphics[width=1.\linewidth]{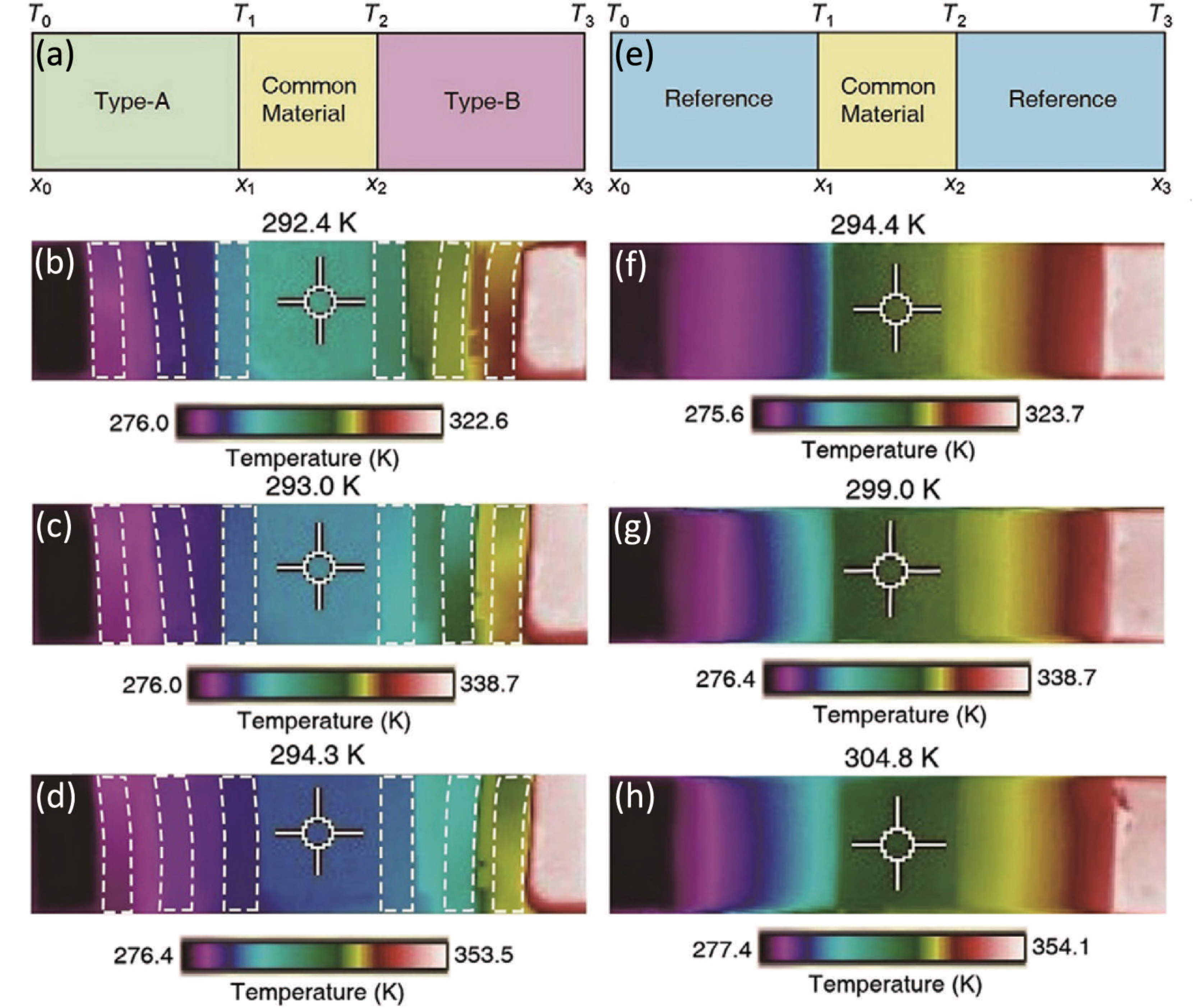}
	\caption{(a) Schematic of a energy-free thermostat to maintain the temperate in the middle yellow region. Type-A and Type-B are nonlinear materials. (b--d) are experimentally observed temperature distributions when the cold source is put on the left and the hot source with three different temperatures is put on the right. The temperatures in the middle of the device are marked. (e) Schematic of a reference in which all the materials are temperature-independent. (f--h) Corresponding experimental results for the reference group. Reproduced with permission from Ref.~\cite{135}. Copyright 2016 American Physical Society.}\label{2-4}
\end{figure}

\begin{figure}[!ht]
	\centering
	\includegraphics[width=.75\linewidth]{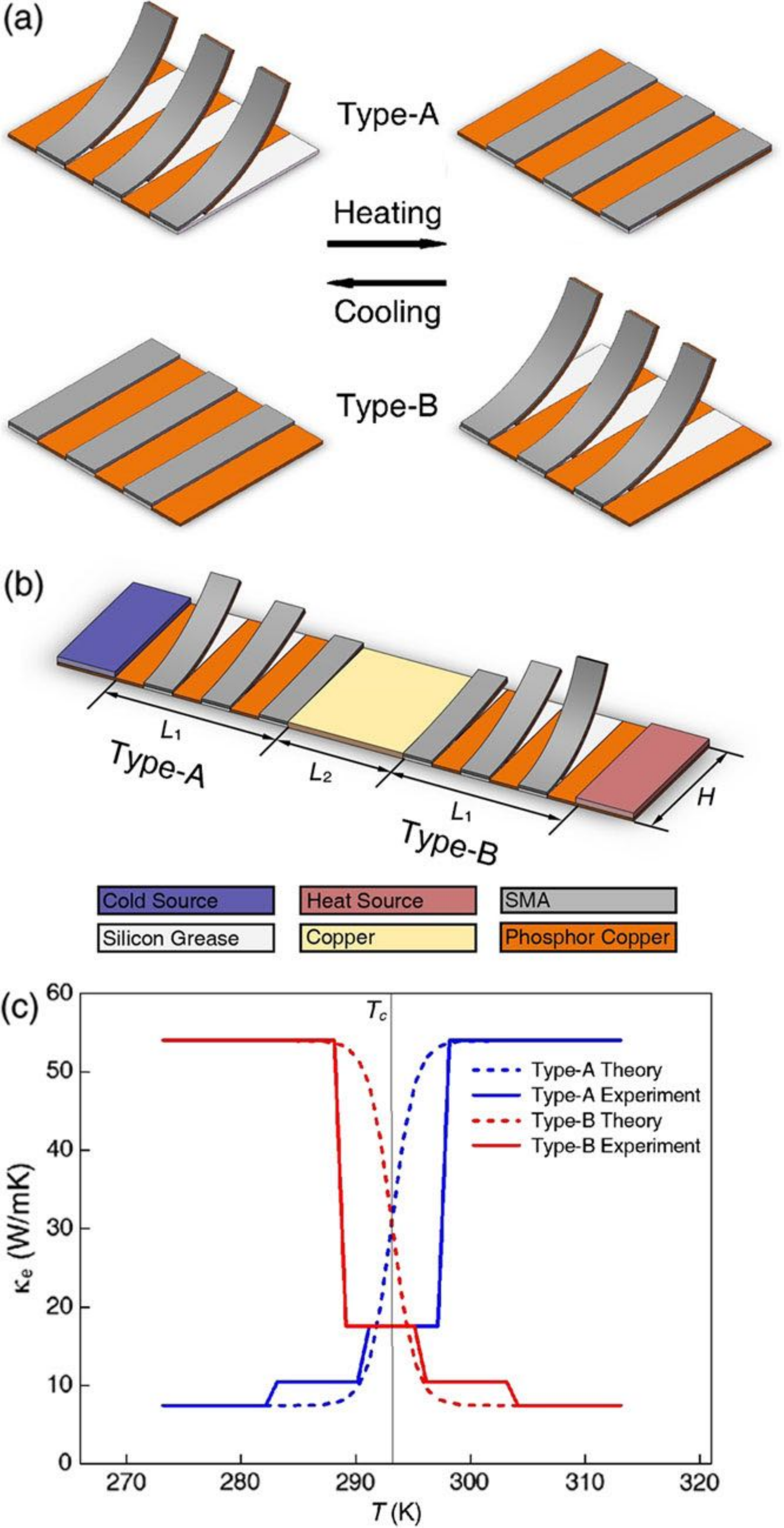}
	\caption{(a) Two types of bimetallic strips made of SMA (in gray) and phosphor copper (in orange). The SMA in Type-A tilt down at high temperatures while Type-B tilt down at low temperatures. (b) Structure of the whole energy-free thermostat. (c) Nonlinear thermal conductivities of Type-A and Type-B. Solid lines represent the realistic values while the dashed lines represent the required values by the temperature trapping theory.  Adapted with permission from Ref.~\cite{135}. Copyright 2016 American Physical Society.}\label{2-5}
\end{figure}
To build a realistic energy-free thermostat, Shen {\it et al.}~\cite{135} developed the temperature trapping theory. They considered an asymmetric three-segment model. As shown in Fig.~\ref{2-4}(a), the region in which they wanted to maintain the temperature is in the middle of such a two-terminal device, made of a common (linear) material with a high $\kappa$. 
The materials on both sides of the common material is highly nonlinear, named Type-A and Type-B respectively, whose thermal conductivities ($\kappa_A(T)$ and $\kappa_B(T)$) can be expressed using the logistic function as Eq.~(\ref{pcm}), writing~\cite{135}
\begin{equation}\label{sma}
{\begin{cases}\kappa_A(T) & =  \delta+ \dfrac{\varepsilon \exp(\lambda T-\lambda T_p) }{1+\exp(\lambda T-\lambda T_p)} ,\\\kappa_B(T) & =  \delta+ \dfrac{\varepsilon  }{1+\exp(\lambda T-\lambda T_p)}.\end{cases}}
\end{equation}
Here is $\delta$ a large thermal conductivity value while $\varepsilon$ is a small value. In addition, $\lambda=1$~K$^{-1}$, which corresponds to $\beta=-1$ and $\beta=1$ in Eq.~(\ref{pcm}) for Type-A and Type-B, respectively. 
As a result, Type-A has a higher thermal conductivity at high temperatures while Type-B behaves the opposite, and the two materials have the same transition temperature $T_P$. By solving the nonlinear heat conduction equation, they proved that the temperature in the middle region only depends on the value of $T_p$, and shouldn't be influenced by the temperatures of two heat sources if the hot one is put on the right side. Of course, the value of $T_p$ must be between the temperatures of the two sources. To fabricate the Tpye-A and Type-B modules, they built two types of bimetallic strips
composed of phosphor copper and different shape memory alloys; see Fig.~\ref{2-5}(a--b). For Type-A, the SMA strips tilt down and contact the phosphor copper strips when being heated, meaning changing into the conductive state. In contrast, the SMA strips in Type-B lie flat together with the phosphor cooper at low temperatures. The thermal conductivities of the two moving contact structures are presented in Fig.~\ref{2-5}(c). Since each type is composed of three copper strips and three SMA strips arranged alternately, th solid lines in Fig.~\ref{2-5}(c) both contains three steps, having a little difference with the required logistic curve. Anyway, the experiment results are also provided in Fig.~\ref{2-4} as a verification. We can see that the temperatures in the middle of the device only have minor changes when the temperature of the hot source increases, whereas the reference group has the same magnitude of change as the hot source. In principle, the SMA used here can be replaced by other two phase change materials tailored with the required thermal conductivities. It can also be found this energy-free thermostat can work like a thermal rectifier. Further, they optimized the structure of a bilayer thermal cloak by incorporating SMA components and simulation results confirm that the area in such a thermal cloak can maintain its initial temperature too~\cite{135}.

In addition, Wang {\it et al.}~\cite{136} considered the case when Type-A and Type-B are thermoelectric material. Since the middle region has a large thermal conductivity, the temperature difference would mainly exit in Type-A and Type-B and be can be utilized to generate thermovoltages. In this way, negative energy consumption can be achived in thermostats in ambient temperatures. What's more, they designed an optimized bilayer thermoelectric cloak in coupled thermal and electric fields, which can also behave as an energy source and a thermostat in addition to shielding the thermal and electric signals of an obstacle in it. Such thermostats accompanied by electricity generation can have valuable potential applications to energy-saving buildings, automobiles, aircraft, and so on.

\section{Transforming the nonlinear heat equation}

\noindent In this section, we turn to another basic method to design nonlinear elements especially the nonlinear thermal metamaterials, i.e., the transformation theory or transforming the nonlinear heat equation. Transformation theory was established and first applied on optics~\cite{137,138}, based on the fact that Maxwell's equations in arbitrary space-time coordinates (frames of reference) is form-invariant if all the physical quantities and differential operators are written in the covariant form. Here the form-invariance actually allows a difference in some coefficients caused by the varying volume elements related to the metric tensor. Similar to the equivalence between gravity and curved space-time, modulated electromagnetic properties of the materials can also bend the light in our daily flat space, and transformation optics is also called ``general relativity in electrical engineering''~\cite{139}. Transformation theory has be quite successful in designing all kinds of metamaterials to control light~\cite{140} and many other physical fields~\cite{141}, like acoustic waves~\cite{142}, quantum mechanics~\cite{143}, (mimicking) celestial mechanics~\cite{144,145}, fluid motion~\cite{146,147}, DC currents~\cite{148}, and of course, heat transfer~\cite{55,56,57,58}. Not every governing equation of a physical system can satisfy the requirement of transformation theory, for example, the conventional elastodynamic equations~\cite{149}. Heat conduction is the first diffusive system on which transformation theory has been established, and such an extension is not trivial. In 2008, Fan {\it et al.}~\cite{61} proposed transformation thermotics under the Fourier's law with a theoretical prediction for thermal cloak started the study of thermal metamaterials. In 2015, Li
{\it et al.}~\cite{62} generalized this theory to steady nonlinear heat conduction and temperature-dependent thermal conductivity began to be incorporated into the design of thermal metamaterials. Later, the 
applicability for transient nonlinear heat conduction was also proved~\cite{150}. In this section, we will first give the basic introduction of transforming the nonlinear Fourier's law. The applications of this nonlinear transformation thermotics are divided into two parts. The passive thermal metamaterials refer to the nonlinear counterpart of previous designs made of linear conductive materials under the same geometric transformation. The intelligent or active metamaterials (and metadevices) usually exploit new structures or new theoretical concepts, and can realize more flexible functions for controlling heat transfer.

\subsection{General theory}

\noindent Though transformation theory seems to have tight relationships with general gravity, we have known that the Fourier's law is even incompatible with special relativity due to the infinite propagating speed of thermal signals.  In addition, the Fourier's law is not frame-invariant because the movement of media should bring a convective term~\cite{151}. Nevertheless, the infinite speed doesn't matter here, and coordinate transformations used most in designs are actually restricted to spatial coordinates. The usual procedure to apply the transformation theory starts from writing the governing equation in the covariant form. If we confirm that Fourier's law is valid in the flat space, Eq.~(\ref{fuliye2}) under any curvilinear coordinate system with a set of contravariant coordinates $\{x^i,x^j,x^k\}$ in the three-dimensional Euclidean space $\mathbb{E}^3$ is~\cite{61}
\begin{equation}\label{trans-fuliye}
\sqrt{g}\rho C {\frac {\partial  T}{\partial t}}-\partial_i\left(\sqrt{g} \kappa^{ij}(T)\partial_j T\right)=0.
\end{equation}
Here $g$ is the determinant of the matrix $\begin{bmatrix}g_{ij}\end{bmatrix}$ composed of the metric tensor's covariant component $g_{ij}$, and $\kappa^{ij}(T)$ is the contravariant components of the nonlinear thermal conductivity tensor $\boldsymbol{\kappa}(T)$.
Writing such a covariant formulation doesn't bring any extra information as it's a trivial result of the definition of tensors. It's crucial to recall the motivation why the transformation theory is used. 

\begin{figure}[!ht]
	\centering
	\includegraphics[width=.85\linewidth]{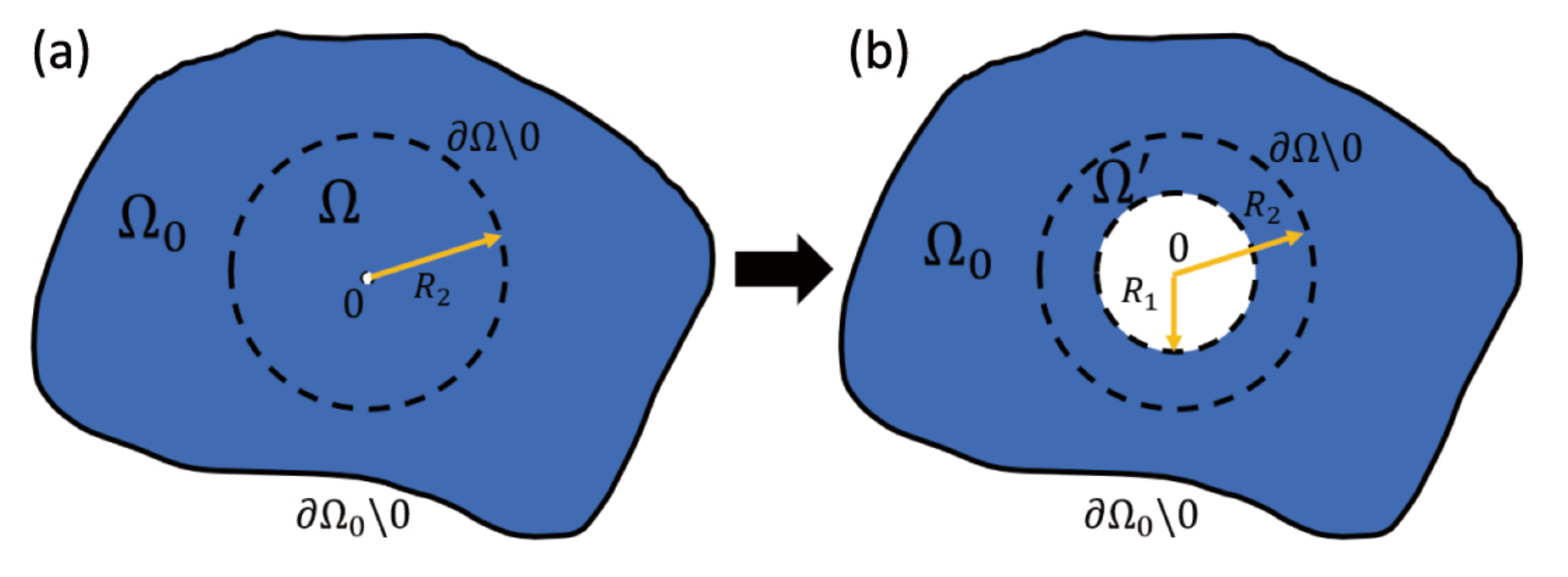}
	\caption{The geometric transformation from (a) to (b) for a cloak. In (a), the domain $\Omega$ is the pre-transformed region framed with a dashed line. In (b), the annular domain $\Omega'$ is th transformed region and the inner region in white disappears in the space.}\label{3-1}
\end{figure}

In Section 2, we mainly talked about how to solve the heat equation directly to find the suitable thermal conductivity $\kappa$ for certain thermal phenomena, whether the solution is exact or approximate. It's indeed the few fortunate situations where the inverse problem can be solved analytically, i.e., how to obtain a certain temperature distribution when the coefficients including $\kappa$ in the Fourier's law are not given? Often the inverse problems can have no analytical or exact solutions so the numerical optimization is needed. The essence of transformation theory is providing an analytical technique to solve some inverse problems. Then the motivation of transformation thermotics can be stated as follows. First, the pre-transformed temperature distribution $T(\mathbf r)$ with its corresponding material parameters and boundary conditions is given. Then the question is how to achieve the target temperature distribution $T'(\mathbf r')$ under the same boundary conditions (both $\mathbf r$ and $\mathbf r'$ are position vectors in $\mathbb{E}^3$). The transformation theory can work when $T(\mathbf r)$ and $T'(\mathbf r')$ can be linked through a geometric transformation: $f: \Omega \to \Omega'$ ($\Omega,\Omega'\subseteq\mathbb{E}^3$; see Fig.~\ref{3-1} as an example), which gives
\begin{equation}\label{mubiao}
T'(\mathbf r')=T(f^{-1}(\mathbf r')).
\end{equation}
Here we can see $f$ should be a bijection. We can write down the heat conduction equation in the transformed area $\Omega'$
with the density $\rho'$, specific heat capacity $C'$ and thermal conductivity tensor $\boldsymbol \kappa$ to be determined
\begin{equation}\label{hehe}
\rho'(\mathbf r')C'(\mathbf r') {\frac {\partial  T'(\mathbf r')}{\partial t}}-\nabla' \cdot \left( \boldsymbol{\kappa}'(T'(\mathbf r'))\nabla' T'(\mathbf r')\right)=0.
\end{equation}
Although $\Omega'$ itself can be a curved space, a part of the global Cartesian coordinates $\{x,y,z\}$ in $\mathbb{E}^3$ can still be used in Eq.~(\ref{hehe}) with another set of notations$\{x',y',z'\}$. The most critical point is that, the geometric transformation $f$ must have the same structure (isomorphism) with a coordinate transformation $h: \Xi \to \Xi'$ with $\Xi,\Xi'\subseteq\mathbb{R}^3$ ($\mathbb{R}^3$ is the three-dimensional real coordinate space), so Eq.~(\ref{trans-fuliye}) can be understood as~\cite{152}
\begin{equation}\label{trans}
\begin{aligned}
\sqrt{g}\rho C {\frac {\partial  T'}{\partial t}}=\partial_{x'}\left(\sqrt{g} \text{J}_{x'x}\kappa^{xy}(T')\text{J}^{\mathrm {T}}_{yy'}\partial_{y'} T'(x',y',z')\right)
\end{aligned}
\end{equation}
through a trick to use coordinates $\{x',y',z'\}$ instead of $\{x^i,x^j,x^k\}$ in form and the relationship $
\kappa^{ij} \text{J}^{-1}_{xi} \text{J}^{-\mathrm {T}}_{jy}=\kappa^{xy}$. In addition, $\text{J}$ is the Jacobian matrix $\begin{bmatrix}\partial x^i/\partial x\end{bmatrix}$ or $\begin{bmatrix}\partial x'/\partial x\end{bmatrix}$
and satisfies an identity with the metric as
$
\sqrt{g}=1/\det \text{J}.
$
The isomorphism between $f$ and $h$ also indicates that the often mentioned virtual space and physical space are diffeomorphic~\cite{139} as long as $f$ is smooth enough. Comparing Eq.~(\ref{hehe}) and Eq.~(\ref{trans}), we can finally obtain the transformed parameters written in the Cartesian coordinates~\cite{153}:
\begin{equation}\label{trans-conduction}
\left\{\begin{array}{ll}\rho'(\mathbf r')C'(\mathbf r')=\rho(f^{-1}(\mathbf r')) C(f^{-1}(\mathbf r'))/\det\text{J}(\mathbf r'),\\
\begin{bmatrix}\kappa'_{x'y'}(T'(\mathbf r'))\end{bmatrix}=\text{J}\begin{bmatrix}\kappa_{xy}(T(f^{-1}(\mathbf r')))\end{bmatrix}\text{J}^{\mathrm {T}}/\det\text{J}.
\end{array}\right.
\end{equation}
All the thermal conductivity tensors are expressed as matrices here to reduce ambiguity in their interaction with the Jacobian matrix, and in the Cartesian coordinate system we don't need to distinguish the covariant and contravariant components.
For a more general case when the pre-transformed parameters takes $\rho(T,\mathbf r)C(T,\mathbf r)$ and $\kappa(T,\mathbf r)$, we can easily check that the conclusions above is still valid. In addition, heat flux is transformed as
\begin{equation}\label{heat-flux}
\begin{aligned}
\mathbf j'(\mathbf r') =-\frac{\text{J}}{\det \text{J}}\left[ \boldsymbol{\kappa} \nabla T(f^{-1}(\mathbf r'))\right]=\frac{\text{J}}{\det \text{J}}\mathbf j(f^{-1}(\mathbf r')).
\end{aligned}
\end{equation}
It may be argued that $T'(\mathbf r')$ can also satisfies another heat equation
if multiplying all coefficients ($\rho'C'$ and $\kappa'$) by a constant. However, it can be proved that only 
Eq.~(\ref{trans-conduction}) can satisfy the interface condition for heat flux. We use the transformation for the cloak in Fig.~\ref{3-1} to give a more detailed explanation. This familiar transformation is~\cite{137}
\begin{equation}\label{tcloak}
\left\{\begin{array}{ll}r'=R_{1}+\frac{R_{2}-R_{1}}{R_{2}}r,\quad {\mbox{if }}0<r<R_{2},\\ \theta'=\theta.\end{array}\right.
\end{equation}
using polar coordinates $(r,\theta)$, and $\Omega$ is the region $0<r<R_2$ excluding the pole ``0''. The larger region $\Omega_0$ also doesn't contain the pole and $\Omega_0 \backslash \Omega$ is the ``background''. The heat sources or other boundary conditions are usually applied on the boundary $\Omega_0\backslash 0$. At the interface ($\Omega\backslash 0$) of the transformed region and background in Fig.~\ref{3-1}(b), the continuity of heat flux finally make 
Eq.~(\ref{trans-conduction}) the only possible solution that we have known. For a cloak, the transformed space doesn't cover the region $r'<R_1$ so what happens in it should have no impact on the outside world $r'>R_2$. Thus, in principle, the cloak and the region in it seem to be an infinite-stability system as the inside temperature could keep its initial value no matter what the boundary condition is. However, we have known such a perfect thermostat can't be realized due to imperfect insulation of realistic materials.   

\subsection{Passive metamaterials}

\noindent Here we can see the transformation rules in Eq.~(\ref{trans-conduction}) have the same form as its linear version by removing $T$ in $\kappa$, so the previous designs based on the linear transformation thermotics can be naturally extended more general cases. However, if we want to realize more different functions than linear thermal metamaterials such as active or adaptive control, only adding temperature-dependence into the pre-transformed material is not enough because the functions are determined by the geometric transformation. 
The usual geometric transformations are all temperature-independent which can induce passive functions like cloaking, concentrating and rotating, and the corresponding devices always exhibit the same function no matter how the ambient temperature changes~\cite{60}. These temperature-independent transformations are sometimes called ``linear transformations''~\cite{154}. By the way, the term ``nonlinear medium transformation'' used in Ref.~\cite{155} has another criterion for nonlinearity, meaning the transformed position vector $\mathbf r'$ has a nonlinear relationship with the pre-transformed $\mathbf r$, i.e., $f(\mathbf r_1+\mathbf r_2)\neq f(\mathbf r_1)+f(\mathbf r_2)$. We can see the transformation for a cloak in Eq.~(\ref{tcloak}) is linear in these two senses.

Anyway, the difficulty to reveal new phenomenon related to nonlinear heat transport with conventional transformations might be the reason why only a few works focus on nonlinear backgrounds. Here the ``background'' means the whole $\Omega_0$ since the pre-transformed materials in $\Omega$ and $\Omega_0\backslash \Omega$ are usually set the same and we will use this concept in the following part for simplicity of statement. In other words, the whole pre-transformed space is thought to be homogeneous (If $\Omega$ and $\Omega_0\backslash \Omega$ are filled with different materials, we must be careful that the transformation rules should be applied to the material in $\Omega$ and have no relationship with the material in $\Omega_0\backslash \Omega$). Sklan and Li~\cite{154} considered transforming Debye solids ($\kappa \propto T^3$ and $C \propto T^3$) to design a thermal cloak-concentrator with different functions within different temperature ranges, but its switchable functions are based on temperature-dependent (or the so-called ``nonlinear''~\cite{154}) transformation parameters which had been used in cases for linear background~\cite{62,150,156}, and we will leave the more detailed discussion in the next subsection about intelligent applications.

\subsection{Intelligent metamaterials}

\begin{figure}[!ht]
	\centering
	\includegraphics[width=.95\linewidth]{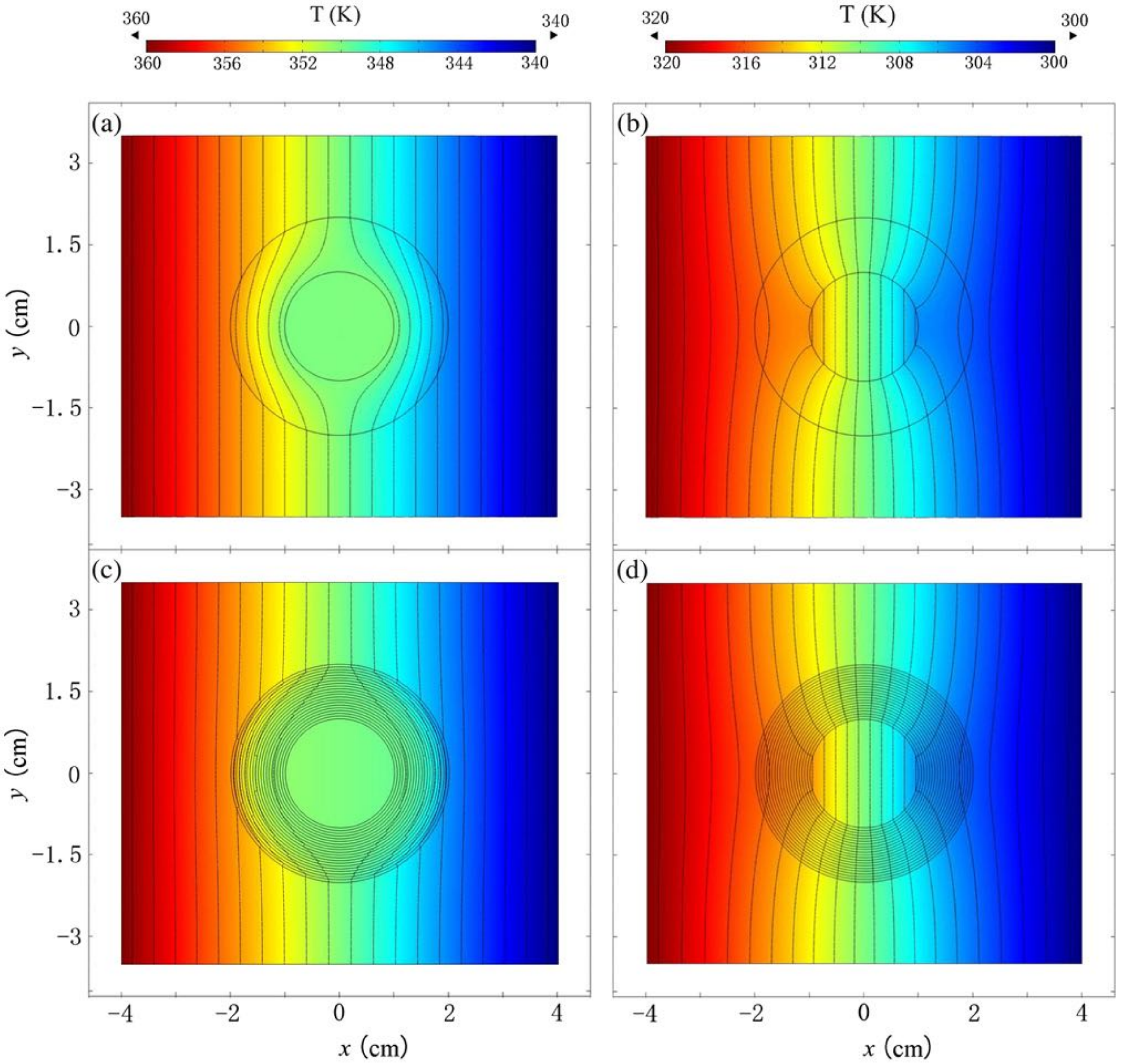}
	\caption{Simulation results of spatial temperature distribution for a Type-A switchable cloak. (a--b) The cloak with exact parameters works when the temperature range is 340--360~K or fails when the temperature range is 300--320~K. The black lines are isotherms. (c) and (d) stand for the results of a multilayered cloak with approximate parameters when the temperature range is also 340--360~K and 300--320~K, respectively. The switchable function is still valid. Adapted with permission from Ref.~\cite{62}. Copyright 2015 American Physical Society.}\label{3-2}
\end{figure}

\noindent Now the problem is how to make a passive metamaterial more intelligent, for example, let it behave as a thermal cloak at high temperatures but lose this ability at low temperatures. Li {\it et al.}~\cite{62} first developed a phenomenological technique that modifies the geometric transformation with a temperature-dependence. For the switchable thermal cloak, Eq.~(\ref{tcloak}) could be rewritten as
\begin{equation}\label{tcloak2}
\left\{\begin{array}{ll}r'=\widetilde{R}_{1}(T)+\frac{R_{2}-\widetilde{R}_{1}(T)}{R_{2}}r, \quad {\mbox{if }}0<r<R_{2},\\ \theta'=\theta.\end{array}\right.
\end{equation}
Here, the logistic function (recall Eq.~(\ref{pcm}) or Eq.~(\ref{sma})) is used in $\widetilde{R}_{1}(T)$. Generally, they designed two types of switchable cloak. Type-A cloak works at high temperatures and they took $\widetilde{R}_{1}(T)=R_1 \left(1-1/e^{\beta(T-T_P)}\right)$.
On the contrary, Type-B cloak works at low temperatures and $\widetilde{R}_{1}(T)=R_1/(1+e^{\beta(T-T_P)})$. Then the transformed thermal conductivity under the polar coordinate system should be 
\begin{equation}\label{tcloak3}
\begin{bmatrix}\kappa_{rr}&\kappa_{r\theta}\\\kappa_{\theta r}&\kappa_{\theta\theta}\\\end{bmatrix}
=\kappa_0\begin{bmatrix}\frac{r'-\widetilde{R}_{1}(T)}{r'}&0\\0&\frac{r'}{r'-\widetilde{R}_{1}(T)}\\\end{bmatrix}
\end{equation}
if the background/pre-transformed material has a linear thermal conductivity $\kappa_0$. We can take the Type-A cloak as an example. When the temperature is higher than $T_p$, the transformed thermal conductivity tends to the conventional value for the function of cloaking. However, when the
temperature is lower than $T_p$, the transformed thermal conductivity would be close to the pre-transformed one. The simulation results for the Type-A cloak in the steady regime are shown in Fig.~\ref{3-2}(a--b). The transition temperature $T_P$ takes 330~K here. We can see the isotherms are straight (though not evenly spaced) in Fig.~\ref{3-2}(a) while disturbed in (b), thus the switchable cloak is validated.
They also used multilayered structures made of two homogeneous isotropic nonlinear materials (Material A and Material B) to obtain the required inhomogeneity and anisotropy of transformed conductivity~\cite{62}. The thermal conductivities for Material A ($\kappa_A$) and Material B ($\kappa_B$) are $\kappa_A=\kappa_a+(\kappa_0-\kappa_a)/(1+e^{\beta(T-T_p)})$ and $\kappa_B=\kappa_b+(\kappa_0-\kappa_b)/(1+e^{\beta(T-T_p)})$ respectively, and meet the requirement $\sqrt{\kappa_a \kappa_b}=\kappa_0$.
Its corresponding temperature distributions are illustrated in Fig.~\ref{3-2}(c--d) with satisfying results. The Type-B cloak can be designed in a similar way.

\begin{figure}[!ht]
	\centering
	\includegraphics[width=.95\linewidth]{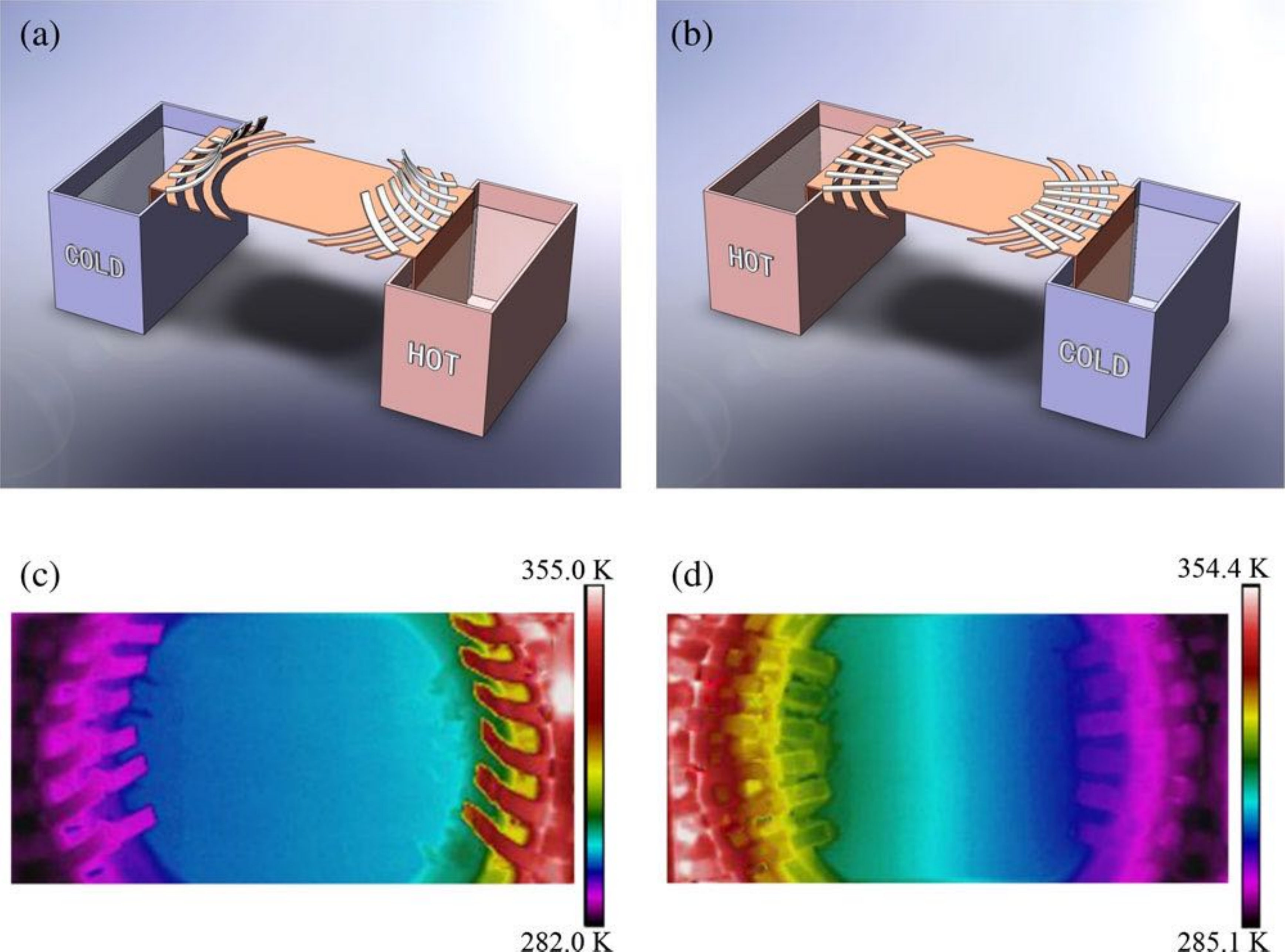}
	\caption{(a--b) Design sketch of a macroscopic thermal diode. Two different bimetallic strips made of SMA and copper as a part of Type-A cloak or Type-B cloak are placed on th right and side of a copper plate, respectively. (a) The reverse mode when the hot source is applied on the right. All the SMAs tilt up and reduce the heat conduction. (b) The forward mode. All the SMAs tilt down, contact the cooper plate and enhance the heat conduction. (c--d) The corresponding experimentally observed temperature distributions for the reverse and forward modes. Adapted with permission from Ref.~\cite{62}. Copyright 2016 American Physical Society.}\label{3-3}
\end{figure}

Based on the structures of Type-A and Type-B cloaks, they further fabricated a macroscopic thermal diode~\cite{62}. Thermal conductivities of logistic function type are realized through two types of SMA which we have introduced when talking about energy-free thermostats. Two different SMA strips are used and they can have mirror-symmetric $\kappa$--$T$ curves (recall Fig.~\ref{2-5}(c)) according to the deformation direction of their thermal expansion. 
See Fig.~\ref{3-3}(a--b). The diode has an asymmetric three-segment structure with a cooper plate in the middle. A  rectangular part cropped from the Type-A cloak is put on the right side while another part of the Type-B cloak is put on the left side. Each part is made of copper and SMA strips with expandable polystyrene (EPS) filling the gaps. Their simulation predicts a high rectification ratio with 0.97, and the experimental results of temperature distributions in the reverse and forward modes are shown in Fig.~\ref{3-3}(a--b). In fact, it might be the first dual--phase-change thermal diode built in labs while its components on the two sides is additionally anisotropic, which is different from other thermal diodes.

In the transient regime, Li {\it et al.}~\cite{150} designed a switchable concentrator using the same technique to incorporate a switching or phase-change factor into the geometric transformation 
\begin{equation}\label{tconcen}
\left\{\begin{array}{ll}r'=\frac{R_{1}}{R_{3}}r,\quad {\mbox{if }} r<R_{3}, \\
r'=\frac{R_{1}-R_{3}}{R_{2}-R_{3}} R_{2}+\frac{R_{2}-R_{1}}{R_{2}-R_{3}}r,\quad {\mbox{if }} R_{3}<r<R_{2},\\\theta'=\theta.\end{array}\right.
\end{equation}
Eq.~(\ref{tconcen}) squeezes the core $r<R_3$ first and then stretches the outer ring $R_3<r<R_2$ to compensate the disappearing area, so the heat flux in $r'<R_1$ can be amplified.
In their work, they chose to replace $R_3$ by $\widetilde R_3(T)=R_{1}+\left(R_{3}-R_{1}\right) /\left(1+\mathrm{e}^{\beta\left(T-T_{p}\right)}\right)$ and the sign of $\beta$ can determine whether the concentrating function turns on at high or low temperatures.
In addition, for transient conduction, the density or the specific heat capacity should also be modulated by the determinant of a temperature-dependent Jacobian matrix. Later, a dual-function device named thermal cloak-concentrator was constructed using SMA again~\cite{156}. This device can automatically change from a cloak to a concentrator or vice versa when the ambient temperature
varies. For steady cases, only the region $R_1<r'<R_2$ needs to be transformed, because the concentrating transformation keeps the thermal conductivity invariant in $r'<R_1$ and the cloaking transformation doesn't care the material properties in it. The key point here is that the transformation by Eq.~(\ref{tcloak2}) using a logistic function should be further modified so it can be approximately equivalent to the cloaking and concentrating transformations above or below the transition temperature respectively~\cite{156}. In other words, we should tune $\kappa_h$ and $\kappa_c$ in Eq.~(\ref{pcm}) to meet this requirement. 

The three devices above all have a linear pre-transformed thermal conductivity. As we have mention before, Sklan and Li also designed a thermal cloak-concentrator but in a nonlinear background~\cite{154}. Similarly, they used the hyperbolic tangent function $\tanh$ to construct a phase-change temperature-dependent transformation. The $\tanh$ function is also a sigmoid function and indeed, a rescaled logistic function.
The two thermal cloak-concentrators have the same switchable function although they have completely different background thermal conductivities, which provides another evidence that the function only depends on the transformation.

\begin{figure}[!ht]
	\centering
	\includegraphics[width=.8\linewidth]{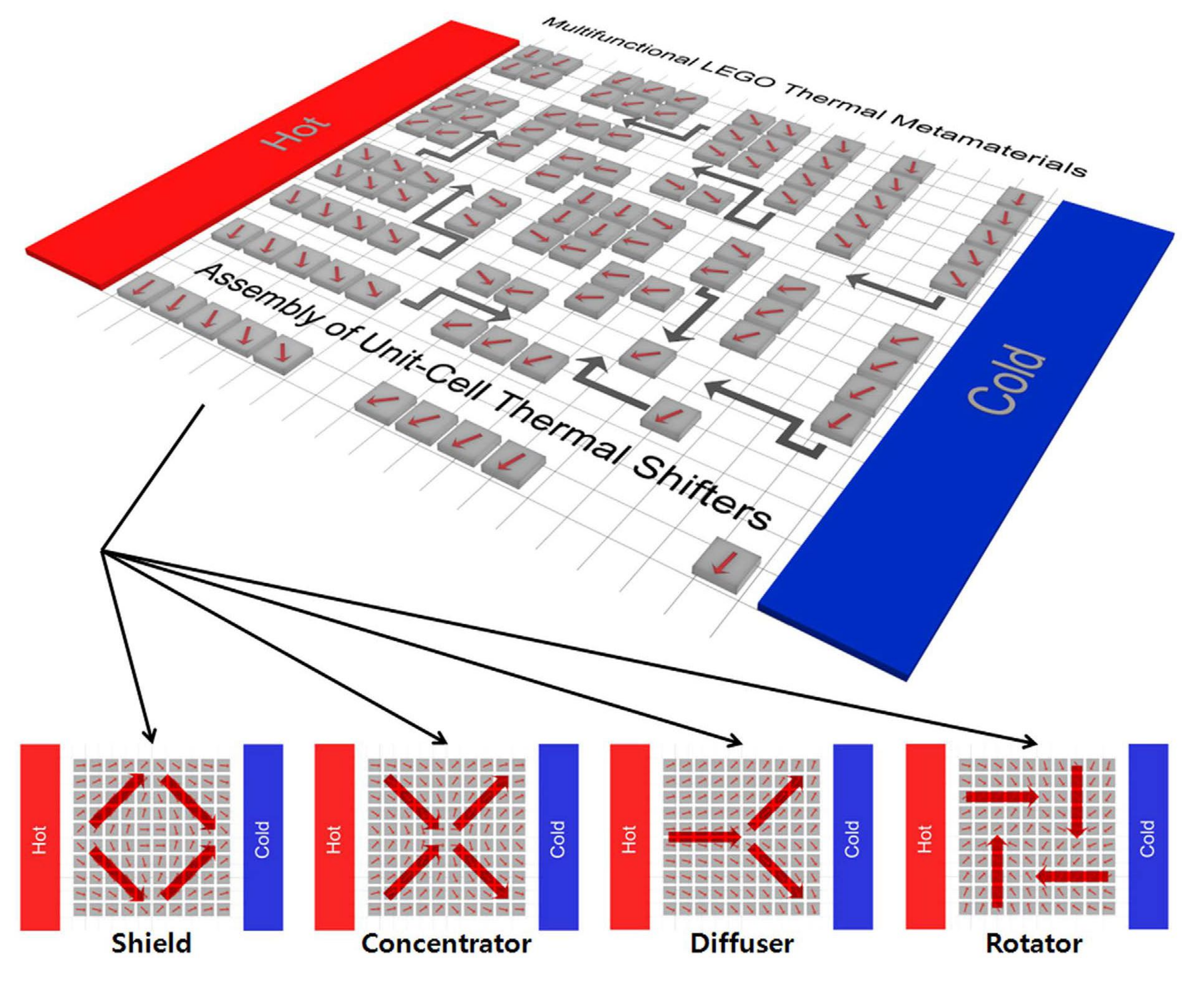}
	\caption{Scheme of multifunctional ``LEGO'' thermal metamaterials composed by unit cells as if playing with LEGO bricks. Hot and cold sources are respectively applied on the left and right boundaries. The red arrow in the unit cell represents its function to shift the direction of local heat flux. The four subplots below are four examples of what functions can be realized on such a modulation platform. When each cell are made of phase change materials, the multifunctional thermal metamaterials can also be temperature-responsive with a switchable function. Adapted from Ref.~\cite{158}. Copyright 2017, The Authors of Ref.~\cite{158}.}\label{3-4}
\end{figure}

Similar to the linear cases, we have seen that fabricating nonlinear metamaterials also needs the design of spatially-discrete structures like the curved multilayered ones in Fig.~\ref{3-2}(c--d) and Fig.~\ref{3-3}(a--b) to approximate the continuous change of thermal properties required by transformation theory. Another method to do a digitization of thermal conductivity distribution is griding the space with squares, which has been applied in linear conduction~\cite{158,159,159bb}.
Recently, inspire by the LEGO bricks, Kang {\it et al.}~\cite{160} built a new type of temperature-responsive metamaterials assembled from tunable unit cells as an extension of their previous work~\cite{158}. Each cell has an anisotropic thermal conductivity according to a geometric transformation $f$:
\begin{equation}
(x',y')=f(x,y)=(x,x\sin\theta+y\cos\theta).
\end{equation}
This transformation means shifting the direction of the heat flux by an angle $\theta$ thus the unit cell is called a thermal shifter.
The combination of shifters with different angle parameters can realize various basic manipulation of heat flux, such as a shield (cloak), a concentrator, a diffuser (splitting the flux into two directions), and a rotator; see Fig.~\ref{3-4}. Again, a phase-change factor related to the temperature is put in the parameter $\theta$ using the Heaviside step function $H$:
\begin{equation}
\theta(T)=\theta_d H(T-T_p).
\end{equation}
Under the transition temperature $T_p$, the unit cell is isotropic and no transformation happens, while the heat flow is turned an angle 
above $T_p$. These tunable unit cells are fabricated using layered structures of stainless steel and phase change nanocomposites which are made of carbon nanotubes and copper powder embedded into n-octadecane. The PCMs here can melt or freeze and thus behave as the switch providing high or low thermal conductivities. Using a $4\times4$ modular structure (16 unit cells) they successfully built a switchable thermal shield (cloak), and used it for thermal management in printed circuit boards as it can insulate the heat or improve heat dissipation when put on different positions surrounded by different environment temperatures~\cite{160}. Also, their method can be feasible to construct other switchable or temperature-responsive metamaterials like the concentrator, diffuser and rotator, and integrate them into a multifunctional
thermal conduction modulation platform.

\section{Field effects as nonlinearity}

\noindent Nonlinearity in heat conduction discussed in the previous sections mainly comes from the temperature-responsive thermal conductivity. If we don't consider nonlinear density or nonlinear specific heat capacity in the transient regime, new mechanisms to introduce nonlinearity in heat transport should ask multiphysics effects for help. Actually, we have talked about some works based on thermo-mechanics in Sections 2--3. The word ``multiphysics'' here can be used to refer to multiple modes of heat transfer like conduction-radiation or convection process, and a thermal filed coupled with other physical fields like elasticity. The methods that solves or transforms the multiphysics equations will be included as long as heat conduction is governed by the Fourier's law.

\subsection{Radiation}

\begin{figure}[!ht]
	\centering
	\includegraphics[width=.65\linewidth]{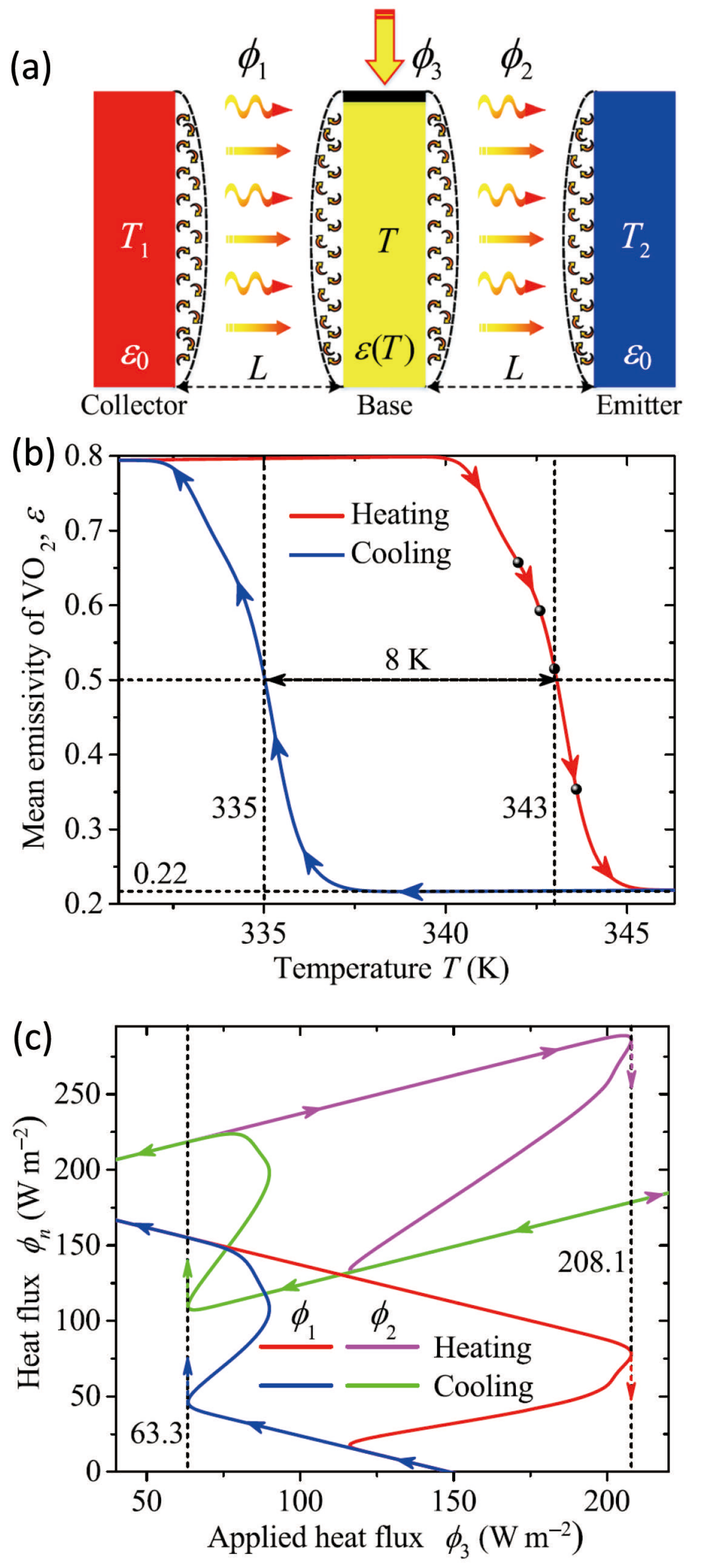}
	\caption{A bipolar thermal transistor. (a) is the schematic diagram. The conductive, convective and radiative heat fluxes are indicated by round, straight, and wavy arrows respectively. Both the collector and emitter are thermostats at temperatures at $T_1$ and $T_2$ while an external flux $\phi_3$ is injected into the base made of VO$_2$. (b) shows the thermal hysteresis loop of how the emissivity of VO$_2$ varies with temperature. (c) plots heat fluxes $\phi_1,\phi_2$ versus the applied flux $\phi_3$ when cooling or heating the base. Reproduced with permission from Ref.~\cite{172}. Copyright 2016 American Physical Society.}\label{4-1}
\end{figure}

\noindent Nonlinear radiative heat transport itself is an important and hot topic such as the photonic thermal rectification~\cite{161,162}. Particularly, by utilizing VO$_2$, the radiative thermal diode~\cite{163,164}, filed effect transistor~\cite{165}, memory elements~\cite{166,167}, memristor~\cite{168}, and shuttling~\cite{169} have all been designed or fabricated, mainly in the near field regime. These contactless devices without the Kapitza resistance can have some advantages compared with their conductive counterparts~\cite{170}. However, here we want to focus on cases in which thermal radiation behaves as an extra nonlinear term in the Fourier's law.  
First we consider the familiar far-field radiation model for a black or gray body. In a steady state, the heat equation for a fin exchanging heat by radiation (Stefan-Boltzmann law), conduction, and convection (Newton's law of cooling) is~\cite{171}
\begin{equation}\label{basic}
\nabla\cdot(\kappa \nabla T)=\frac{h C_p}{ A}\left(T-T_{\text {env }}\right)+\frac{\varepsilon \sigma C_p}{ A}\left(T^{4}-T_{\text {sur }}^{4}\right).
\end{equation}
Here $\sigma$ is the Stefan's constant, $\varepsilon$ is the emissivity, $h$ is the heat transfer coefficient, $A$ is the surface area of heat flow, $T_{\text {env }}$ is the environment temperature of ambient fluids, and $T_{\text {sur }}$ is the temperature of radiating surroundings. The radiative term brings a nonlinearity since it includes $T^4$. However, it might need more detailed designs to realize nonlinear elements under such a regime. Ordnez-Miranda {\it et al.}~\cite{172} designed a bipolar transistorlike thermal device using PCM VO$_2$. The three parallel components of the device includes a collector (temperature $T_1$), an emitter (temperature $T_2$), and the base (temperature $T$; made of VO$_2$) between them; see Fig.~\ref{4-1}(a). Also the convection and conduction of the intracavity gas between the three components are considered, so the heat flux exchange for base-collector/base-emitter ($n$ takes 1 or 2 respectively) is~\cite{172}
\begin{equation}
\phi_n=(-1)^{n}\left[G\left(T-T_{n}\right)+\varepsilon_e(T) \sigma\left(T^{4}-T_{n}^{4}\right)\right].
\end{equation}
The conductive effect of the intracavity gas is absorbed in $G=(h+\kappa/L)^{-1}$, taking $\nabla_z T=(-1)^n(T_n-T)/L$. The effective emissivity $\varepsilon_e(T)=(1/\varepsilon_0+1/\varepsilon(T)-1)^{-1}$ is temperature-responsive due to the similar property of $\varepsilon(T)$ for VO$_2$, while the emissivity $\varepsilon_0$ for the surface of the collector/emitter is temperature-independent. The working principle for such a transistorlike device is the change of transition temperature of VO$_2$ under cooling and heating driven by a external flux $\phi_3=\phi_2-\phi_1$, meaning the $\varepsilon$--$T$ curve shows a hysteresis loop (shown in Fig.~\ref{4-1}(b)). Further, the amplification function of the transistor means $\alpha=\partial \phi_2/\partial \phi_3>1$, while the amplification ratio $\alpha$ is given by~\cite{172}
\begin{equation}\label{bipolar}
\alpha=\frac{1}{2}\frac{\varepsilon_{e}^{\prime}(T)\left(T^{4}-T_{2}^{4}\right)+4 \varepsilon_{e}(T) T^{3}+G / \sigma}{\varepsilon_{e}^{\prime}(T)\left(T^{4}-T_{e}^{4}\right)+4 \varepsilon_{e}(T) T^{3}+G / \sigma},
\end{equation}
where $T_{e}=\sqrt[4]{\left(T_{1}^{4}+T_{2}^{4}\right) / 2}$ and $\varepsilon_{e}^{\prime}(T)=d \varepsilon_{e}(T) / d T$. 
It is easy to see that $\alpha$ can be maximized when the denominator approaches zero. For a bipolar transistor, the base temperature $T$ is modulated by the applied flux $\phi_3$, and the solution $T_p$ (more than one value) for a zero denominator in Eq. (\ref{bipolar})  corresponds to the condition that $\partial T/\partial\phi_3$ is divergent. Soon this divergence can be related to the phase change of VO$_2$, and $T_p$ is just the transition temperature. Finally, the fluxes $\phi_1,\phi_2$ as the functions of $\phi_3$ are calculated and shown in Fig. (c) and we can see flux amplification exists.
In fact,  $\varepsilon(T)$ adds another nonlinearity to heat transfer, and the competition of $T^4$ (radiation) and $\varepsilon(T)$ ($\text{d} \varepsilon_{e}(T) / \text{d} T<0$ for a single heating or cooling process) can induce thermal bistability. As a result, we can see bi-valued $\phi_1,\phi_2$ in Fig.~\ref{4-1}(c).
What's more, the temperature on the base could suffer a jump at the phase transition points of VO$_2$, and this phenomenon can be used to efficiently heat or cool down something. Also, the device is marcoscaled and can work in a normal working condition (300--350~K).

\begin{figure}[!ht]
	\centering\includegraphics[width=1.\linewidth]{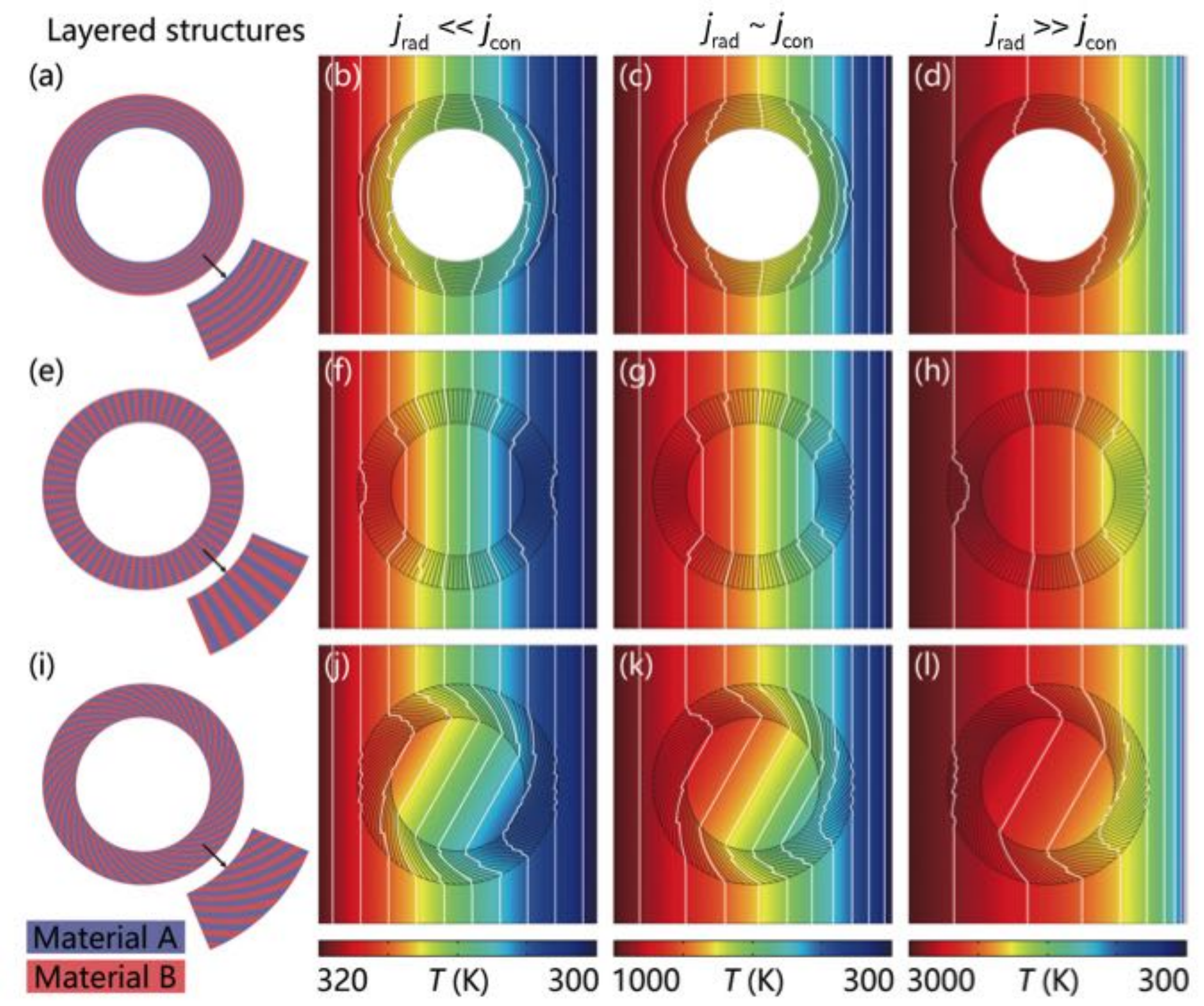}
	\caption{Thermal metamaterials for controlling heat conduction and Rosseland radiation model. (a) shows the multilayered structures using to get for a themral cloak. (b--d) respectively stand for the spatial temperature distributions solved by FEM for a cloak when applying different thermal bias. The relative role of conductive flux $j_{\text{con}}$ and radiative flux $j_{\text{rad}}$ are different in (b--d), meaning $j_{\text{rad}}\ll j_{\text{con}}$, $j_{\text{rad}}\sim j_{\text{con}}$ and $j_{\text{rad}}\gg j_{\text{con}}$ in turn. (e--h)/(i--l) show the structures and simulation results for a concentrator/rotator.  Reproduced with permission from Ref.~\cite{186}. Copyright 2020 American Physical Society.}\label{4-2}
\end{figure}

Now we turn to another far-field radiation model called radiative Fourier law or Rosseland diffusion approximation~\cite{173,174}. When the radiative process slightly deviates from from Stefan-Boltzmann law in optically-thick media, the radiative heat flux density $\mathbf j_{\text{rad}}$ is proportional to the temperature gradient, writing~\cite{173}
\begin{equation}
\mathbf j_{\text{rad}}=-\frac{16}{3}{\beta}^{-1}{n}^2\sigma T^3\cdot{\nabla}T.
\end{equation}  
Here $n$ is the refractive index, and $\beta$ is the Rosseland mean opacity.
Rosseland diffusion approximation has been widely used for studying the thermal transport properties in high-temperature environments, such as
the mantle in Earth and other planets~\cite{175,176,177}, and porous insulation materials like fibers and aerogels~\cite{178,179,180,181,182,183,184,185}.
Here, the radiative conductivity $\kappa_{\text{rad}}=\frac{16}{3}{\beta}^{-1}{n}^2\sigma T^3$ can be seen as a nonlinear term in thermal conductivity so the nonlinear transformation thermotics can work in radiation-conduction systems under Rosseland approximation. Xu {\it et al.}~\cite{186} gave a series of meatmaterial including thermal cloaks, concentrators, and rotators when the background is filled by linear conductive materials with a nonlinear radiative thermal conductivity. Also, they used two isotropic homogeneous materials to meet the required anisotropy for both the conductive and radiative conductivities, which follows the multilayered structures from previous conductive thermal metamaterials; see Fig.~\ref{4-2}. The scattering cancellation method can also work here. For example, Xu and Huang~\cite{187} designed circular/elliptical core-shell structured metamaterials with functions like thermal transparency, cloaking (the shell or cloak can be single-layered or bilayered) and expanding within such a conduction-radiation system. In addition, these methods have been generalized to omnithermotics in which thermal convection is taken into consideration as well as conduction and diffusive radiation~\cite{188,189}.

The relationship $\kappa_{\text{rad}}\propto T^3$ between radiative conductivity and temperature can also be obtained directly from Stefan-Boltzmann law~\cite{73}. If we write the term $T^4-T^4_{\text{sur}}$ in Eq.~(\ref{basic}) as $(T-T_{\text{sur}})(T^3+T^2T_{\text{sur}}+TT^2_{\text{sur}}+T^3_{\text{sur}})$ and take $(T-T_{\text{sur}})/L$ as the negative temperature gradient ($L$ is the distance to the radiating surroundings), the high temperature limit for the effective radiative conductivity is proportional to $T^3$.
Besides the simple expression as the third power of temperature, radiative conductivity can have more complicated relationships on temperature~\cite{190} and even be anisotropic~\cite{191}. In these more general cases, transformation theory is universally effective. Su {\it et al.}~\cite{192} also applied the scattering cancellation method to materials whose total conductivities are polynomials of temperature. They further designed an intelligent device which concentrates the heat flux at low temperatures while shields it at high ones, using the basic principle that the constant part competes with the temperature-dependent part of conductivity when the temperature varies.

\subsection{Convection}

\noindent The last basic mechanism of heat transfer we haven't talked about carefully is convection since the Newton's law of cooling mentioned above can be seen as an application of Fourier's law in the normal direction of the solid-liquid surface. Convection actually includes the heat transfer driven by both the movement of media (advection) and spatially varying temperature (conduction).The heat equation for convection or advection-conduction is the same thing as Fourier's law in a sense. Replacing $\frac {\partial }{\partial t}$ in Eq.~(\ref{fuliye2}) by the material derivation ${\frac {\mathrm {D} }{\mathrm {D} t}}\equiv {\frac {\partial }{\partial t}}+\mathbf {u} \cdot \nabla  $, we can directly obtain the equation for heat transfer of moving media:
\begin{equation}
\rho C\frac {\partial  T}{\partial t}-\nabla\cdot\left(\kappa \nabla T\right)+ \rho C \mathbf {u}  \nabla T=0.
\end{equation}
Here $\mathbf {u}$ is the velocity of media. 

\begin{figure}[!ht]
	\centering
	\includegraphics[width=.8\linewidth]{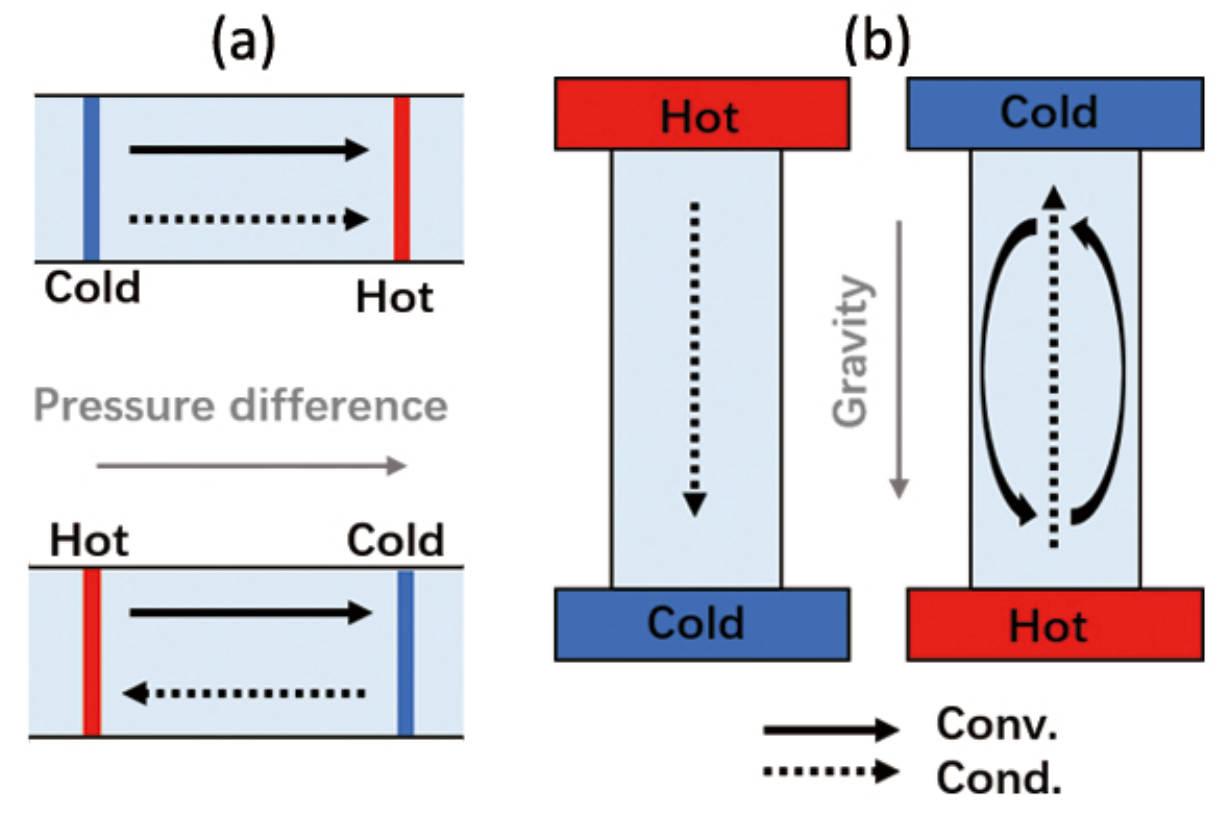}
	\caption{Thermal rectification induced by (a) forced convection and (b) Rayleigh–Bénard convection. The arrow with a solid line represents the convective (conv.) heat flux while the arrow with dotted lines represents the conductive (cond.) one. Usually the upper subplot of (a) and the left subplot of (b) correspond to the forward mode while the left are reverse modes. In particular, the chamber in (b) has closed boundaries for fluid flows while the channel in (a) is open-bounded and the velocity is set to the same everywhere.}\label{4-3}
\end{figure}

If $\mathbf {u},\rho$ and $C$ are functions without an explicit variable of the temperature, the advection-conduction equation is still linear and the solution of temperature satisfies the superposition property.
However, the local constitutive relationship of heat flux density doesn't show a linear response to the temperature gradient anymore, i.e., 
\begin{equation}
\mathbf j=-\kappa\nabla T+\rho C_p T\mathbf u.
\end{equation}
Here, for simplicity, the reference temperature of the environment $T_{\text{env}}$ is neglected for all $\left(T-T_{\text{env}}\right)$ above. More specifically, the constitutive relationship of heat flux density along the $x$ axis is $j_x=-\kappa\nabla_x T+\rho C_p T u_x$ since we only reverse the temperature bias in this direction. When the conductive flux is much larger than the advective one, there should be no rectification. 
In the opposite case when advection dominates the heat transfer (meaning large Prandtl numbers Pr), isotherms of the temperature will be squeezed at the heat ($T_{h}$) or cold ($T_{c}$) source when exchanging two heat sources, which forms quasi-uniform temperature distributions and thus gives a rectification ratio $\gamma$ estimated as $(T_{h}+T_{c})/T_{h}$. This estimation comes from the trick that the heat flux density is uniform in a quasi-one-dimensional channel without intrinsic heat sources. If we write back the environment temperature, the rectification ratio can be expressed more precisely as 
\begin{equation}
\gamma\approx\frac{\left|T_{h}+T_{c}-2T_{\text{env}}\right|}{\text{Max}\left\{\left|T_{h}-T_{\text{env}}\right|,\left|T_{c}-T_{\text{env}}\right|\right\}}.
\end{equation}
The key point here is $u_x$ is invariant in the forward and backward modes, and changing the speed of media (or Pr) can tune the rectification ratio in a certain range. So, this can be an example of making a thermal diode in a linear system or with temperature-independent materials. Another possible technique to realize rectification under linear conduction is using asymmetric thermal conductivity tensor~\cite{23}. When the heat fluxes in the two modes have the same direction, $\gamma$ must be larger than 1, and at the same time, there must exist a mode in which the heat is carried from the cold source to the heat source, showing a trivial example of heat pump or shuttling.  We must point out that our conclusion are based on uniform velocity distribution (see the channel in Fig.~\ref{4-3}(a) with open boundaries) so the analysis based on local constitutive relationship of heat flux can work for the global effect of rectification. For more complicated velocity distributions especially in two/three-dimensional systems which can't degenerate to quasi-one-dimensional cases, the motion of media might fail to generate thermal rectification~\cite{23,193}. Recently, some works found convection's influence on temperature distribution can be mimicked through spatiotemporal modulation of the thermal conductivity and density (or specific heat capacity) for wave-like temperature profiles~\cite{194,195}, but the crucial asymmetric heat flux for the claimed thermal rectification/nonreciprocity is still lacking~\cite{194}.

Now we turn to a more comprehensive case. The convection state described by $(T,\mathbf u,\rho)$ is totally determined by advection-conduction equation, Navier-Stokes equations and the law of continuity~\cite{18}. 
It's known that natural convection (e.g., the Rayleigh–Bénard convection in a plane horizontal layer~\cite{197}; see see Fig.~\ref{4-3}(b)) is driven by gravity/buoyancy and vertical density difference when thermal expansion exists. Thermal expansion means the density varies with temperature and thus the heat equation is nonlinear. In addition, Navier-Stokes equations are also famous strongly nonlinear equations, and the nonlinear fluid dynamics can be quite complicated in thermal convection~\cite{198,199,200,201}. For example, in history, chaos or strange attractor was first found just in atmospheric convections~\cite{202}. Nevertheless, rectification can also be realized in natural convection through basic qualitative analysis. Roberts and Walker~\cite{22} considered the fluids under gravity with heating/cooling sources on the top/bottom side or vice versa. The Rayleigh–Bénard convection can only happens when being heat from below (denoted as the forward mode) since the fluids usually are lighter at high temperatures, so thermal rectification should happen when exchanging the sources. The rectification effect is major determined by the Rayleigh number Ra measuring the relative contribution of buoyancy-induced heat transfer and thermal conduction. 

Wong {\it et al.}~\cite{203} further fabricated a more asymmetric structure called water-vapor chamber thermal diode. In the chamber, water is only partially filled in the bottom and the air occupies the upper area. Also, the Rayleigh–Bénard convection can only happen in the forward mode for the two fluids and the convection can be dominant as the two fluids both have large Ra higher than the critical value. What's more, the vaporation of water at the liquid-air interface is considered as a vapor layer, whose convective effect is significant when the temperature bias is larger than 30~K and can enhance the heat transfer performance in the forward mode. Through theoretical analysis of the total thermal resistance/conductance from conduction, convection and liquid-gas interface latent heat transfer (related to the enthalpy of vaporization) as well as experimental verification, they found the rectification ratio increases with increasing thermal bias $\Delta T$ (larger Ra) and is also influenced by the water-air volume ratio. The highers rectification ratio they obtained is 1.43, taking  $\Delta T=50$~K and filling half of the chamber with water. Soon after, Pugsley {\it et al.}~\cite{204} realized a similar structure called horizontal planar liquid-vapor thermal diode. The planar here is flatter than the previous axial chamber, so the vapor from the hotter bottom in the forward mode can arrive and condensate at the upper side (a colder plate), and thus transfer more heat. They got a high rectification ratio close to 1.

From Refs.~\cite{203,204}, we can see phase changes can also be introduced in convective systems to design nonlinear elements, such as vaporation/condensation~\cite{205,206} and melting/solidification~\cite{207}. Any way, the basic working principle is still looking for different thermal conductance in different modes and the conductance should be somewhat temperature-related, whether the relationship is apparent or implicit in the direction of convection and various types of thermally-induced phase changes. As a summary of this part, we have mainly introduced convective thermal diodes based on forced convection or natural convection. For forced convection, the conclusions are also valid for moving solid medium so the discussion here is not limited to fluid thermal diodes. Some engineering designs 
for gas/liquid thermal diodes and other devices like switches can have more complicated structures like heat pipes, which are beyond our concern in this part and can refer to Refs.~\cite{208,209}.

\subsection{Other field effects}

\noindent Besides the radiation (electromagnetic field) and convection (mass transport), other multi-physical effects can be incorporated into the Fourier's law to behave as or induce nonlinearity. Though some important mechanisms like thermoelectrics~\cite{41} and thermo-optic effects~\cite{210,211} have been widely researched in mesoscopic and nanoscale structures, it is still worth continuing to exploiting the field effects at the phenomenological level. For example, the works based on the shape change of SMA~\cite{62,95,135} or the thermal expansion of Si~\cite{112} or mercury~\cite{99} can be seen as thermo-mechanical effects as we have mentioned in Section 2. What's more, thermal convection is actually a special thermo-mechanical process as well.

\begin{figure}[!ht]
	\centering
	\includegraphics[width=.6\linewidth]{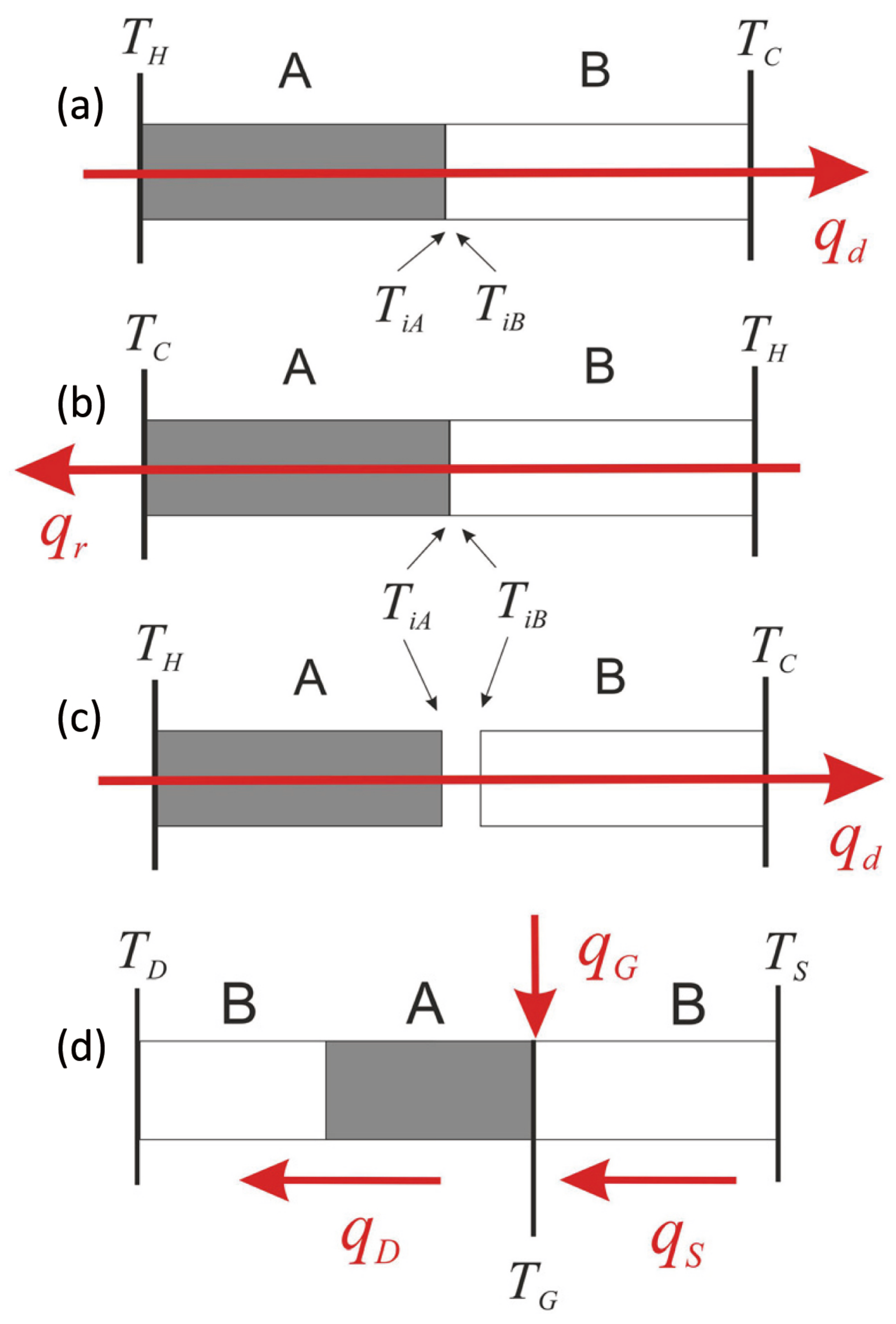}
	\caption{(a--c) are the sketch of a thermoelastic heat switch. Material A (in gray) is Si and Material B (in white) is Ge. $T_H$ or $T_C$ is the temperature of heat/cold source, and $T_{iA}, T_{iB}$ are the interface temperatures which are the same in the absence of a gap. In (a--b) there is no gap between the bars while in situations like (c) the gap can reduce the heat transfer. As a result, (a) and (c) can be the on and off states respectively. (d) shows a thermoelastic heat transistor. This field effect transistor ia made of a Si bar (gate) sandwiched between two Ge bars (drain and source). The gate temperature $T_G$ is variable while both the source temperature $T_S$ and drain temperature $T_D$ are fixed. The change of $T_G$ can influence the flux into the drain by $q_D=q_G-q_S$. Flux amplification requires $\left|\partial q_D/\partial q_G\right|>1$. Reproduced with permission from Ref.~\cite{212}. Copyright 2017 AIP Publishing.}\label{4-4}
\end{figure}

More complicated functional devices can be realized if more than two physical fields are considered simultaneously.
Criado-Sancho and Jou~\cite{212} designed a thermoelastic heat switch and field effect transistor under the regime of conduction, radiation and thermal expansion of solid materials. The heterostructured device made of Ge and Si bars is placed juxtaposed to each other in the initial state (see Fig.~\ref{4-4}). The two materials have different thermal conductivities (both are temperature-dependent) and different thermal expansion coefficients. With the asymmetric thermal dilatation/contraction of the hotter/colder bars, a gap (and a temperature jump) appears between the bars and the heat transfer equation should be the Fourier's law (with an interface resistance) plus the radiation across the gap. The radiation term plays the key role here and its transport mode can be modulated by the gap width. The Stefan-Boltzmann law is valid when the gap is larger than the dominant thermal radiation wavelength (taking about 10~$\mu$m at room temperature). Near-field effect like photon tunneling by coupled evanescent waves must be considered, which is a violation of Stefan-Boltzmann law and usually can enhance the heat transport greatly~\cite{213}. The  mode transition of radiation (absence/far field/near filed) opens up the opportunity for 
flexible conductance modulation to design nonlinear devices. The authors~\cite{212} found that, in the Stefan-Boltzmann approximation, the flux--bias
curve can show the characteristic of a thermal switch for a certain range of $\Delta T$. The on and off states account for the absence and presence of the gap respectively. 
When near filed radiation is dominant, which can be estimated by a phenomenological heat transfer coefficient $h_i$, a sudden drop in the flux--bias curve occurs, meaning the NDTR exits. Further a thermal transistor model was proposed~\cite{212}, whose schematic diagram is shown in Fig.~\ref{4-4}(d), and the flux into the drain can be amplified compared with the flux from the gate. Here, NDTR requires that the two materials' thermal expansion coefficients only have a tiny difference, which echos the NDTR in a homojunction~\cite{112} mentioned in Section 2.
In addition, the radiative flux between the gap (at the interface) is $h_{i}(T_{iA}-T_{iB})$ and we can see $h_i$ behaves just like the reciprocal of ITR ($R_i$). Actually, the radiative heat transfer coefficient $h_i$ here satisfies~\cite{212}
\begin{equation}\label{ritr}
h_i^n=A\exp(-L_i^n/C)+B
\end{equation}
where $n,A,B$ and $C$ are phenomenological parameters ($n<1$), and $L_i$ is the distance between the bars. Eq.~(\ref{interface}) and Eq.~({\ref{ritr}) both indicate the interface resistance increases with a larger $L_i$ and has an implicit relationship with the temperature. We can see two types of sigmoid functions appear in Eq.~(\ref{interface}) and Eq.~({\ref{ritr}), i.e., the hyperbolic tangent function and the logistic function respectively. These may reveal some generalities in the design of NDTR.
		
		Similarly, thermal bistability and thermal memory have also been designed based on the transform between far field and near field modes in thermo-mechanical systems including a stem with thermal expansion, while the radiative flux is calculated more carefully through fluctuational electrodynamics~\cite{213}. Also, Reina {\it et al.}~\cite{214} used a bendable cantilever to realize thermal bistability. The bilayerd cantilever, attached to a wall and governed by Euler-Bernoulli equation, is made of SiO$_2$ and the widely-used PCM VO$_2$, and in near field radiative interaction with a substrate. The phase transition of the cantilever enables two possible stable states for this conductive-radiative heat transfer coupled with elasticity. Though the separation distance between the radiators can be mirco/nano-scaled and the first-principle analysis of near field radiation is often needed, the size of the bars/stems/cantilever can be in millimeter or centimeter level~\cite{212,213,214}.

		\section{Effective medium theory for nonlinear composites}
		
		\noindent Since the transformed thermal conductivity is usually inhomogeneous and anisotropic that can't be found in naturally occurring materials, we have seen examples in which multilayered structures made of common materials are used instead~\cite{62,186}. Also, parameters solved out directly from the heat equation often can't be matched simply by existing materials. All these problems increase the demand to predict the effective properties of structured composites accurately (i.e., homogenization~\cite{216b}), especially the effective thermal conductivities, for both linear and nonlinear elements. The common case is that the heat equation in inhomogeneous materials is difficult to be solved out analytically and exactly, and we have to sacrifice some precision. Effective medium theory (EMT) or effective medium approximations (EMA) might be most widely-used analytical method to calculate the effective or average properties of systems including the electromagnetic, elastic, and thermal domains~\cite{216b,216,217}. In heat transfer especially linear heat conduction, EMT has been applied to core-shell structure~\cite{152}, porous structure~\cite{217}, heterostructures~\cite{218}, and even rotating structures~\cite{219}, in addition to the layered structures. Though EMT is usually concerned with the mesoscopic and macroscopic scales, it has also been found feasible in some models for heat conduction in nanofluids when the Fourier's law might even break down~\cite{220,221}.
		
		Here, we focus on some recent works studying the calculation of effective nonlinear thermal conductivity for different structured composites, particularly, the nonlinearity enhancement phenomenon. We will talk about the extension of three basic models in EMT to nonlinear heat conduction, including the Maxwell Garnett (M\&G) theory, the Bruggeman theory, and the Rayleigh method.  Because the steady Fourier's law and electrical conduction equation has the same Laplace type (the temperature or electric potential behave as the general potential), the effective medium theory for thermal and electrical 
		conductivities can share similar conclusions in the linear regime. However, the nonlinear electrical conductivity is usually affected by the electric field or the gradient of potential, whereas its thermal counterpart depends on the potential directly. This difference would make the calculation in nonlinear heat conduction quite complicated or even invalid if we directly apply the methods in nonlinear electrical conduction~\cite{275}.  
		
		\begin{figure}[!ht]
			\centering
			\includegraphics[width=.8\linewidth]{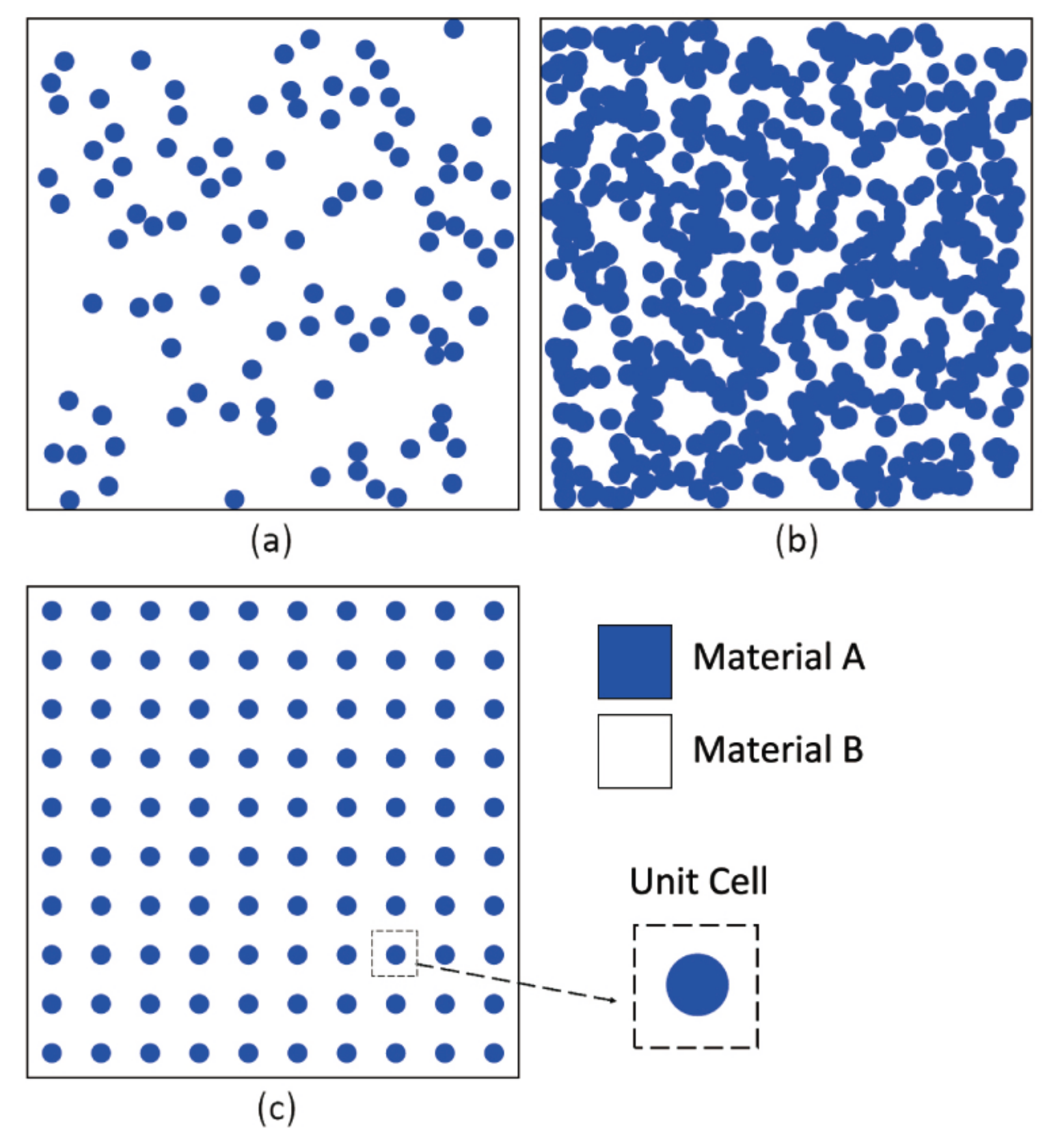}
			\caption{Three basic models of binary composites corresponding to (a) the M\&G theory, (b) the Bruggeman theory, and (c) the Rayleigh method. In (a) and (b), the identical circles (Material A) are randomly put into the host (Material B), with a difference in whether the inclusions can tough. In (c), a square array of identical circles are embedded into the host.}\label{5-1}
		\end{figure}

		\subsection{M\&G and Bruggeman theories}
		
		\noindent The M\&G~\cite{223} and Bruggeman theories~\cite{216} are the most famous models in EMT with some essential differences. Sometimes the EMT specifically refers to the Bruggeman theory under the ``self-consistency'' assumption~\cite{216b,216}. For two-component linear composites consisting of Material A and Material B in the two-dimensional space, the M\&G theory (Eq.~(\ref{e1a})) and the Bruggeman theory (Eq.~(\ref{e1b})) can be respectively expressed as~\cite{222}
		\begin{subequations}\label{e1}
			\begin{align}
			\frac{\kappa_{e}-\kappa_{B}}{\kappa_e+\kappa_{B}} &= f_{A} \frac{\kappa_{ A}-\kappa_{B}}{\kappa_{A}+\kappa_{B}}, \label{e1a}\\
			f_{A} \frac{\kappa_e-\kappa_{A}}{\kappa_e+\kappa_{A}} &+ f_{B} \frac{\kappa_e-\kappa_{B}}{\kappa_e+\kappa_{B}}=0.\label{e1b}
			\end{align}
		\end{subequations}
		Here $\kappa_A$ and $\kappa_B$ ($f_A$ and $f_B$) are respectively the thermal conductivities (area fractions) of Material A and Material B, and there must have $f_A+f_B=1$. In principle, the effective thermal conductivity $\kappa_e$ is defined through the Fourier's law over the whole composite:
		\begin{equation}\label{dengxiaordl}
		-\kappa_e\langle\nabla_x T\rangle=\langle j_x\rangle.
		\end{equation}
		Here we take $\kappa_e$ as a scalar for simplicity and only care about the $x$-direction component of the spatial average flux $\langle j_x\rangle$ and temperature gradient $\langle\nabla_x T\rangle$, if the thermal bias is put along the $x$ direction. It's well known that the Bruggeman theory is symmetric for Material A and Material B~\cite{216}, saying that the formula keeps invariant if exchanging the subscripts in Eq.~(\ref{e1b}), so we can't tell which material is the host and which material is the inclusions.  In this sense, the M\&G theory is asymmetric with a clear picture where the inclusions are embedded into the host. In addition, the M\&G theory is usually considered only applicable to small $f_A$ if Material A represents the inclusions~\cite{216}. It's also the exact solution when there exists only one circular particle in an infinite host. However, for the Bruggeman theory, there is no such a strict restriction on the area fraction, and percolation due to the clusters of (thermal) conductors can exist in the Bruggeman composites~\cite{216}.

		Based on many-particle structures described by the M\&G and Bruggeman theories, researchers have successfully designed some linear thermal metamaterials with functions like thermal illusion or thermal camouflage~\cite{224,225,226,227}.
		Dai {\it et al.}~\cite{228} considered a nonlinear conductive composite model with different structures produced by computer-aided design. In their binary composites, circular inclusions are made of Material A and would be randomly put into a rectangular host made of Material B. If all be inclusions don't overlap, this case obviously corresponds to M\&G theory. On the contrary, when inclusions can have overlapping areas during the process of filling (though they can't really overlap in the physical space), the structure generated with many connected domains (clusters of inclusions) satisfies the Bruggeman theory as the clusters make it difficult to distinguish the roles of the two materials. Firstly, their simulation results validated the feasibility of this model for linear composites, and percolation did happens in overlapping structures. 
		Then the question is how to incorporate nonlinearity into this model and corresponding effective medium theories. They assumed that $\kappa_A(T)$ and $\kappa_B(T)$ can be generally written as
		\begin{subequations}
			\begin{align}
			\kappa_A(T)=\kappa_{A0}+\chi_A T^{n_A}, \\
			\kappa_B(T)=\kappa_{B0}+\chi_B T^{n_B},
			\end{align}
		\end{subequations}
		where $\kappa_{A0},\kappa_{B0}$ are constants and $\chi_{A},\chi_{B}$ are the nonlinear coefficients. Here a reference temperature $T_{\text{ref}}$ can also be added into thermal conductivities, meaning replacing $T$ by $(T-T_{\text{ref}})$, which wouldn't change the final results. Then they directly took these expressions of conductivities into Eq.~(\ref{e1}). If the temperature is the same everywhere, this procedure is strict.  Anyway, a temperature-dependent effective thermal conductivity $\kappa_e$ can thus be obtained. However, if we want to check the nonlinearity enhancement, whose optical counterpart is quite important~\cite{222}, $\kappa_e$ should have the form:
		\begin{equation}
		\kappa_e(T)=\kappa_{e0}+\chi_e T^{n_e}+...
		\end{equation}
		Here $\kappa_{e0}$ is the constant or linear part of $\kappa_e(T)$, and $\chi_e$ is the effective nonlinear coefficient. In addition, if the higher order terms of temperature could be ignored, a further assumption of weak or perturbative nonlinearity is needed. Now, the problem is well-posed when only one material is nonlinear, and $\chi_e$ can be calculated through series expansion. For example, when $\chi_B=0$ (denoted as Case I), they got~\cite{228}
		\begin{equation}
		\frac{\chi_e}{\chi_{A}}  \approx\frac{\partial \kappa_e}{\chi_{A}T^{n_A}\partial \chi_{A}}=\frac{4 f_{A}}{\left(1+\frac{\kappa_{A0}}{\kappa_{B0}}+f_{A}-f_{A}\frac{\kappa_{A0}}{\kappa_{B0}}\right)^2}
		\end{equation}
		from the M\&G theory. When $\chi_A=0$ (i.e., Case II), the M\&G theory gives 
		\begin{equation}
		\begin{aligned}
		\frac{\chi_e}{\chi_{B}}=\frac{(1-f_{A}^2)\left[1+\left(\frac{\kappa_{A0}}{\kappa_{B0}}\right)^2\right]+2(1-f_{A})^2\frac{\kappa_{A0}}{\kappa_{B0}}}{\left(1+\frac{\kappa_{A0}}{\kappa_{B0}}+f_{A}-f_{A}\frac{\kappa_{A0}}{\kappa_{B0}}\right)^2}.
		\end{aligned}
		\end{equation}
		Similarly, the corresponding results for nonlinear coefficients they obtained from the Bruggeman theory are
		
		\begin{widetext}
			\begin{subequations}\label{bru-c}
				\begin{align}
				\frac{\chi_e}{\chi_{A}} &= \frac{1}{2} \left[\frac{\left(2 f_{A}-1\right) \left(2 f_{A} -2 f_{A} \frac{\kappa_{B0}}{\kappa_{A0}}-1+\frac{\kappa_{B0}}{\kappa_{A0}}\right)+2 \frac{\kappa_{B0}}{\kappa_{A0}}}{ \sqrt{\left(2 f_{A} -2 f_{A} \frac{\kappa_{B0}}{\kappa_{A0}}-1+\frac{\kappa_{B0}}{\kappa_{A0}}\right){}^2+4 \frac{\kappa_{B0}}{\kappa_{A0}}}}+2 f_{A}-1\right], \quad {\mbox{if only Material A is nonlinear,}}\label{bru-c-1}\\
				\frac{\chi_e}{\chi_{B}} &= \frac{1}{2}  \left[\frac{\left(2 f_{A}-1\right) \left(2 f_{A} -2 f_{A} \frac{\kappa_{A0}}{\kappa_{ B0}}-1+\frac{\kappa_{A0}}{\kappa_{B0}}\right)+2 \frac{\kappa_{ A0}}{\kappa_{B0}}}{ \sqrt{\left(2 f_{A} -2 f_{A} \frac{\kappa_{ A0}}{\kappa_{B0}}-1+\frac{\kappa_{A0}}{\kappa_{ B0}}\right){}^2+4 \frac{\kappa_{A0}}{\kappa_{ B0}}}}-2 f_{A}+1\right],\quad {\mbox{if only Material B is nonlinear.}}\label{bru-c-2}
				\end{align}
			\end{subequations}
		\end{widetext}
		
		\noindent We can see the two formulas in Eq.~(\ref{bru-c}) are also symmetric when exchanging the subscripts, and the value $\chi_e/\chi_A$ or $\chi_e/\chi_B$ measuring the nonlinearity enhancement only relies on the area fraction $f_A$ and the ratio of linear parts $\frac{\kappa_{A0}}{\kappa_{ B0}}$.
		In addition, $\kappa_{e0}$ is just the solution when taking $\kappa_{A0}$ and $\kappa_{B0}$ into Eq.~(\ref{e1}).
		The predictions of effective nonlinear coefficients are plotted in Fig.~\ref{5-2} for different values of $\kappa_{A0}/\kappa_{B0}$. The condition for nonlinearity enhancement in the M\&G and Bruggeman composites are quite different. For M\&G theory, nonlinearity enhancement happens when the host (Material B) is nonlinear and $\kappa_{A0}/\kappa_{B0}$ larger than 1. For the symmetric Bruggeman theory, nonlinearity enhancement can exist in both the cases. Let's turn back to the debate that the temperature varies in space so the results should not be strict. Here the argument is that the nonlinearity is weak so temperature-dependent conductivity can approximately satisfies M\&G and Bruggeman's equations.

		\begin{figure}[!ht]
			\centering
			\includegraphics[width=1.\linewidth]{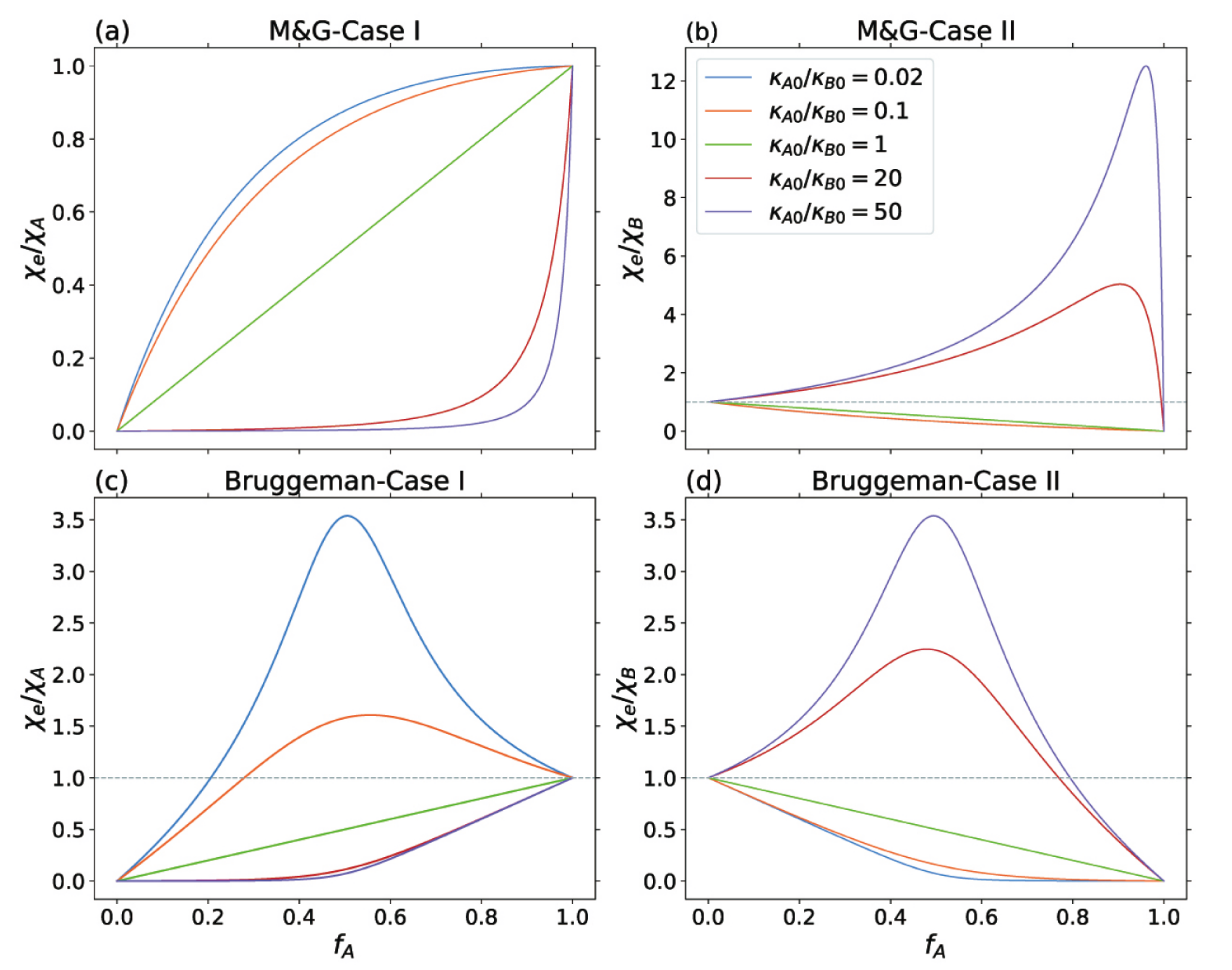}
			\caption{$\chi_e/\chi_A$ or $\chi_e/\chi_B$ against the area fraction $f_A$ given by two EMT models. (a) The M\&G composites with nonlinear inclusions and a linear host. (b) The M\&G composites with linear inclusions and a nonlinear host. (c) The Bruggeman composites with nonlinear inclusions and a linear host. (d) The Bruggeman composites with linear inclusions and a nonlinear host. Reproduced from Ref.~\cite{229}.}\label{5-2}
		\end{figure}

		\subsection{The Rayleigh method}
		
		\noindent The structures above treated with the M\&G and Bruggeman theories are disordered as the inclusions are embedded randomly. Due to the mean filed effect, these theories actually demonstrate the discrete dipole approximation representing the interaction between different materials~\cite{223}. Another important structure in metamaterials is the periodic lattices, which can be exploited to realize novel functions such as thermal transparency and Janus thermal illusion~\cite{226,227}, or to construct a thermocrystal~\cite{264}. However, we have learned from periodic electromagnetic composites that the M\&G and Bruggeman theories would give biased estimations in this situation due to the significant multipole interaction between the close inclusions~\cite{216}.

		The Rayleigh method was developed specifically to handle this dilemma, and has been used in photonic, phononic, and electric systems~\cite{216,230,231,232}. If we consider two-dimensional infinite lattices in which circular inclusions (denoted as Material A again) are periodically arranged in the form of square lattices with a uniform  thermal field (temperature gradient)
		applied along the $x$ direction, then each unit cell containing one inclusion in the center could be treated equally under linear conduction as 
		the temperature distribution in them should differ only by a constant. Thus solving the periodic heat equation can be simplified as classical single particle problem with some unique boundary conditions. When the particle is small enough, we can see the problem is reduced to the M\&G theory. However, in general, the conventional boundary conditions at infinity in the dilute limit are not valid for general inclusion concentration because both the host (also denoted as Material B) and the inclusions can appear at infinity and we can't know the corresponding thermal filed there. Instead, the Rayleigh identity is used to measure how other unit cells, especially the inclusions in them, interact with a chosen cell. The Rayleigh identity or the field identity is a conclusion of the superposition property  of linear differential equations~\cite{231,232}. If we choose one unit cell (named as the 0-th cell) and establish the polar coordinates $(r_0,\theta_0)$ with the pole in its center, the temperature solution in the inclusion ($T_0^A$) can be generally expressed using a simple version of the Bessel-Fourier series~\cite{232}
		\begin{equation}\label{inclusion}
		\begin{aligned}
		T^A_0(r,\theta)&=C_{00}+\sum_{m=1}^{\infty}\big[C^2_{0m} r_0^m \cos(m\theta_0)\\&+C^1_{0m} r_0^m \sin(m\theta_0)\big].
		\end{aligned}
		\end{equation}
		Similarly, we can also write the general solution in the host ($T_0^B$) as~\cite{232}
		\begin{equation}\label{host}
		\begin{aligned}
		T^B_0(r,\theta)&=A_{00}+\sum_{m=1}^{\infty}\big[(A^2_{0m} r_0^m+B^2_{0m} r_0^{-m}) \cos(m\theta_0)\\&+(A^1_{0m} r_0^m+B^1_{0m} r_0^{-m}) \sin(m\theta_0)\big].
		\end{aligned}
		\end{equation}

		Then Rayleigh's inference is based on which factors could influence the solution of $T_0^B$ in the host~\cite{231}. When no inclusions exist, $T_0^B$ should be evenly spaced as $-x(T_L-T_R)/L+\text{Constant}$. Here $L$ is the length of the whole composite though we use the assumption of infinite lattices, $T_L$ and $T_R$ are heat source put on he left or right along the $x$ direction, and $(T_L-T_R)/L$ is the applied thermal field. But such a uniformly varying distribution should be disturbed by all the inclusions. The terms with $r_0^{-m}$ in Eq.~(\ref{host}) are divergent at $r_0=0$, so they're related to the inclusion in the 0-th cell. The scattering effect from other inclusions should have the similar form with translational symmetry, so the general solution for the host in the 0-th cell can have another expression~\cite{232,233}
		\begin{equation}\label{sub}
		\begin{aligned}
		T^B_0 &= \sum_{k=0}^{\infty} \sum_{m=1}^{\infty} r_k^{-m}\big[B^2_{0m} \cos(m\theta_k)+B^1_{0m}\sin(m\theta_k)\big]\\&-\frac{T_L-T_R}{L}x.
		\end{aligned}
		\end{equation}
		Here the coordinates $(r_k,\theta_k)$ are established in the $k$-th cell with the pole in the center, and the two parts of Eq.~(\ref{sub}) represent the influence of the inclusions and the host itself respectively. Each term in the sum $\sum_{k=0}$ is the local expansion in different unit cells and all of them can satisfy the conduction equation in the absence of the boundary conditions. 
		
		Then the Rayleigh identity can be obtained from comparing Eq.~(\ref{host}) and Eq.~(\ref{sub}) as they must give the same result, 
		and finally all the non-zero coefficients in Eqs.~(\ref{inclusion}-\ref{host}) can be solved with this identity and the interface condition for two materials in the 0-th cell~\cite{232}. 
		Thus the Rayleigh method is a kind of first principle approach as we are intended to solve out the temperature distribution based on the governing equation, while there are still three approximations that should be made, i.e., truncating the coefficients $A_{m0},B_{m0}$ and $C_{m0}$, doing numerical calculations for the sum of an infinite number of unit cells, and using the self-consistent mean field method to obtain the effective properties. Though the value of $C_{00}$ in Eq.~(\ref{inclusion}) can't be determined without giving extra information, the effective linear thermal conductivity can still be calculated, given by~\cite{232,233}
	
		\begin{widetext}
			\begin{equation}\label{r}
			\kappa_e =\kappa_{B0} \frac{(-\beta_{1}+\beta_{1}f_A+f_A^4)\kappa_{B0}^2-2 (\beta_{1}+f_A^4)\kappa_{B0}\kappa_{A0}+(-\beta_{1}-\beta_{1}f_A+f_A^4)\kappa_{A0}^2}{(-\beta_{1}-\beta_{1}f_A+f_A^4)\kappa_{B0}^2-2(\beta_{1}+f_A^4)\kappa_{B0}\kappa_{A0}+(-\beta_{1}+\beta_{1}f_A+f_A^4)\kappa_{A0}^2}.
			\end{equation}
		\end{widetext}
		
		\noindent Here $\beta_1=3.31248$ and it comes from a lattice sum over all unit cells. As the inclusions are circular, here the area fraction $f_A$ has a upper limit with $\pi/4$ and Eq.~(\ref{r}) would be divergent if $f_A\geq\pi/4$ accounting for the overlapping inclusions. This formula has the same form as the one in linear electrical conduction which is not strange as they are both derived from periodic Laplace equations~\cite{232,233}.
		
		For electrical conduction with weak nonlinearity, the perturbation method and high-order Rayleigh identities can be developed to find the approximate analytical solution of temperature's high order small quantities~\cite{232}. However, this technique isn't suitable for nonlinear heat conduction because the uncertainty of $C_{00}$ (or the presence of the reference temperature $T_{\text{ref}}$) and the direct dependence of temperature in thermal conductivity can make the solving procedure difficult to proceed. Inspired by the truncated series expansion in the M\&G and Bruggeman theories~\cite{238}, Dai and Huang~\cite{233} gave the effective nonlinear coefficient for periodic composites for the cases when only one material is nonlinear, saying
		
		\begin{widetext}
			\begin{subequations}\label{reuili}
				\begin{align}
				\frac{\chi_e}{\chi_A} &= \frac{4 \beta_{1}  f_{A} \left[\beta_{1}  (\lambda+1)^2+f_{A}^4
					(\lambda-1)^2\right]}{\left[\beta_{1}  (\lambda+1) (f_{A}
					-f_{A}\lambda+\lambda+1)-f_{A}^4
					(\lambda-1)^2\right]^2},\quad {\mbox{if only Material A is nonlinear,}}\label{ruili1}\\
				\frac{\chi_e}{\chi_B}&=\frac{-\beta_{1} ^2 (f_{A}-1) (\lambda+1)^2 \left[(f_{A}+1) \lambda^2-2
					(f_{A}-1) \lambda+f_{A}+1\right]-2 \beta_{1} 
					f_{A}^4 (\lambda-1)^2 \left[2 (f_{A}+1) \lambda+\lambda^2+1\right]
				}{\left[\beta_{1}  (\lambda+1) (f_{A}
					-f_{A}\lambda+\lambda+1)-f_{A}^4
					(\lambda-1)^2\right]^2}\notag\\&+\frac{f_{A}^8 (\lambda-1)^4}{\left[\beta_{1}  (\lambda+1) (f_{A}
					-f_{A}\lambda+\lambda+1)-f_{A}^4
					(\lambda-1)^2\right]^2}, \quad {\mbox{if only Material B is nonlinear.}}\label{ruili2}
				\end{align}
			\end{subequations}
		\end{widetext}
		
		\noindent Here the ratio $\lambda=\kappa_{A0}/\kappa_{B0}$ is used for simplification, the thermal conductivities of the two materials have the same form as those used in the M\&G and Bruggeman theories, and the nonlinear conductivity is still assumed to be much smaller than the linear part.
		For Eq.~(\ref{ruili1}), the inclusion (Material A) is nonlinear (Case I) while the host (Material B) is nonlinear (Case II) for Eq.~(\ref{ruili2}). In Fig.~\ref{5-3}, we illustrate the predictions of the Rayleigh method for the two cases. We can see nonlinearity enhancement can happen in both cases. Particularly, in Case I, nonlinearity enhancement requires a small $\lambda$ with a threshold close to $1/3.5$, whereas it corresponds to a large $\lambda$ in Case II, approximately more than 2.5~\cite{229}.
		Also, they validated the predictions with simulation results through FEM~\cite{233}.

		\begin{figure}[!ht]
			\centering
			\includegraphics[width=1.\linewidth]{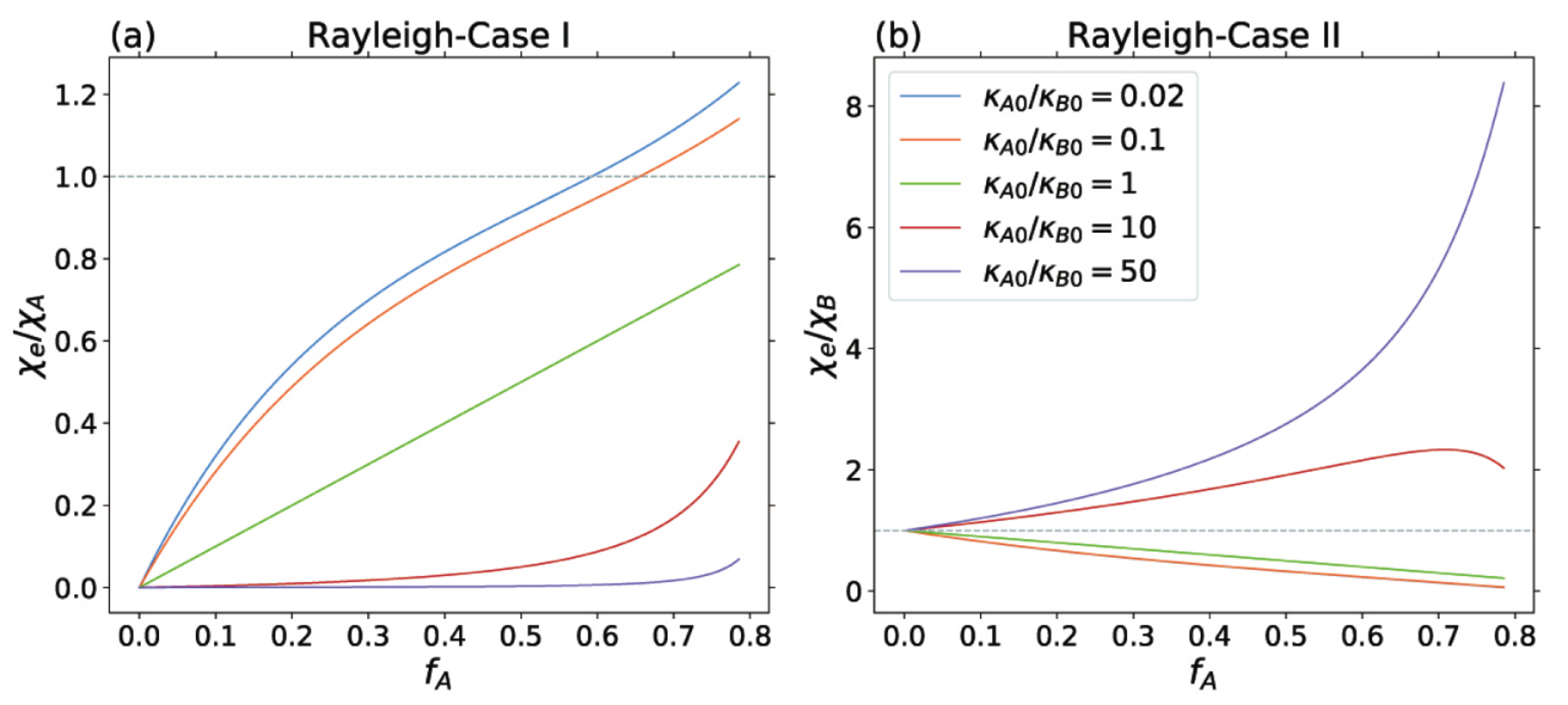}
			\caption{$\chi_e/\chi_A$ or $\chi_e/\chi_B$ against $f_A$ given by the Rayleigh method. (a) Periodic composites with nonlinear inclusions and a linear host. (b) Periodic composites with linear inclusions and a nonlinear host. Reproduced from Ref.~\cite{229}.}\label{5-3}
		\end{figure}

		In Fig.~\ref{5-4}, the three basic EMT models are compared under different parameters for periodic composites. We can see the Rayleigh method is the most accurate for the whole range of $f_A$, the M\&G theory performs well if $f_A$ is not large, and Bruggeman theory only works when $f_A$ is very small. Actually, the conclusion is the same for linear conductivity. Periodic structures are of course not symmetric. In addition, as we should consider the impact of all the inclusions on the host which are also interacting with each other, the mean effect can never be as simple as a dipole and the M\&G is not valid for large $f_A$. In fact, the Rayleigh method is also called a multipole method~\cite{230,231}. Especially, when $f_A\to 0$, the M\&G theory and the Rayleigh method can have the same dilute limit. However, it's quite interesting that, when $f_A$ is approaching critical volume fraction $\pi/4$, the inclusions begin to touch each other and the curves standing for the Rayleigh method in Fig.~\ref{5-3} or Fig.~\ref{5-4} show some percolation-like behaviors. For example, the curve for $\kappa_{A0}/\kappa_{B0}=50$ rises sharply near $f_A=\pi/4$ in Fig.~\ref{5-3}(b), and the curves for $\kappa_{A0}/\kappa_{B0}=$0.02 or 0.1 can break through the horizontal line representing ``1'' in Fig.~\ref{5-3}(a) and Fig.~\ref{5-4}(b), while the M\&G theory fails to predict this nonlinearity enhancement in these cases. From this perspective, the Rayleigh method and Bruggeman theory are a bit like.
		
		\begin{figure}[!ht]
			\centering
			\includegraphics[width=1.\linewidth]{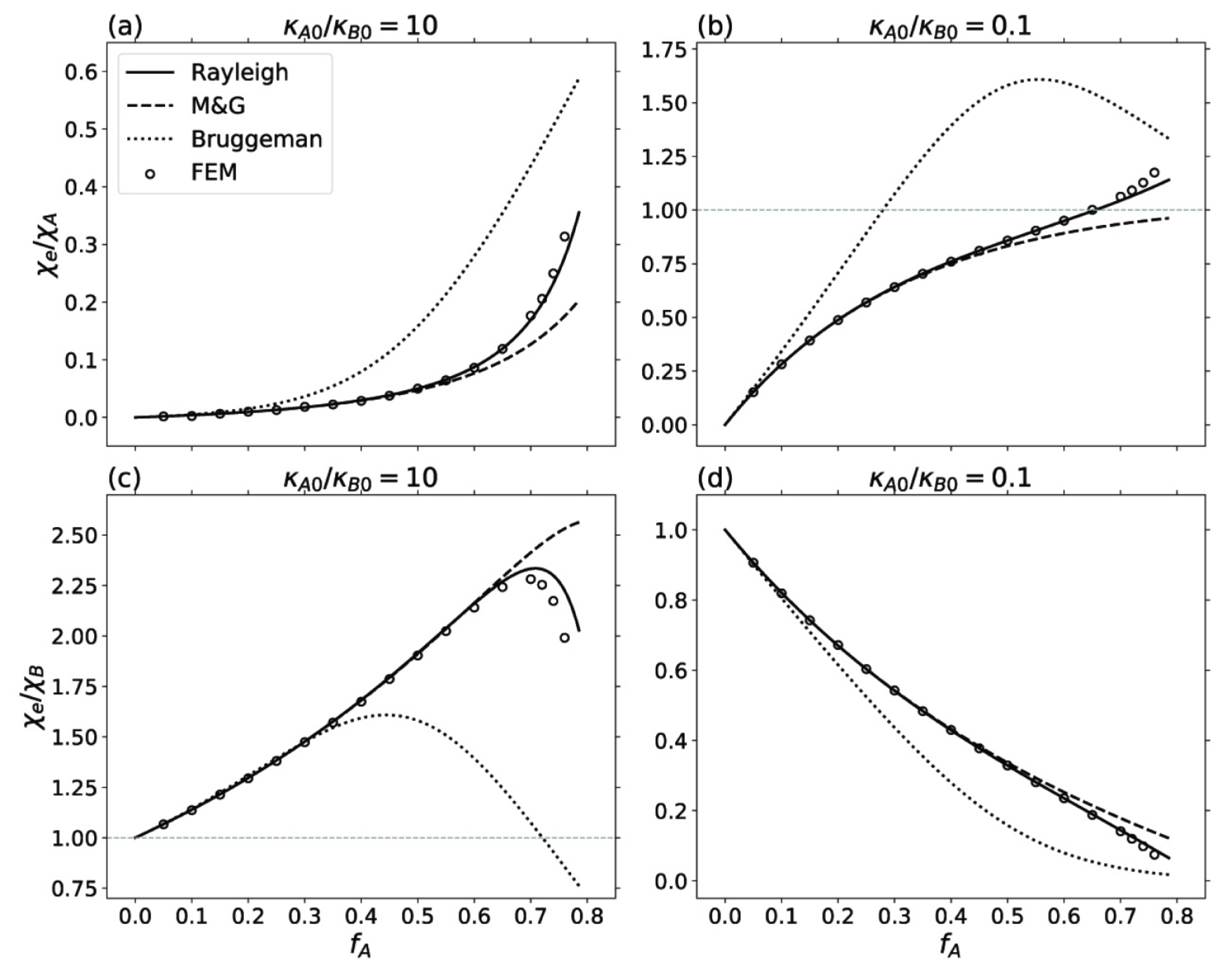}
			\caption{Simulation results of $\chi_e/\chi_A$ or $\chi_e/\chi_B$ against $f_A$ given by FEM for periodic composites. (a--b) Nonlinear inclusions embedded into a linear host. (c--d) Linear inclusions embedded into a nonlinear host. The predictions from three basic EMT models are also plotted. Reproduced with permission from Ref.~\cite{233}. Copyright 2019 Elsevier Ltd.}\label{5-4}
		\end{figure}
		
		Finally in this section, we would like to give some remarks on the above works dealing with the effective conductivities of nonlinear composites. The M\&G and Bruggeman theories can describe the mixing effect of composites with random or disordered features. Generally, their equations can be applied to inclusions with other shapes and extended to cases when more than two materials are mixed~\cite{216}. The Rayleigh method is especially feasible for periodic composites, but the formulas derived should be different when the symmetry of the lattice or the shape of the inclusions changes~\cite{235a,235b,235c,236}.
		Also, the works we have introduced are based on weak nonlinearity and some kind of perturbation method is used if we have known the results in linear models, which can be feasible in the core-shell structures as well~\cite{234}. To obtain a nonlinear coefficient, another solvable model is that the two materials both have overwhelming nonlinearity, and the effective nonlinear coefficient just behaves like the effective linear conductivity under the three basic effective medium theories. Though the temperature-responsive conductivity can have drastical spatial changes, through FEM simulations, e.g., for periodic composites with, we can see the EMT can still have accurate predictions and an explanation can be given borrowing Kirchhoff transformation~\cite{229,233}.

		\section{Summary and perspectives}
		
		\noindent To summarize, we have presented the recent progress of designing nonlinear thermal devices and metamaterials under the Fourier's law in the current review. The nonlinearity here mainly refers to a temperature-dependent thermal conductivity. We talked about two basic designing methods, i.e., solving the heat equation directly and the transformation thermotics. Also, exploiting the nonlinear phenomena in multiphysics and the effective medium theory for nonlinear conductive composites were included in our discussion. We can see nonlinear heat transport can provide thermal metamaterials and devices with more active and tunable functions just as their counterpart in other domains~\cite{237,238,239,240,241,242}. These works might be concluded as a solid phenomenological road towards the nonlinear thermotics, an emerging discipline that deserves more concern to improve people's ability to manipulate heat transport.
		
		In particular, radiation, convection and other field effects like thermal expansion can enrich the mechanisms of nonlinear behaviors in addition to modulating the conduction only. Most of the works we have introduced focus on the direct analysis from the equations. How to transform the multiphysics equations, especially the nonlinear coupled ones, is still a problem as we have known some important equations in continuum mechanics are not form-invariant under arbitrary coordinate (geometric) transformations, although transformation theory has been extended to some coupled dual-physics cases such as thermal convection in the porous media~\cite{248,249} and thermoelectric Seebeck effect~\cite{250} whose governing equations can be linear or nonlinear. The importance of considering multiphysics effects is also reflected in that the real environment is always in the interaction of multiple physical fields, and the performance and robustness of the elements can be enhanced with more detailed and comprehensive designs.

		What's more, the methods of solving or transforming a equation can be regarded as a whole to deal with an inverse problem seeking for suitable physical quantities for a certain thermal phenomenon. Actually, transformation theory has an intrinsic relationship with the famous inverse problem (Calderón's problem) to hide an object in electrical impedance tomography~\cite{245,246,247}. We haven't covered numerical optimization in this review, the general and powerful tool in inverse problems, which have been used in inverse design of linear thermal metamaterials and other functional devices~\cite{251,252,253,254,255,255b,255c}. Some works have also studied multi-physical~\cite{253b,253c} cases including nonlinear thermo-mechanical metadevices~\cite{253c}. Since it's usually not easy to do analytical calculations in nonlinear equations and the transformation theory might have restrictions on the forms of equations and target solutions, numerical optimization based on various algorithms, whether they're gradient-based~\cite{252,255b,255c,253c} or black boxes~\cite{253,254,255,253b}, can have potential applications to more flexible designs of nonlinear thermal elements.
		
		Last but not least, we want to talk about the possible heat waves in nonlinear thermal elements. Tuning the wave nature of heat conduction is often difficult and even controversial~\cite{16,17}, and might require a multi-scale analysis from macroscopic heat conduction to its microscopic heat carriers. 
		On one hand, the wave nature itself in the absence of nonlinearity is worth studying as researchers have accumulated a lot of experience on how to use artificial structures to control the propagation of waves.
		On the other hand, the diffusive heat conduction usually need very strong nonlinearity to generate corresponding phenomena like bistability and switchable functions. Conversely, nonlinear metamaterials in conventional wave systems like optics and acoustics can utilize more general nonlinear feature such as high harmonic generation, which can be induced by a perturbative nonlinearity. 
		Phenomenological non-Fourier heat conduction were first widely studied to reveal heat waves~\cite{16,17}, and some models can even reveal the memory effect~\cite{17}. 
		Although some recent works began to tailor thermal metamaterials in heat wave systems like the Maxwell-Cattaneo-Vernotte model~\cite{256,257}, nonlinear heat transfer hasn't been involved yet. Nonlinearity in such non-Fourier models can simply be incorporated by temperature-dependent quantities like specific heat capacity, thermal conductivity or relaxation times, or by introducing extra nonlinear terms through extended irreversible thermodynamics~\cite{258}, phonon hydrodynamics~\cite{260}, and other filed effects~\cite{259}. Another aspect of tuning the wave nature is focusing on the ballistic transport of low-frequency phonons~\cite{261,262,263}, especially at low temperatures. In this way, heat conduction is sometimes essentially no different from elastic waves and can be modulated by phononic crystals~\cite{264,265,266}. It's interesting that designing nonlinear thermal elements in these systems would turn back to the nonlinear lattices models but in a continuum. Recently, wave-like temperature profile in convection (advection-conduction) systems is drawing people's attention~\cite{23,267,268,269,269b}. In addition to breaking the reciprocity~\cite{23}, physics related to exceptional degeneracies like anti--parity-time (APT) symmetry breaking~\cite{267,268} have also been revealed in linear equations. When nonlinearity also considered, some new phenomena might be found in APT systems or more general convection-reaction-diffusion systems.
		
		\acknowledgements{The author thanks Jun Wang for beneficial discussions.}

		
		\raggedend
		\begin{small}
			
		\end{small}
	\end{document}